\documentclass[pre,superscriptaddress,floatfix,showpacs]{revtex4-1}
\usepackage[utf8]{inputenc}
\usepackage{amsmath}
\usepackage{amssymb}
\usepackage{graphicx}
\usepackage{hyperref}

\makeatletter
\usepackage{graphics}
\usepackage{epsfig}
\usepackage{color}

\newcommand{\be}{\begin{equation}} \newcommand{\ee}{\end{equation}}
\newcommand{\bea}{\begin{eqnarray}} \newcommand{\eea}{\end{eqnarray}}

\begin{document}

\title{Synaptic shot-noise triggers
fast and slow global oscillations \\  in balanced neural networks}
\date{\today}

     \author{Denis S. Goldobin}
      \affiliation{Institute of Continuous Media Mechanics, Ural Branch of RAS, Acad.\ Korolev street 1, 614013 Perm, Russia}
\affiliation{Institute of Physics and Mathematics, Perm State University, Bukirev street 15,
614990 Perm, Russia}
 \author{Maria V. Ageeva}
      \affiliation{Institute of Continuous Media Mechanics, Ural Branch of RAS, Acad.\ Korolev street 1, 614013 Perm, Russia}
\author{Matteo di Volo }
\affiliation{Universit\'e Claude Bernard Lyon 1, Institut National de la Sant\'e et de la Recherche M\'edicale, Stem Cell and Brain Research Institute U1208, Bron, France}
\author{Ferdinand Tixidre}
 \affiliation{Laboratoire de Physique Th\'eorique et Mod\'elisation, CY Cergy Paris Universit\'e, CNRS, UMR 8089, 95302 Cergy-Pontoise cedex, France}
                \author{Alessandro Torcini}
                \email[corresponding author: ]{alessandro.torcini@cyu.fr}
                 \affiliation{Laboratoire de Physique Th\'eorique et Mod\'elisation, CY Cergy Paris Universit\'e, CNRS, UMR 8089, 95302 Cergy-Pontoise cedex, France}
        \affiliation{CNR - Consiglio Nazionale delle Ricerche - Istituto dei Sistemi Complessi, via Madonna del Piano 10, 50019 Sesto Fiorentino, Italy}
          \affiliation{INFN Sezione di Firenze, Via Sansone 1, 50019 Sesto Fiorentino, Italy}

\date{\today}

\begin{abstract} 
Neural dynamics is determined by the transmission of discrete synaptic pulses
(synaptic shot-noise) among neurons. However, the neural responses are usually obtained
within the diffusion approximation modeling synaptic inputs as continuous Gaussian noise.
Here, we present a rigorous mean-field theory that encompasses synaptic shot-noise for 
sparse balanced inhibitory neural networks driven by an excitatory drive.
Our theory predicts new dynamical regimes, in agreement with numerical simulations,
which are not captured by the classical diffusion approximation.
Notably, these regimes feature self-sustained global oscillations emerging 
at low connectivity (in-degree) via either continuous or hysteretic transitions
and characterized by irregular neural activity, as expected for balanced dynamics.
For sufficiently weak (strong) excitatory drive (inhibitory feedback)
the transition line displays a peculiar re-entrant shape
revealing the existence of global oscillations at low and high in-degrees, separated
by an asynchronous regime at intermediate levels of connectivity.
The mechanisms leading to the emergence of these global oscillations are distinct:
drift-driven at high connectivity and cluster activation at low connectivity.
The frequency of these two kinds of global oscillations can be varied from slow ($\sim 1$ Hz) 
to fast ($\sim 100$ Hz), without altering their microscopic and macroscopic features, by adjusting the excitatory 
drive and the synaptic inhibition strength in a prescribed way.
Furthermore, the cluster-activated oscillations at low in-degrees could correspond to the $\gamma$
rhythms reported in mammalian cortex and hippocampus and attributed to ensembles of
inhibitory neurons sharing few synaptic connections [{\it
G. Buzs\'aki and X.-J. Wang, Annual Review of Neuroscience 35, 203 (2012)}].
\end{abstract}

\maketitle

%%\beginsupplement

\section{Introduction}

Shot-noise (or Poissonian noise) is a model of discontinuous noise characterized by a sequence of discrete pulses occurring
at Poisson-distributed times. It is a ubiquitous process in Nature, observable
in electrical circuits and in optics due to the discrete nature of charges and photons \cite{sc1918,henry1996}, 
as well as in population dynamics due to the discreteness of individuals \cite{goncalves2023}.
Furthermore, the inclusion of shot-noise is fundamental to fully characterize the considered phenomena in many fields of physics ranging from granular systems \cite{kanazawa2017,lucente2023,lucente2025} to active matter \cite{fodor2018}, and from mesoscopic conductors \cite{blanter2000} to Anderson localization \cite{frisch1960}.

Neural dynamics is also intimately linked to the discrete nature of the synaptic events (post-synaptic potentials (PSPs))
regulating the firing activity of the neurons. In the cortex, the PSPs received by a neuron are usually assumed to be of small 
amplitude and to have high arrival rates, due to the high number of pre-synaptic connections. Furthermore, 
since the number of pre-synaptic neurons is small with respect to the large number of neurons present in the cortex, the PSPs stimulating a specific neuron are assumed to be uncorrelated. Therefore, the synaptic inputs have been usually treated as white noise (as a continuous Gaussian process) giving rise to the so-called Diffusion Approximation (DA) for the mean-field description of the neural dynamics \cite{capocelli1971,tuckwell1988}.

However, experimental results challenge the hypotheses at the basis of the DA.
First of all, there is clear evidence that synaptic weight distributions display
heavy tails towards large values \cite{miles1990,song2005,buzsaki2014}.
Moreover, rare large-amplitude synaptic connections are likely to contribute strongly to reliable information processing \cite{lefort2009} 
and the synaptic weight distributions can be deeply modified by plasticity during learning processes and ongoing activity
\cite{bi1999,barbour2007}. A second important aspect is that neural circuits
can display a connectivity definitely lower than expected. Indeed, networks of inhibitory neurons with 
very low number of pre-synaptic connections (in-degree $K \simeq 30-80$) have been found in the cat visual cortex
\cite{kisvarday1993} and in the rat hippocampus \cite{sik1995}, the latter population of interneurons
is believed to be responsible for the emergence of global oscillations (GOs) in the $\gamma$-band ($30-100$ Hz) \cite{buzsaki2012}.
Along this direction goes a recent study reporting that primate excitatory and inhibitory
neurons receive 2-5 times fewer excitatory and inhibitory synapses than similar mouse neurons \cite{wildenberg2021}.

These experimental indications call for the investigation of the dynamical regimes
emerging in highly diluted neural networks under the effect of synaptic shot-noise.
Several works have been devoted to the development of mean-field approaches 
describing the population dynamics of Integrate-and-Fire neurons subject to
synaptic shot-noise, though mostly focused on asynchronous regimes \cite{richardson2010, iyer2013, olmi2017,angulo2017,droste2017,richardson2018,richardson2024}.

However, oscillations are extremely relevant for the brain functioning, since they play a fundamental role
in orchestrating cognitive functions and are often disrupted in neurological disorders \cite{buzsaki2006,singer2018}.
Oscillations in the brain are typically characterized by irregular neural firing, where neurons fire at frequencies much lower 
than those of the GOs both {\it in vivo} \cite{csicsvari1998} and in {\it in vitro} \cite{fisahn1998}.
Several works have shown that this type of oscillations can arise in balanced 
excitatory-inhibitory sparse networks \cite{bal1}, due to the interplay between 
endogenous fluctuations and neural coupling \cite{brunel2000,brunel2003,geisler2005,bi2021}.
In particular, inhibition has been shown to play a critical role for the emergence of GOs, both in 
experimental data and in neural network models \cite{whittington2000,mann2007}. Therefore, 
GOs have often been analyzed within the simplified context of purely inhibitory networks, where the balance occurs
between an external excitatory drive and the recurrent inhibition \cite{brunel1999, bal4, wolf, matteo, noi}.
The characterization of the dynamical regimes exhibited by these models has been usually performed 
at a mean-field level within the context of the DA \cite{brunel1999,brunel2000}.
Only quite recently, some of the authors of this article have introduced in \cite{prl2024}
a mean-field approach incorporating shot-noise and sparsity in the connectivity
to describe the emergence of GOs in balanced inhibitory neural networks.
 
In the present paper, we report in details the derivation of the {\it complete mean-field approach} (CMF)
introduced in \cite{prl2024} for the Quadratic Integrate-and-Fire (QIF) neurons \cite{ermentrout1986,gutkin2022}.
Furthermore, we will extend the analysis in \cite{prl2024} to the derivation of the diffusion (DA) and third order approximation (D3A),  
obtained by expanding the source term in the continuity equation to the second and third order, respectively.
A weakly non-linear analysis of the stationary solutions displayed by the CMF allows to identify 
hysteretic and non-hysteretic transitions towards GOs with associated region of coexistence of GOs and
asynchronous states. These forecasts are in good agreement with accurate numerical simulations,
as shown via finite size analysis of the observed transitions. On the other hand, DA and D3A 
fail in reproducing the network dynamics for sufficiently low excitatory drive (high inhibitory feedback).
A peculiarity of the bifurcation diagram found with the CMF approach is a re-entrant Hopf bifurcation line
separating GOs observable at low and large in-degree $K$ from asynchronous irregular regime at intermediate $K$.
These two kinds of GOs, induced by synaptic shot-noise, are characterized, as the intermediate asynchronous state,
by neurons spiking irregularly, but they emerge due to two different mechanisms.
Moreover, both types of GOs can be observed at low frequencies (in the $\delta$-band ($0.5-3.5$ Hz))
and at high frequencies (in the $\gamma$-band) by properly rescaling the external excitatory drive
and the inhibitory synaptic strength, while the mean frequency of the irregularly firing 
neurons is always definitely slower.

The paper is organized as follows. Section II is devoted to the introduction of the network model, of the 
simulation protocols and of the indicators employed to characterize microscopic and macroscopic dynamics. The mean-field
formulation of the model is reported in Section III. In particular, we derive the macroscopic evolution equations for the CMF,
the diffusion and third order approximation, together with the corresponding linear and weakly non-linear stability analysis of the
asynchronous regime. The mean-field results are compared with network simulations in Sect. IV. This section reports
a characterization of the observed transitions and macroscopic regimes, as well as the explanation of the two mechanisms that are at the
origin of the two classes of GOs, induced  by discrete synaptic events, observable at small and large in-degree. A summary
and discussion of the results can be found in Sect. V. A detailed presentation of the integration methods employed for 
the network model and for the Langevin and continuity equations can be found in Appendix A-C. Finally Appendix D 
compares the results for the macroscopic dynamics obtained by network simulations
and mean-field analysis for non-instantaneous synapses.

\section{Model and Methods}

\subsection{The Network Model}

As a prototype of a dynamically balanced system we consider a sparse inhibitory network made of $N$
pulse-coupled QIF neurons \cite{wolf,matteo,noi} whose membrane potentials evolve according to the equations
\begin{equation}\label{eq:1}
\dot{v}_{i}(t) = v_{i}^2 + I - g \sum_{j=1}^{N} \sum_{n} \varepsilon_{ji} \delta(t - t_{j}^{(n)}) \quad v \in ]-\infty,+\infty[ ;
\end{equation}
where $I$ represents an external DC current, $g$ the synaptic coupling, and the last term on the rhs the inhibitory synaptic current. 
The latter is the linear superposition of instantaneous inhibitory PSPs emitted at times $t_{j}^{(n)}$ from
the pre-synaptic neurons connected to neuron $i$. $\varepsilon_{ji}$ is the adjacency matrix of the random network with entries $1$ $(0)$
if the connection from node $j$ to $i$ exists (or not), and we assume the same in-degree $K =\sum_j\varepsilon_{ji}$ for all neurons.

We consider the quadratic integrate-and-fire (QIF) neuron \cite{ermentrout1986}, which is a current-based model of class I excitability. In particular, whenever the membrane potential ${v}$ reaches infinity,
a $\delta$-spike is emitted and instantaneously delivered to the post-synaptic neurons and ${v}$ is reset to $-\infty$. In the absence of synaptic coupling, the QIF model displays excitable (oscillatory) dynamics for $I <0$ ($I>0$). Since we consider a purely inhibitory network in order to have
non-trivial collective dynamics the neurons should be supra-threshold, i.e. with $I > 0$,
in this case in absence of synaptic coupling the single neuron firing frequency is given by
\begin{equation}
\nu_0 = \frac{\sqrt{I}}{\pi} = \frac{1}{T_0}
\label{free}
\end{equation}
In the excitable case ($I<0$)  the neuron has a stable and an unstable fixed points
located at $V^s = - \sqrt{-I}$ and $V^u =  \sqrt{-I}$, respectively.

The DC current and the synaptic coupling are assumed to scale as
$I= i_0 \sqrt{K}$ and  $g=g_0/\sqrt{K}$ as usually done in order to ensure
a self-sustained balanced state for sufficiently large $K$ \cite{bal1,wolf,matteo}.

The times (frequencies) are reported in physical units by
assuming as a time scale a membrane time constant $\tau_m = 10$ ms.

\subsection{Simulation Protocols}

The simulation of the QIF network model \eqref{eq:1} has been performed thanks to an exact
event-driven method (see Appendix A for more details), this allowed us to reach very large system
sizes from $N=10000$ up to $N=80000$ with $ 10 \le K \le 1000$, to integrate for long time spans
up to $50-100$ secs and to average over 5-20 different network realizations.

The simulations performed to identify the emergence of collective
oscillations are usually based on two quasi-adiabatic protocols, where $i_0$ ($K$) is
varied in steps $\Delta i_0$ ($\Delta_K$) by maintaning all the other parameters constant.
The simulations are performed for some value of $i_0$ ($K$) where the quantities of interest
are evaluted over a certain time interval $t_E \simeq 20 -30$ s after discarding a transient of durations
$t_T \simeq 20 -30 $ s. At the end of the simulation the final values of the variables are stored and used to initialize the next simulation step with a parameter value $i_0 \pm \Delta i_0$ ($K \pm \Delta K$), whose duration will be again $t_T+t_E$. These steps are repeated until the final parameter value is achieved, to test for hysteresis the procedure can be reversed by decreasing (increasing) the parameter value in small steps to recover the initial value.

\subsection{Indicators}

The microscopic evolution of the neurons is typically measured in terms of the
mean inter-spike interval (ISI) of each neuron $\langle T_i \rangle$ and of the corresponding firing rate
$\nu_i = 1/\langle T_i \rangle$, where $\langle \cdot \rangle$ denotes a time average.
The neural variability of each neuron is customarily measured via the so-called
coefficient of variation 
$$cv_i = \frac{\sqrt{\langle T^2_i \rangle - \langle T_i \rangle^2}}{\langle T_i \rangle} $$
which is the ratio between the standard deviation and the mean of the ISIs of the neuron $i$.
A value $cv_i=0$ ($cv_i=1$) corresponds to a periodic (Poissonian) activity of the considered
neuron \cite{tuckwell1988}.

The activity of the neurons in the network can be therefore characterized in terms of the
ensemble averages of the firing rates ${F} = \overline{\nu_i} = \frac{1}{N} \sum_{i=1}^N \nu_i$,
of the coefficient of variations $CV = \overline{cv_i}$ and of the corresponding probability distribution functions
PDFs.

The population activity can be also analyzed in terms of the population firing rate $\nu(t)$
\begin{equation}
\nu(t) =  \frac{N_{sp}(T_W)}{T_W N} \quad .
\label{nnu}
\end{equation}
where $N_{sp}(T_W)$ is the number of spikes emitted by the
$N$ neurons in the network in a time interval $T_W$, which
we typically consider of the order of $0.005$.

The macroscopic evolution of the network can be characterized by the
indicator introduced in \cite{golomb}
\begin{equation}
\rho = \left( \frac{\sigma^2_V}{\overline{\sigma_i^2}}\right)^{1/2} \quad {\rm where} \quad
\sigma_i^2 = \langle v_i^2 \rangle - \langle v_i \rangle^2 \quad 
\label{indicator}
\end{equation}
and $\sigma_V$ is the standard deviation of the mean membrane potential $V(t) = \overline{v_i(t)}  = \sum_{i=1}^N v_i(t)/N$. The actual value of $\rho$ is related to the level of synchronization among the neurons:
asynchronous evolution (perfect synchrony) corresponds to $\rho \simeq {\cal O} (1/\sqrt{N})$
($\rho \equiv 1$). The presence of asynchronous dynamics or GOs can be inferred from the finite size scaling of
$\rho$ as explained in the following.

The global oscillations differentiated in terms of their frequency $f_G$ and of the following two ratios:
\begin{equation}
R_G =  \frac{f_G}{F} \qquad , \qquad  R_0 = \frac{f_G}{\nu_0} \quad ;
\label{ratios}
\end{equation}
where $R_G$ represents the level of locking of the single neuron activity
with respect to the population bursts and $R_0$ can be employed to understand the origin of GOs
and the influence of inhibitory PSPs. As a general remark we have found that $R_G \ge 2$
in all simulations, indicating that in the studied model the firing rate of the neurons is always
slower than the frequency of the GOs \cite{clusella2024}.

\begin{figure}
\includegraphics[width=0.45\textwidth]{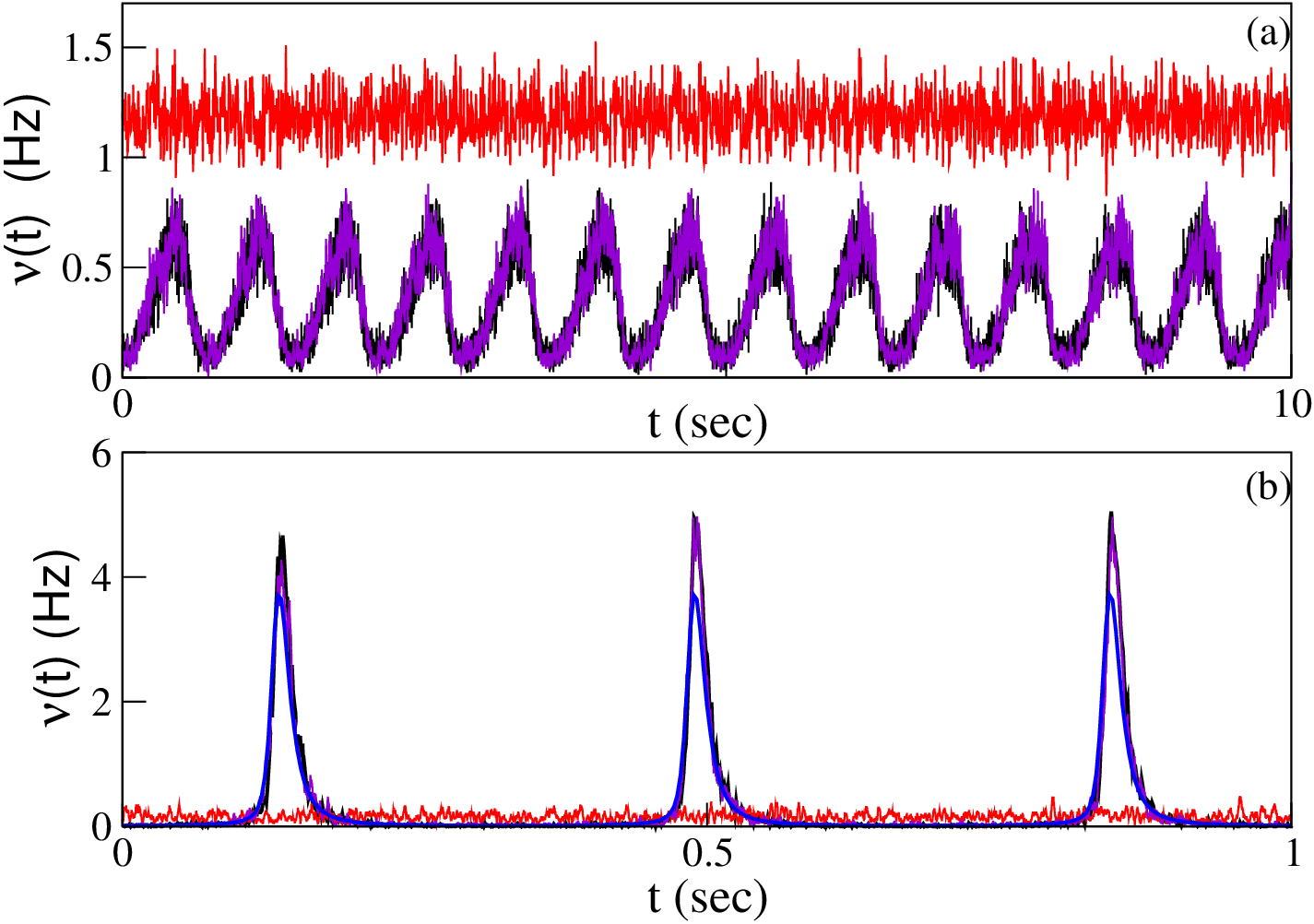}
\includegraphics[width=0.45\textwidth]{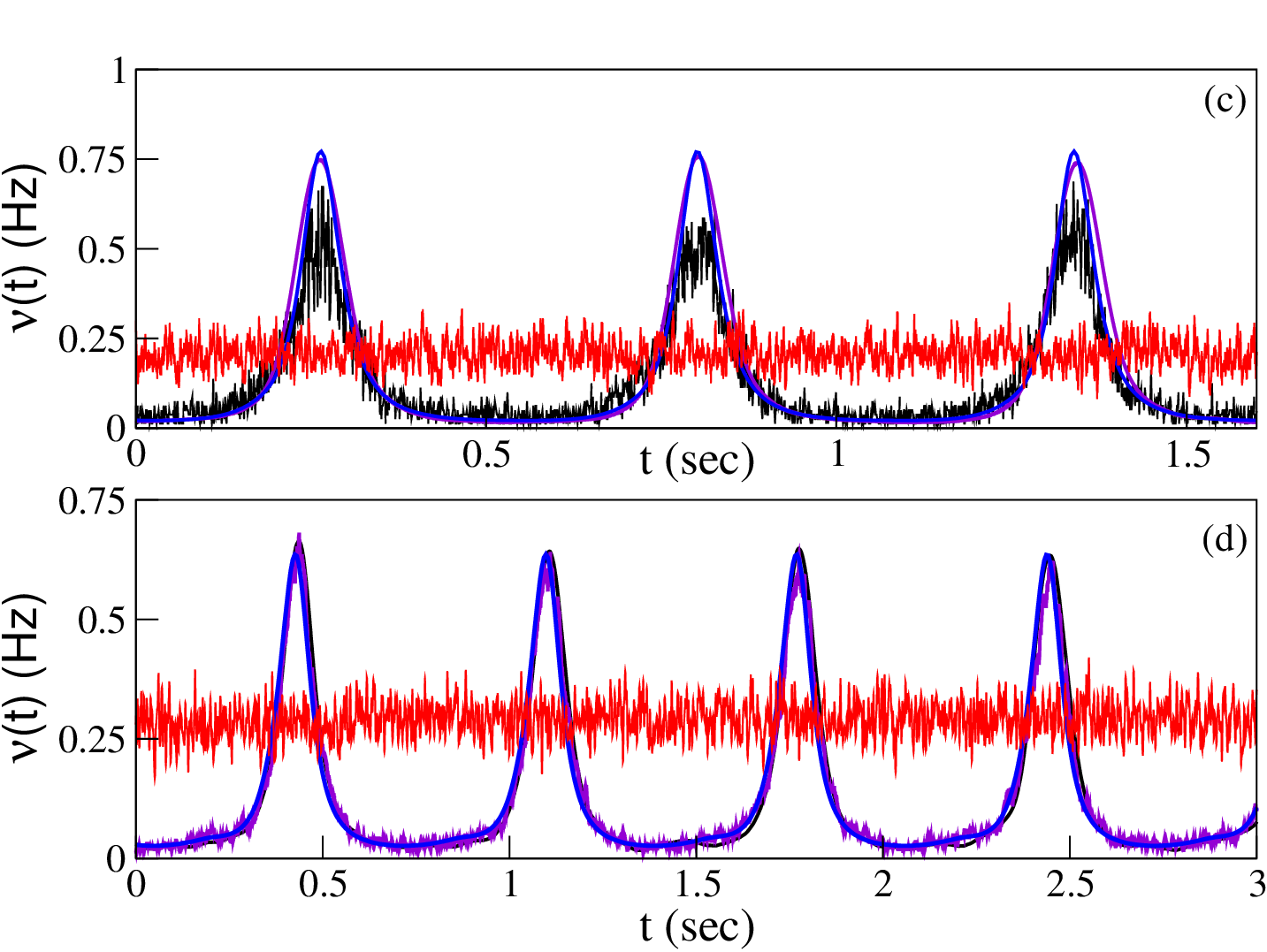}
\caption{Population firing rates $\nu(t)$ versus time for
$i_0=0.00055$:  $K=10$ (a) and $K=210$ (b); $i_0=0.00033$ and $K=100$ (c)
and $i_0=0.00027$ and $K=60$ (d). The black lines refer to network simulations with $N=80000$
neurons, the violet (red) lines to Langevin simulations with shot-noise (DA) with $N=20000$ (a-b) and $N=80000$ (c-d).
The blue solid lines denote the results of the CMF approach (\ref{eqGP03},\ref{eqGP12}).
All data refer to $g_0=1$. Details on the integration of the Langevin equations are reported in Appendix B.}
\label{fig1}
\end{figure}

\section{Mean-field description}

For a sufficiently sparse network with $K \ll N$, the spike trains emitted by $K$ pre-synaptic neurons
can be assumed to be uncorrelated and Poissonian \cite{brunel1999}, therefore
the mean-field dynamics of a QIF neuron can be represented in terms of the following Langevin equation:
\begin{equation}\label{eq:langevin}
\dot{V}(t) = V^2 + I - g S(t)
\end{equation}
where $S(t)$ is a Poissonian train of $\delta$-spikes with rate $R(t) = K \nu(t)$,
and $\nu(t)$ is the population firing rate self-consistently estimated.
Usually the Poissonian spike trains are approximated within
the DA \cite{capocelli1971,tuckwell1988} as
$S(t)  =  R(t) + \sqrt{R(t)} \xi(t)$, where $\xi(t)$ is a normalized Gaussian white noise signal.

However, the DA can fail in reproducing the neural dynamics. Indeed,
as shown in Fig. \ref{fig1} for sparse networks with different sets of parameters
by employing the DA in Eq. \eqref{eq:langevin} one obtains an asynchronous dynamics (red curve), while
the network evolution is characterized by GOs (black lines), that can be recovered only by
explicitly taking into account the Poissonian spike trains in Eq. \eqref{eq:langevin} (violet lines).
As shown in panel (b), the DA is unable to reproduce simulation results even with quite large in-degree ($K=210$) for sufficiently small $i_0$.
On the other hand, the Langevin mean-field data with shot-noise are in good agreement with the simulations in all considered cases indicating the validity of this approach.

In the mean-field framework the population dynamics is usually described in terms of
the membrane potential probability density function (PDF) $P(V,t)$,
whose time evolution for the QIF model is given (according to \eqref{eq:langevin})
by the continuity equation
\begin{equation}
\label{CE}
{\partial_t P}(V,t) + \partial_V[(V^2 + I)P(V,t)] = R(t) \Delta P(V,T)
\end{equation}
with boundary condition $\lim_{V \to \infty} V^2 P(V,t) = \nu(t)$ and
where $\Delta P(V,T) = [P(V^+,t)-P(V,t)]$ with $V^+=V + g$.
The right-hand side term in \eqref{CE} represents the time variation of $P(V,t)$
due to the arrival of the Poissonian trains of instantaneous inhibitory kicks of amplitude $g$
from $K$ pre-synaptic neurons.
By assuming that $g$ is sufficiently small we can expand the latter term as
\begin{equation}
\Delta P(V,t) = \sum_{p=1}^\infty \frac{g^p}{p !} \partial_V^p P(V,t)   \enskip ;
\label{expansion}
\end{equation}
and by limiting to the first two terms in this expansion we recover the
DA corresponding to the following Fokker-Planck Equation (FPE) \cite{haskell2001}
\begin{equation}
\label{fpe}
{\partial_t P(V,t)} + {\partial_V}[(V^2 + A(t))P(V,t)] =
D(t) {\partial^2_{V} P(V,t)}
\end{equation}
where $A(t) = \sqrt{K}[i_0 - g_0 \nu(t)]$ is the effective input current and $D(t) = g_0^2 \nu(t)/2$ represents the
diffusion coefficient.

In the following we will consider also a generalized version of the FPE (GFPE) obtained
by considering up to the third order term in the expansion \eqref{expansion},
which gives the following evolution equation for the $P(V,t)$
\begin{equation}
\label{gfpe}
{\partial_t P(V,t)} + {\partial_V}[(V^2 + A(t))P(V,t)] =
D(t) {\partial^2_{V} P(V,t)} + H(t) {\partial^3_V P(V,t)} \quad ,
\end{equation}
where $H(t) = \frac{g_0^3 \nu(t)}{3! \sqrt{K}}$ .

\begin{figure}
\includegraphics[width=0.45\textwidth]{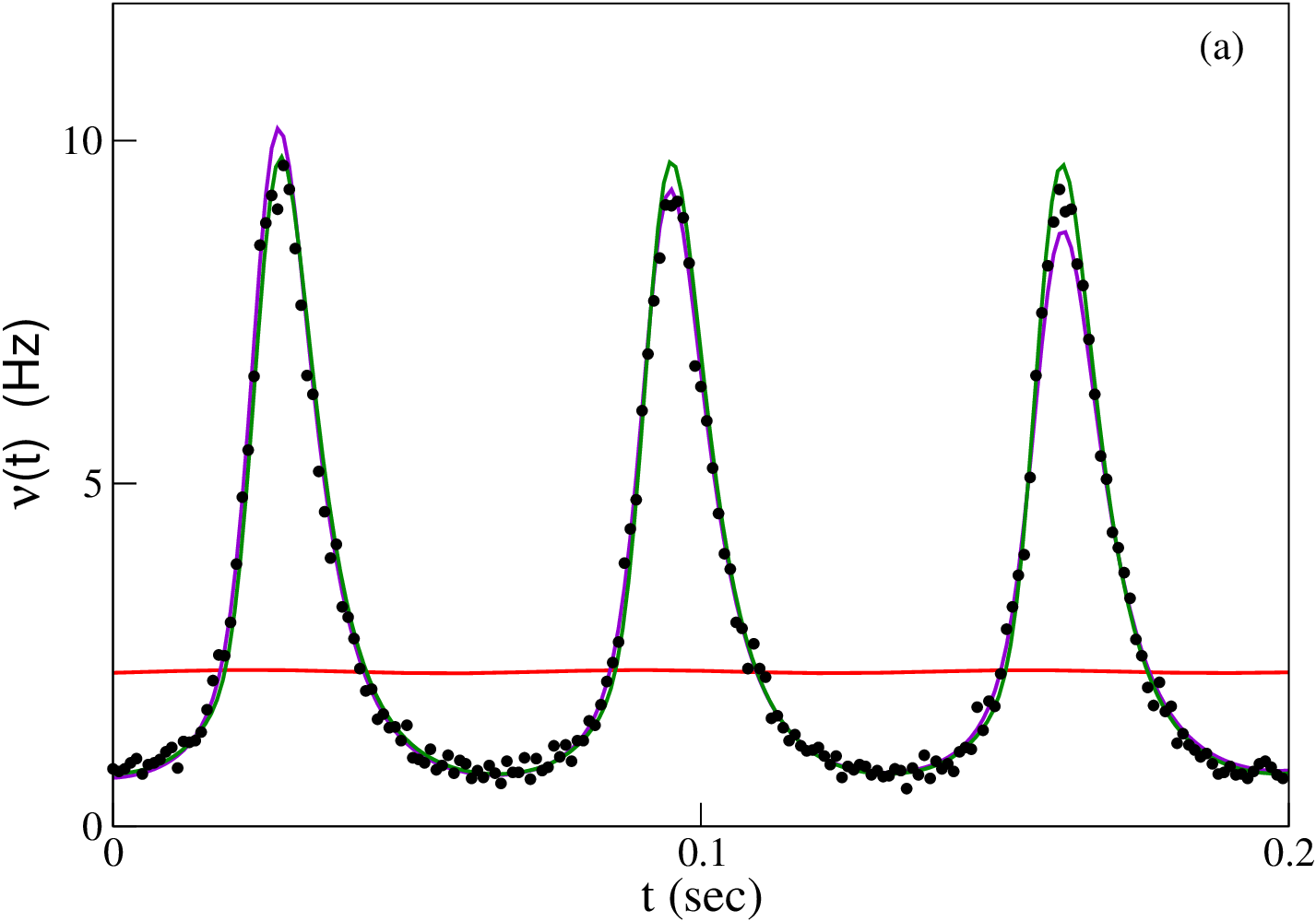}
\includegraphics[width=0.45\textwidth]{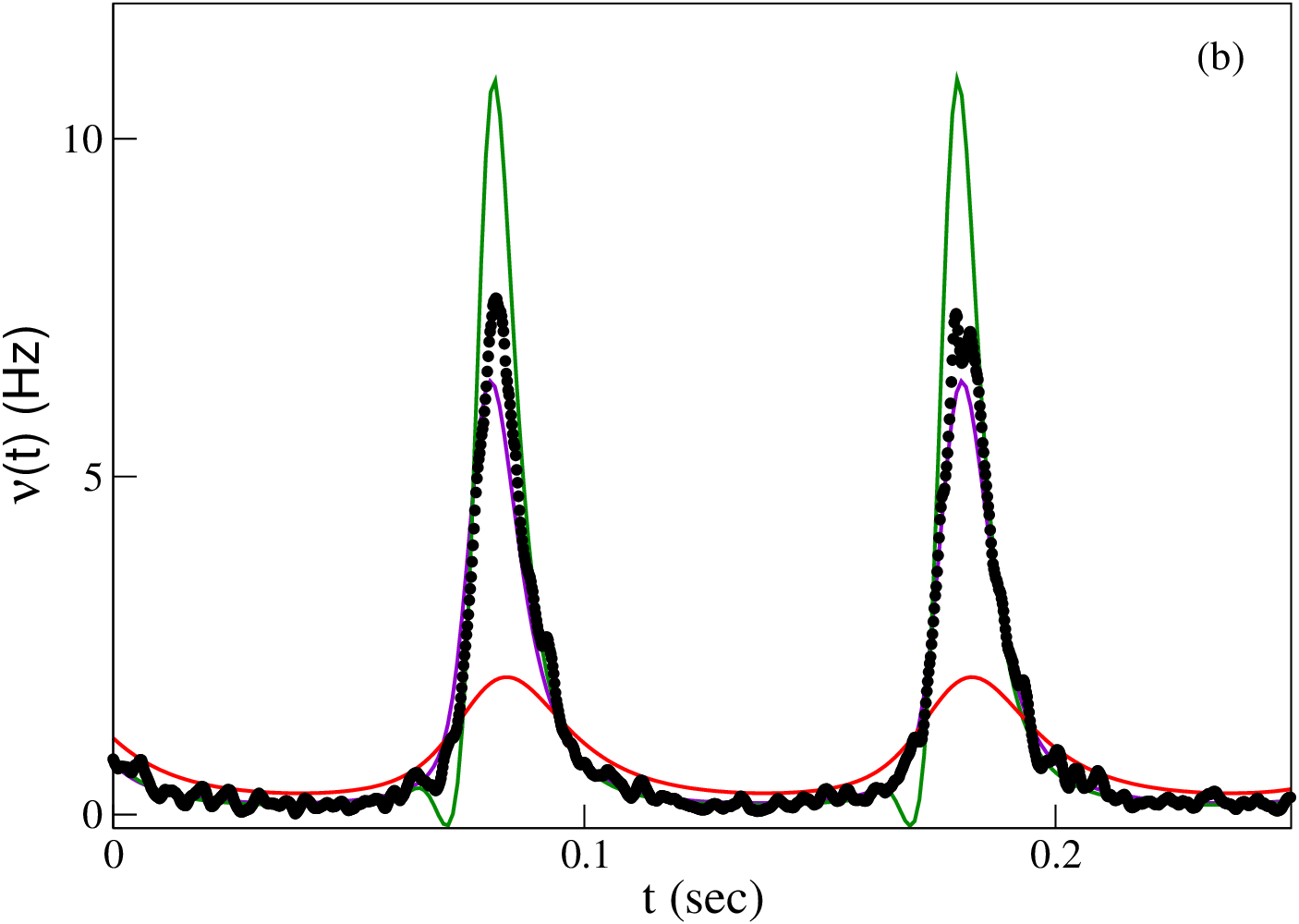}
\caption{Population firing rates $\nu(t)$ versus time for
$i_0=0.02$,  $K=200$ (a) and $i_0=0.006$, $K=400$ (b). The black circles refer to network simulations with $N=80000$ (a)
and $N=20000$ (b), the violet (red) solid lines to integration of the continuity equation \eqref{CE} (FPE \eqref{fpe}).
The green solid lines denote the results of the integration of the GFPE including third order terms \eqref{gfpe}.
All data refer to $g_0=1$, the numerical integration of the partial differential equations has been performed by
employing time splitting pseudo-spectral methods with a time step $\Delta t = 10^{-4}$ ms and with 128--512 Fourier modes
(for more details see Appendix C).}
\label{fig2}
\end{figure}

Comparisons among the network simulations (black circles) and the results obtained by  integrating
the partial differential equations \eqref{CE} (violet line), \eqref{fpe} (red lines) and \eqref{gfpe} (green lines)
via time splitting pseudo-spectral methods (as explained in details in Appendix C)  are shown in
Fig. \ref{fig2}. These results show that DA can give incorrect predictions even at the level of population dynamics.
Indeed, as shown in Fig. \ref{fig2} (a) the network dynamics
is oscillatory with $f_G \simeq 15$ Hz (black circles). This evolution is correctly captured by the continuity equation
\eqref{CE} (violet line). On the contrary, the FPE \eqref{fpe} (red line)
converges to a stable fixed point corresponding to asynchronous dynamics.
In this case, the inclusion of the third order term of the expansion \eqref{expansion}
in the evolution equation for $P(V,t)$ \eqref{gfpe} gives a stable solution (green solid line) capturing quite well
the network dynamics. Unfortunately, this is not always the case due to the intrinsic instability
of the third order expansion~\cite{Marcinkiewicz-1939,Gardiner-1997}, that e.g. does not guarantee that the sign  of the PDF remains positive, as observable
in the case reported in panel (b). In this latter situation, the DA is able to reproduce the
oscillatory behaviour displayed by the network evolution, however the amplitudes of
such oscillations are definitely better captured by the continuity equation \eqref{CE} (violet line).

In summary, to correctly reproduce the collective dynamical regimes observable in the network
it is necessary to consider the complete continuity equation \eqref{CE}, even if in some cases the third
order expansion \eqref{gfpe} could be sufficiently accurate. As a general remark the time splitting
pseudo-spectral integration methods here employed suffer of numerical instabilities in particular for small $K$, that render
the approach not particularly reliable. Therefore, as we will explain in the following we have developed an
accurate and stable formalism encompassing synaptic shot-noise
to identify the various possible regimes displayed by \eqref{CE} and to characterize their linear and non-linear stability.

\subsection{Derivation of the macroscopic evolution equations}

As already mentioned, the dynamical evolution of the QIF model can be
transformed in that of a phase oscillator, the $\theta$-neuron \cite{ermentrout1986,ermentrout2008},
by introducing the phase variable $\theta = 2 \arctan{V}$.
However, this transformation renders quite difficult to distinguish asynchronous
from partially synchronized states  \cite{kralemann2007,dolmatova2017}.
A more suitable phase transformation, able
to take into account correctly the phase synchronization phenomena, is the following
\[
\psi=2\arctan\frac{V}{\sqrt{I}} \in [-\pi,\pi] \;,\qquad
V=\sqrt{I}\tan\frac{\psi}{2} \in ]-\infty, +\infty[ \;,
\]
where $I=i_0\sqrt{K}$. Indeed, for $I>0$ and  in the absence of incoming pulses
the phase $\psi $ is uniformly rotating with angular velocity $2 \sqrt{I}$
at variance with the $\theta$-neuron where even for uncoupled oscillators the
velocity depends on the phase.

The probability distribution function (PDF) $w(\psi)$ for the phase variables $\psi$ is related to the one of the membrane potentials $P(V,t)$
via the relation
$$
w(\psi,t) = P(V,t) \frac{I+V^2}{2 \sqrt{I}}
$$
due to the probability conservation under (non-linear) transformations of stochastic variables.
Therefore, the continuity equation \eqref{CE} for $P(V,t)$ :
\begin{align}
&\frac{\partial P(V,t)}{\partial t}
=-\frac{\partial}{\partial V}\Big[\big(I+V^2\big)P(V,t)\Big]
%\nonumber
%\\[5pt]
%&\quad
%{}
+K\nu(t)\Big[P(V+g,t)-P(V,t)\Big]\;,
%\quad
%a=\frac{g}{\sqrt{K}}\;.
\label{eqSN201}
\end{align}
can be recast as follows for $w(\psi,t)$
\begin{align}
&\frac{\partial w(\psi,t)}{\partial t}
=-\frac{\partial}{\partial\psi}\left[2\sqrt{I}w(\psi,t)\right]
%\nonumber
%\\[5pt]
%&\quad
%{}
+K\nu(t)\left[\frac{I+V^2}{I+(V+g)^2}w(\psi_+,t)-w(\psi,t)\right]\;,
\label{eqGP01}
\end{align}
where $\psi_+$ is the shifted phase:
\[
V+g=\sqrt{I}\tan\frac{\psi_+}{2}\;,
\qquad
\tan\frac{\psi_+}{2}=\alpha+\tan\frac{\psi}{2}\;,
%\]
%\[
%\psi_a=2\arctan\left(\alpha+\tan\frac{\psi}{2}\right)\;,
\qquad \alpha\equiv\frac{g}{\sqrt{I}}=\frac{g_0}{\sqrt{i_0}K^{3/4}}\;.
\]

By making all the terms in the right-end side of Eq.~(\ref{eqGP01}) explicit in terms of the variable $\psi$
\begin{align}
\frac{I+V^2}{I+(V+g)^2}=\frac{1+\tan^2\frac{\psi}{2}}{1+\left(\alpha+\tan\frac{\psi}{2}\right)^2} \nonumber
%\\
=\frac{1}{1+\frac{\alpha^2}{2}+\alpha\sin\psi+\frac{\alpha^2}{2}\cos\psi}\;,
\nonumber
\end{align}
we can finally rewrite Eq.~(\ref{eqGP01}) as
\begin{align}
&\frac{\partial w(\psi,t)}{\partial t}
=-\frac{\partial}{\partial\psi}\left[2\sqrt{I}w(\psi,t)\right]
%\nonumber
%\\[5pt]
%&\quad
%{}
+K\nu(t) \left[\frac{w(\psi_+,t)}{1+\frac{\alpha^2}{2}+\alpha\sin\psi+\frac{\alpha^2}{2}\cos\psi} -w(\psi,t)\right]\;.
\label{eqGP02}
\end{align}

In Fourier space the PDF $w(\psi,t)$ can be expressed as
\[
w(\psi,t)=\frac{1}{2\pi}\sum\limits_{n=-\infty}^{+\infty}z_ne^{-in\psi},
\]
with $z_0=1$ and $z_{-n}=z_n^\ast$, where the complex coefficients $z_n$ are the
so-called Kuramoto--Daido order parameters $z_n$ \cite{kuramoto2012, daido1992}.
In terms of these coefficients the continuity equation (\ref{eqGP02}) becomes
\begin{align}
&\dot{z}_n=i2n\sqrt{I}z_n +K\nu(t)\left[\sum_{m=-\infty}^{+\infty} I_{nm}z_m-z_n\right]\;,
\label{eqGP03}
\end{align}
where
\begin{equation}
I_{nm}\equiv\frac{1}{2\pi}\int\limits_0^{2\pi} \frac{e^{in\psi}\left(e^{-i\psi_+}\right)^m\mathrm{d}\psi}{1+\frac{\alpha^2}{2}+\alpha\sin\psi+\frac{\alpha^2}{2}\cos\psi}\;.
\label{eqGP04}
\end{equation}

In order to calculate the integrals $I_{nm}$, we need to express explicitely the term
$e^{-i\psi_+}=\cos\psi_+-i\sin\psi_+$ entering in Eq. \eqref{eqGP04}.
In particular, by noticing that
\[
\cos\psi_+
=\frac{1-\tan^2\frac{\psi_+}{2}}{1+\tan^2\frac{\psi_+}{2}}\;,
\qquad
\sin\psi_+
=\frac{2\tan\frac{\psi_+}{2}}{1+\tan^2\frac{\psi_+}{2}}\;,
\]
and therefore that
\begin{align}
&e^{-i\psi_+} =\cos\psi_+-i\sin\psi_+
 =\frac{1-i\tan\frac{\psi_+}{2}}{1+i\tan\frac{\psi_+}{2}}
=-\frac{\tan\frac{\psi}{2}+\alpha+i}{\tan\frac{\psi}{2}+\alpha-i}\;.
\nonumber
\end{align}
Moreover by expressing $\zeta \equiv  e^{i\psi}$ as follows
\[
\zeta \equiv
e^{i\psi}=-\frac{\tan\frac{\psi}{2}-i}{\tan\frac{\psi}{2}+i}\;,
\quad\mbox{ and hence }\;
\tan\frac{\psi}{2}=i\frac{1-\zeta}{1+\zeta}\;,
\]
one finally finds
\begin{align}
e^{-i\psi_+}&=-\frac{\alpha +2i +\alpha \zeta}{\alpha +(\alpha-2i) \zeta}\;.
\label{eqGP05}
\end{align}

By employing the above expression  the integral~(\ref{eqGP04}) can be rewritten as
\begin{align}
I_{nm}&=\int\limits_0^{2\pi}
 \frac{(2\pi)^{-1}e^{in\psi}\left[-\frac{\alpha+2i+\alpha e^{i\psi}}{\alpha +(\alpha-2i)e^{i\psi}}\right]^m\mathrm{d}\psi}
 {1+\frac{\alpha^2}{2}-\frac{i\alpha}{2}(e^{i\psi}-e^{-i\psi})+\frac{\alpha^2}{4}(e^{i\psi}+e^{-i\psi})}
%\nonumber\\
%&
=\oint\limits_{|\zeta|=1}
 \frac{(i2\pi)^{-1}\zeta^{n-1} \left[-\frac{\alpha+2i+\alpha\zeta}{\alpha+(\alpha-2i)\zeta}\right]^m\mathrm{d}\zeta}
 {1+\frac{\alpha^2}{2} -\frac{i\alpha}{2}(\zeta-\frac{1}{\zeta}) +\frac{\alpha^2}{4}(\zeta+\frac{1}{\zeta})}
\nonumber\\
%&=\frac{1}{2\pi i}\oint\limits_{|\zeta|=1}
% \frac{\zeta^n}{(\alpha-2i)^m}\left[-\frac{\alpha+2i+\alpha\zeta}
% {\zeta+\frac{\alpha}{\alpha-2i}}\right]^m
%\nonumber\\
%&{}\qquad
%\times
% \frac{\mathrm{d}\zeta}{(\frac{\alpha^2}{4}-\frac{i\alpha}{2})\zeta^2+(1+\frac{\alpha^2}{2})\zeta +\frac{\alpha^2}{4}+\frac{i\alpha}{2}}
%\nonumber\\
%&=\frac{1}{2\pi i}\oint\limits_{|\zeta|=1}
% \frac{4\zeta^n}{\alpha(\alpha-2i)^{m+1}}\left[-\frac{\alpha+2i+\alpha\zeta}
% {\zeta+\frac{\alpha}{\alpha-2i}}\right]^m
%% \nonumber\\
%%&{}\qquad
%%\times
% \frac{\mathrm{d}\zeta}{(\zeta+\frac{\alpha}{\alpha-2i})(\zeta+\frac{\alpha+2i}{\alpha})}
%\nonumber\\
&=\frac{1}{2\pi i}\oint\limits_{|\zeta|=1}
 \frac{4(-\alpha)^m\zeta^n(\zeta+\frac{\alpha+2i}{\alpha})^{m-1}}
 {\alpha(\alpha-2i)^{m+1}(\zeta+\frac{\alpha}{\alpha-2i})^{m+1}}\mathrm{d}\zeta\;.
\label{eqGP06}
\end{align}

To evaluate the complex integral in \eqref{eqGP06} we employ the residue theorem and
therefore we should identify the poles of the integrand.
We can limit our consideration to $n\ge1$ since $z_0=1$, and $z_{-n}=z_n^\ast$, for
$m \le 0$ the integrand has the following pole
\[
\zeta_2=-\frac{\alpha+2i}{\alpha}\;,
\]
which is always beyond the unit circle since $|\frac{\alpha+2i}{\alpha}|^2=1+4/\alpha^2>1$ and, therefore, does not contribute to the integral $I_{nm}$.
However, for $m \ge 0$ the integrand displays another pole located at
\[
\zeta_1=-\frac{\alpha}{\alpha-2i} \quad .
\]
This pole is always within the integration contour $|\zeta|=1$ since $|\frac{\alpha}{\alpha-2i}|^2=\frac{\alpha^2}{\alpha^2+4}<1$ and therefore it contributes to the integral \eqref{eqGP06}.

%%%%%%%%%%%%%%%%%%%%%%%%%%%%%%%%%%%%%%%%%%%%%%%%%%%%%%%%%%%%
%%%%%%%%%%%%%%%%%%%%%%%%%%%%%%%%%%%%%%%%%%%%%%%%%%%%%%%%%%%%
\begin{figure}[!t]
\centerline{
\includegraphics[width=0.55\textwidth]{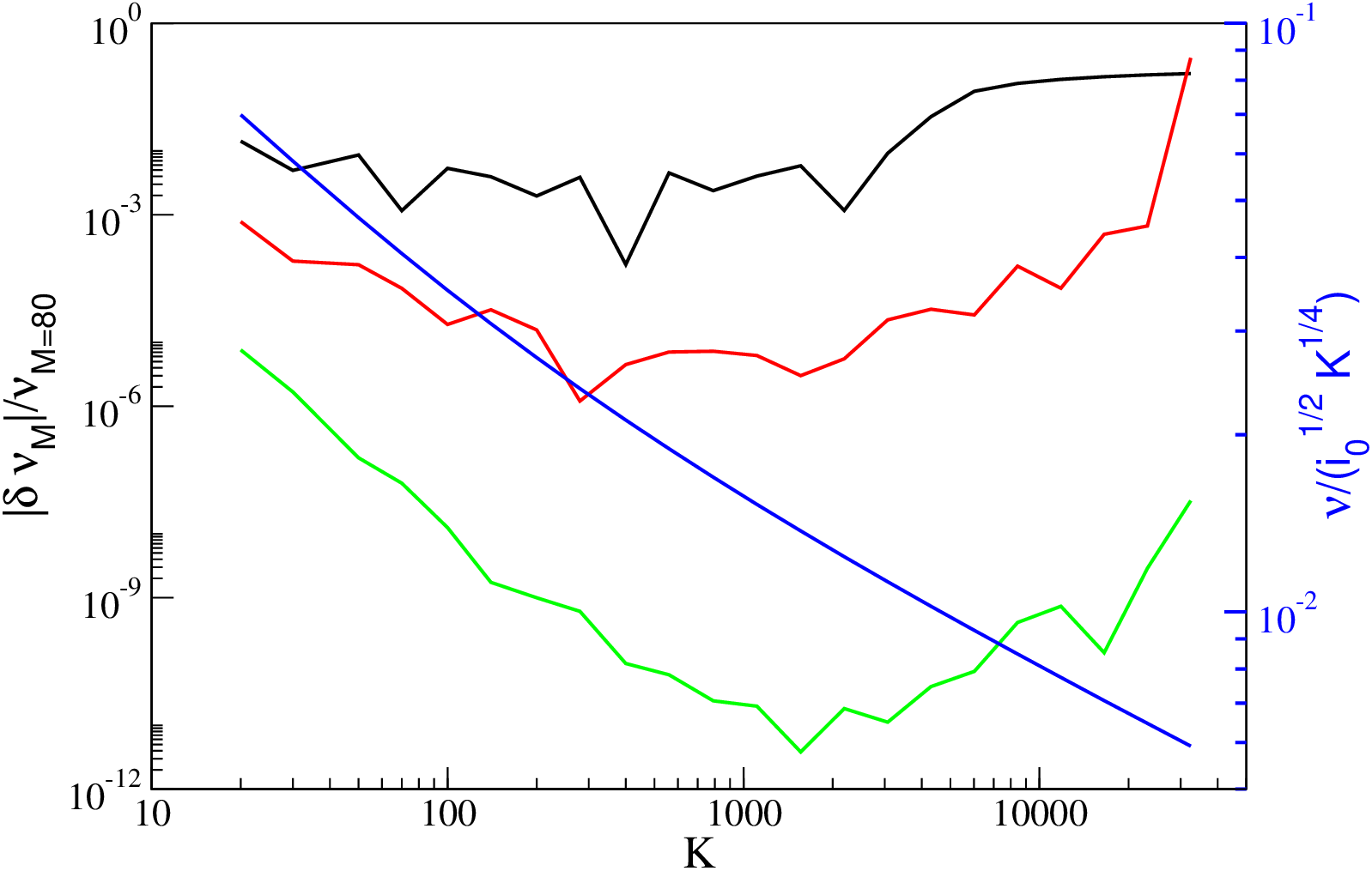}}
\caption{Relative error in the estimation of the firing rate $\nu$ for a stationary situation (asynchronous dynamics) computed with (\ref{eqGP03})--(\ref{eqGP12}) plotted versus the in-degree $K$ for $M=10$ (black line), $20$ (red line), $40$ (green line). The reference firing rate was estimated by employing $M=80$ and it is shown in rescaled units as a blue line. Parameters: $(i_0,g_0)=(0.006,1)$.
}
  \label{fig-acc}
\end{figure}
%%%%%%%%%%%%%%%%%%%%%%%%%%%%%%%%%%%%%%%%%%%%%%%%%%%%%%%%%%%%
%%%%%%%%%%%%%%%%%%%%%%%%%%%%%%%%%%%%%%%%%%%%%%%%%%%%%%%%%%%%

We employ the residue theorem to evaluate the following integral:
\begin{align}
&\frac{1}{2\pi i}\oint\limits_{|\zeta|=1}
 \frac{\zeta^n(\zeta+\frac{\alpha+2i}{\alpha})^{m-1}}
 {(\zeta+\frac{\alpha}{\alpha-2i})^{m+1}}\mathrm{d}\zeta
%\nonumber\\
%&\qquad
=\left\{
\begin{array}{cr}
\frac{1}{m!}\left.\frac{\mathrm{d}^m}{\mathrm{d}\zeta^m} \left(\zeta^n\left(\zeta-\zeta_2\right)^{m-1}\right)\right|_{\zeta=\zeta_1}\;,
& m\ge0\;;\\
0\;,
& m\le-1
\end{array}
\right.
\nonumber\\
&\qquad
=\left\{
\begin{array}{cc}
\sum\limits_{j=0}^{\min(n,m)-1}\Big({m-1 \atop j}\Big)\Big({n+m-1-j \atop m}\Big)
%&
%\\
%\qquad\qquad\qquad
%{}\times
\zeta_1^{n-1-j}\left(-\zeta_2\right)^{j}\;,
&m\ge 1\;;\\[5pt]
\frac{\zeta_1^n}{\zeta_1-\zeta_2}\;,
&m=0\;;\\[5pt]
0\;,
&m\le-1
\end{array}
\right.
\nonumber
\end{align}
where $\min(n,m)$ returns the minimal of two values, the binomial coefficients are defined as $\left(n\atop m\right)=\frac{n!}{m!(n-m)!}$. In particular, for obtaining the result reported in the latter line we considered
separately the cases for $1\le n\le m$ and for $n>m$. Hence,
%\[
%I_{nm}\equiv\frac{4}{\alpha(\alpha-2i)}\mathcal{I}_{nm}\;,
%\]
\begin{align}
I_{nm}(\alpha)=\left\{
\begin{array}{cc}
\sum\limits_{j=0}^{\min(n,m)-1}
\frac{(\zeta_1-\zeta_2)(n+m-1-j)!\cdot\zeta_1^{m+n-1-j}\left(-\zeta_2\right)^{j}}{m\cdot j!\cdot(m-1-j)!\cdot(n-1-j)!},
&m\ge 1\,;\\[5pt]
\zeta_1^{n}\;,
&m=0\,;\\[5pt]
0\;,
&m\le-1\,.
\end{array}
\right.
\label{eqGP07}
\end{align}
After the substitution of the values of $\zeta_1$ and $\zeta_2$ in \eqref{eqGP07}, the coefficients
$I_{nm}(\alpha)$ take the form :

\begin{equation}
I_{nm}(\alpha)\equiv\frac{1}{2\pi}\int\limits_0^{2\pi} \frac{e^{in\psi}\left(e^{-i\psi_+}\right)^m\mathrm{d}\psi}{1+\frac{\alpha^2}{2}\alpha\sin\psi+\frac{\alpha^2}{2}\cos\psi}
\label{eqFP04}
=\!\left\{
\begin{array}{cr}
\big(\frac{\alpha}{2i-\alpha}\big)^n\;,
&m=0\,;\\[5pt]
\sum\limits_{j=1}^{\min(n,m)}
\frac{4(-1)^j(n+m-j)!\cdot\alpha^{m+n-2j}(4+\alpha^2)^{j-1}} {m(j-1)!\cdot(m-j)!\cdot(n-j)!\cdot(2i-\alpha)^{m+n}},
&m\ge 1\,.
\end{array}
\right.
\end{equation}

In particular, for $n \ge 1$, one has $I_{nm}=0$ for $m<0$;
therefore, the matrix $I_{nm}$ has non zero elements only for $n=1,2,3,...$ and $m=0,1,2,3,...$\,:
\begin{align}
&\left(\frac{I_{nm}}{\zeta_1-\zeta_2}\right)=
%\nonumber\\
%&
%\frac{4}{\alpha(\alpha-2i)}
\left(
\begin{array}{cccccllllll}
\frac{\zeta_1}{\zeta_1-\zeta_2} & \zeta_1 & \zeta_1^2 & \zeta_1^3
%& \zeta_1^4
 &\dots
\\[5pt]
\frac{\zeta_1^2}{\zeta_1-\zeta_2} & 2\zeta_1^2 & 3\zeta_1^3-\zeta_1^2\zeta_2
 & 4\zeta_1^4-2\zeta_1^3\zeta_2
%& 5\zeta_1^5-3\zeta_1^4\zeta_2
 &\dots
\\[5pt]
\frac{\zeta_1^3}{\zeta_1-\zeta_2} & 3\zeta_1^3 & 6\zeta_1^4-3\zeta_1^3\zeta_2
& 10\zeta_1^5-8\zeta_1^4\zeta_2+\zeta_1^3\zeta_2^2
%& 15\zeta_1^6-15\zeta_1^5\zeta_2+3\zeta_1^4\zeta_2^2
&\dots
\\[5pt]
\frac{\zeta_1^4}{\zeta_1-\zeta_2} & 4\zeta_1^4 & 10\zeta_1^5-6\zeta_1^4\zeta_2 & 20\zeta_1^6-20\zeta_1^5\zeta_2+4\zeta_1^4\zeta_2^2
%& 35\zeta_1^7-45\zeta_1^6\zeta_2+15\zeta_1^5\zeta_2^2-\zeta_1^4\zeta_2^3
&\dots
\\[5pt]
&\dots
\end{array}
\right).
\nonumber
\end{align}

In the continuity equation~(\ref{eqGP01}) appears the population firing rate $\nu(t)$ that should be evaluated self-consistently during the time evolution. The instantaneous firing rate is given by the flux at the firing threshold,
therefore
\begin{align}
\nu(t)&= \lim_{V \to \infty} V^2 P(V,t) = 2 \sqrt{I}\,w(\pi,t)
%\nonumber\\
%&
=\frac{\sqrt{I}}{\pi}\mathrm{Re}\big(1-2z_1+2z_2-2z_3+2z_4-\dots\big)\;.
\label{eqGP12}
\end{align}

The dynamical system~(\ref{eqGP03},\ref{eqGP12}) has been numerically integrated with the exponential time differencing method~\cite{Cox-Matthews-2002,Permyakova-Goldobin-2025} for time-dependent regimes and numerically solved (in the Maple analytical calculation package) for time-independent regimes, stability and weakly non-linear analyses by truncating the
series $z_n$ at $n=M$ terms. Usually we fixed $M=100$ and, where necessary, $M$ was increased in order to keep the relative simulation error below $10^{-12}$ \cite{maple} (see Fig.~\ref{fig-acc}).

\subsection{Diffusion and third order approximations}

Consider the continuity equation (\ref{gfpe}), which yields the DA with $H(t)$ set to zero and the D3A for $H(t)=g_0^3\nu(t)/(3!\sqrt{K})$, but recast it as
\begin{equation}
\label{DA23:01}
{\partial_t P(V,t)} + {\partial_V}[(V^2 + I)P(V,t)] =
K\nu(t)\left[g\partial_{V}P(V,t) + \frac{g^2}{2}\partial^2_{V} P(V,t) + \frac{g^3}{3!}\partial^3_{V}P(V,t) \right] \quad .
\end{equation}
For the probability density $w(\psi,t)$ of the phase variable $\psi=2\arctan(V/\sqrt{I})$, continuity equation (\ref{DA23:01}) yields a modified version of (\ref{eqGP01})\;:
\begin{align}
&\frac{\partial w(\psi,t)}{\partial t}
=-\frac{\partial}{\partial\psi}\left[2\sqrt{I}w(\psi,t)\right]
%\nonumber
%\\[5pt]
%&\quad
%{}
+K\nu(t)\left[\frac{g}{\sqrt{I}}\mathcal{Q}w(\psi,t) + \frac{g^2}{2I}\mathcal{Q}^2w(\psi,t) + \frac{g^3}{3!I^{3/2}}\mathcal{Q}^3w(\psi,t) \right] \quad ,
\label{DA23:02}
\end{align}
where the operator
\[
\mathcal{Q}(\dots)\equiv\frac{\partial}{\partial\psi}\Big[(1+\cos\psi)(\dots)\Big] \, .
\]
Hence, for the Kuramoto--Daido order parameters $z_n$, one finds a modified version of equation system (\ref{eqGP03}):
\begin{align}
&\dot{z}_n=i2n\sqrt{I}z_n +K\nu(t)\sum_{m=0}^{+\infty} I_{nm}^{[\mathrm{tr}]}z_m\;,
\quad n=1,2,3,\dots,
\label{DA23:03}
\end{align}
where $z_0=1$; for DA, the ``truncated'' matrix
$$
I_{nm}^{[\mathrm{tr}]}=I_{nm}^{[\mathrm{DA}]}\equiv\alpha Q_{nm} +\frac{\alpha^2}{2}(\mathcal{Q}^2)_{nm}
$$
and, for D3A,
$$
I_{nm}^{[\mathrm{tr}]}=I_{nm}^{[\mathrm{D3A}]}\equiv\alpha Q_{nm} +\frac{\alpha^2}{2}(\mathcal{Q}^2)_{nm}+\frac{\alpha^3}{3!}(\mathcal{Q}^3)_{nm}.
$$
In Fourier space, operator $\mathcal{Q}$ has the following matrix form:
\[
Q_{nm}=\left(
\begin{array}{ccccccc}
0 &0 &0 &0 &0 &\dots\\
-\frac{i}{2} & -i & -\frac{i}{2} & 0 &0 &\dots\\
0 & -i & -2i & -i & 0 &\dots\\
0& 0& -\frac{3i}{2} & -3i & -\frac{3i}{2} &\dots\\
%0& 0& 0& 2i & 4i & 2i &\dots\\
&\dots&&
\end{array}
\right)\,,
\]
where the numbers of rows and columns run from $n=0$ and $m=0$, respectively; and matrices $\mathcal{Q}^2$, $\mathcal{Q}^3$ can be easily calculated via conventional matrix multiplication.
By substituting the matrix coefficients, one can recast Eq.~(\ref{DA23:03}) as follows:
\begin{align}
\dot{z}_n&\textstyle
=i2n\sqrt{I}z_n +K\nu(t)\Big\{\frac{n\alpha}{i}\left(z_n+ \frac{z_{n \pm 1}}{2}\right)
-\frac{n\alpha^2}{2}\left[\frac{3n}{2}z_n
+\left(n\pm\frac12\right)z_{n\pm1}+\frac{n\pm1}{4}z_{n\pm2}\right]
\nonumber\\
&\textstyle
+
\frac{in\alpha^3}{6}\left[\frac{5n^2+1}{2}z_n
+\frac{15n(n\pm1)+6}{8}z_{n\pm1}
+\frac{3(n\pm1)^2}{4}z_{n\pm2} +\frac{(n\pm1)(n\pm2)}{8}z_{n\pm3}\right]
\Big\} \quad ,
\label{DA23:04}
\end{align}
%\begin{align}
%\dot{z}_n&\textstyle
%=i2n\sqrt{I}z_n +K\nu(t)\Big\{\frac{n\alpha}{i}\left(\frac{z_{n-1}}{2}+z_n+\frac{z_{n+1}}{2}\right)
%\nonumber\\
%&\textstyle\quad
%-\frac{n\alpha^2}{2}\left[\frac{n-1}{4}z_{n-2}+\left(n-\frac{1}{2}\right)z_{n-1}
%\right.
%\nonumber\\
%&\textstyle\qquad\qquad
%\left.
% +\frac{3n}{2}z_n
%+\left(n+\frac12\right)z_{n+1}+\frac{n+1}{4}z_{n+2}\right]
%\nonumber\\
%&\textstyle
%+
%\frac{in\alpha^3}{6}\left[\frac{(n-1)(n-2)}{8}z_{n-3}+\frac{3(n-1)^2}{4}z_{n-2}
%\right.
%\nonumber\\
%&\textstyle
%\qquad
%+\frac{15n(n-1)+6}{8}z_{n-1} +\frac{5n^2+1}{2}z_n
%+\frac{15n(n+1)+6}{8}z_{n+1}
%\nonumber\\
%&\textstyle
%\qquad
%\left.
%+\frac{3(n+1)^2}{4}z_{n+2} +\frac{(n+1)(n+2)}{8}z_{n+3}\right]
%\Big\} \quad ,
%\label{DA23:04}
%\end{align}
where the terms with $\pm$ are not alternative, but they should be considered as both present in the sum
of the terms on the right-hand side and the diffusion approximation can be obtained by omitting the $\alpha^3$-term.
By employing the system of equations~(\ref{DA23:03}) (or (\ref{DA23:04}))
we can deal with the DA and the D3A in technically the same way as done for the CMF model (\ref{eqGP03}),
once replaced the matrix $\left(I_{nm}(\alpha)-\delta_{nm}\right)$ with $I_{nm}^{[\mathrm{tr}]}(\alpha)$.

\subsection{Linear stability of the asynchronous regime}
\label{linear_bif}

In order to analyzie the stability of the asynchronous dynamics and the
emergence of global oscillations in the macroscopic evolution,
we rewrite the system~(\ref{eqGP03},\ref{eqGP12}) as follows :
\begin{align}
\frac{\mathrm{d}z_n}{\mathrm{d}t}&=F_n(z_1,z_2,z_3,\dots)
\nonumber\\
&=\sqrt{i_0}K^\frac14\Bigg(i2nz_n
%\nonumber
%\\
%&\qquad
+K\widetilde{\nu}\Big[I_{n0}(\alpha)+\sum_{m=1}^{+\infty} I_{nm}(\alpha)\,z_m-z_n\Big]\Bigg)\,,
\label{eqGPD01}
\end{align}
with
\begin{equation}
\widetilde{\nu}= \frac{\nu}{\sqrt{I}} = \frac{1}{\pi}\mathrm{Re}(1-2z_1+2z_2-2z_3+2z_4-\dots)\,.
\label{eqGPD02}
\end{equation}

As a first aspect we notice that the dynamical evolution of the system~(\ref{eqGPD01})--(\ref{eqGPD02}) is controlled
by two dimensionless parameters, namely
\[
K \quad\mbox{ and }\quad\alpha=\frac{g_0}{\sqrt{i_0}K^{3/4}}\,,
\]
while the time can be rescaled by the factor $\sqrt{I} = \sqrt{i_0}K^{1/4}$.
Therefore a bifurcation diagram in the plane $(K,i_0/g_0^2)$ is sufficient to
capture all possible dynamical regimes observable for the system  (\ref{eqGPD01})--(\ref{eqGPD02}).

In order to analyze the stability of the stationary regimes in this plane, we should linearize the system
(\ref{eqGPD01})--(\ref{eqGPD02}) around a fixed point solution $\{ {z}_n^{(0)} \}$.
In particular, since $F_n$ depend on $\widetilde{\nu}$, which is non analytic function of $z_m$,
the linearization of (\ref{eqGPD01}) should be performed by considering as independent variables the
real and imaginary parts of the Kuramoto--Daido order parameters : namely, $z_n^\mathrm{re}=\mathrm{Re}(z_n)$ and $z_n^\mathrm{im}=\mathrm{Im}(z_n)$ (here and hereafter, superscripts ``$\mathrm{re}$'' and ``$\mathrm{im}$'' denote the real and imaginary parts).

Therefore the linearization of (\ref{eqGPD01}) can be written as
\begin{eqnarray}
{\delta \dot x}_{2n-1} &=& \frac{\partial F_n^\mathrm{re}}{\partial z_m^\mathrm{re}} {\delta x}_{2n-1}  +
\frac{\partial F_n^\mathrm{re}}{\partial z_m^\mathrm{im}} {\delta x}_{2n}
\label{lina}\\
{\delta \dot x}_{2n} &=& \frac{\partial F_n^\mathrm{im}}{\partial  z_m^\mathrm{re}} {\delta x}_{2n-1}  +
\frac{\partial F_n^\mathrm{im}}{\partial z_m^\mathrm{im}} {\delta x}_{2n}
\label{linb}
\end{eqnarray}
where  ${\delta x}_{2n-1}$ and ${\delta x}_{2n}$ are the infinitesimal perturbations of the real and imaginary parts of $z_n$,
respectively. Moreover, the other terms entering in \eqref{lina}--\eqref{linb} can be explicitly written as
\begin{subequations}
\label{eqGPD03}
\begin{align}
\frac{\partial F_n^\mathrm{re}}{\partial z_m^\mathrm{re}}&=\sqrt{i_0}K^\frac14\bigg(K\widetilde{\nu}\big(I_{nm}^\mathrm{re}-\delta_{nm}\big)
+\frac{4n(-1)^m}{\pi\widetilde{\nu}}\big[{z}^{(0)}_n\big]^\mathrm{im}
\bigg)\,,
\label{eqGPD03a}\\
 \frac{\partial F_n^\mathrm{im}}{\partial z_m^\mathrm{re}}&=\sqrt{i_0}K^\frac14\bigg(2n\delta_{nm}+K\widetilde{\nu}I_{nm}^\mathrm{im}
-\frac{4n(-1)^m}{\pi\widetilde{\nu}}\big[{z}^{(0)}_n\big]^\mathrm{re}
\bigg)\,,
\label{eqGPD03b}\\
\frac{\partial F_n^\mathrm{re}}{\partial z_m^\mathrm{im}}&=\sqrt{i_0}K^\frac14\left(-2n\delta_{nm}-K\widetilde{\nu}I_{nm}^\mathrm{im}\right)\;,
\label{eqGPD03c}\\
\frac{\partial F_n^\mathrm{im}}{\partial z_m^\mathrm{im}}&=\sqrt{i_0}K^\frac54\widetilde{\nu}\big(I_{nm}^\mathrm{re}-\delta_{nm}\big)\;.
\label{eqGPD03d}
\end{align}
\end{subequations}

The linear stability analysis of time-independent solutions ${\bf z}^{(0)} = \{ {z}_n^{(0)} \}$ of system~(\ref{eqGPD01})--(\ref{eqGPD02}) can be performed by solving the eigenvalue problems associated to Eqs. \eqref{lina}--\eqref{linb}. In particular, by truncating the system~(\ref{eqGPD01})--(\ref{eqGPD02}) to the $M$-th Kuramoto--Daido order parameter the eigenvalue
problem can be solved for a  $(2M\times2M)$ matrix with real-valued elements given
by (\ref{eqGPD03}) to find the associated complex eigenvalue spectrum $\{\lambda_j\} \quad  j=1,\dots,2M$.
The stationary solution is stable whenever $\mathrm{Re}\enskip \lambda_j < 0 \quad \forall j $.

\begin{figure}
\begin{center}
\includegraphics[width=0.8 \linewidth]{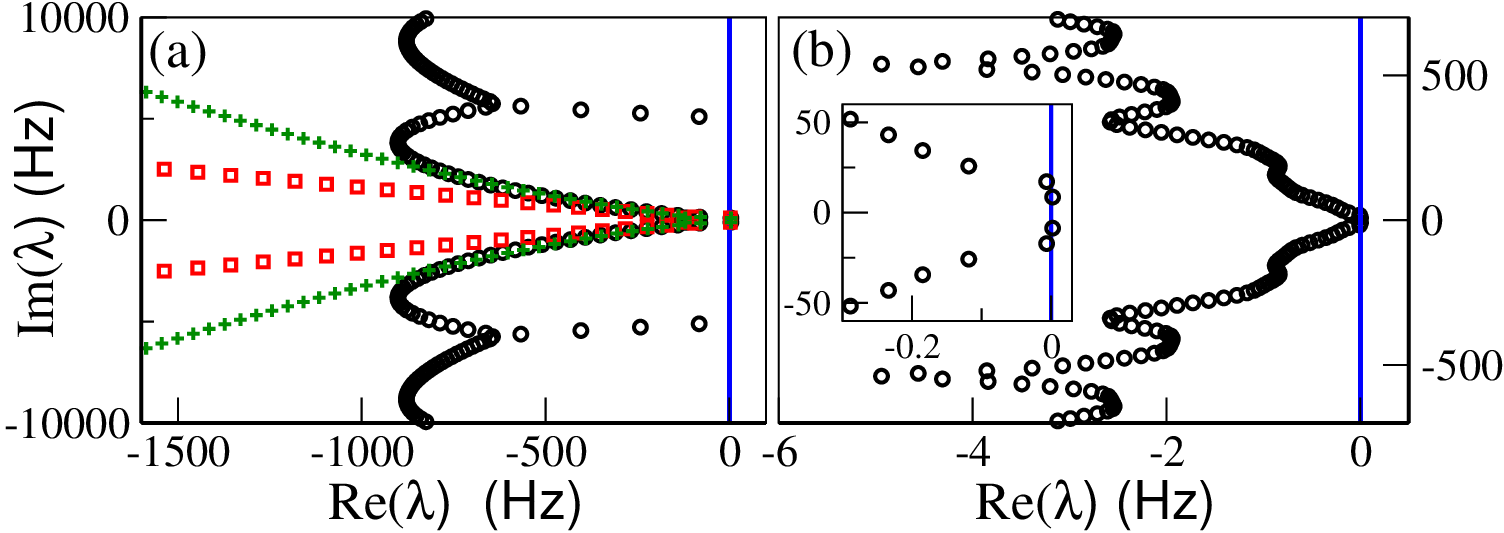}
\end{center}
\caption{Spectrum of the eigenvalues $\{\lambda_i\}$ for a stationary solution of the system~(\ref{eqGP03},\ref{eqGP12}) for $(i_0/g_0^2,K)=(0.02,400)$ (a), and $(0.00055,10)$ (b). An enlargement is reported in the inset in (b).
Black circles refer to the CMF, the red squares to the DA and the green pluses to the D3A.
}
\label{fig3s}
\end{figure}

Let us compare the spectra obtained within the CMF, the DA and the D3A to better understand the origin of the instabilities leading to oscillatory dynamics in the presence of microscopic shot-noise. As a first remark, we observe that the DA and D3A spectra are characterized besides
the most unstable modes, which can give rise to the oscillatory instability, by modes that are strongly
damped as shown in Fig. \ref{fig3s} (a). The case shown in  Fig. \ref{fig3s} (a) refers to a situation
where the dynamics is well reproduced already within the DA (red circles), in this case the DA and D3A (green pluses)
eigenvalues corresponding to small ${\rm Im} \enskip \lambda_j$ in proximity of the Hopf instability approximate quite well the CMF spectrum (black circles). However, while the CMF eigenvalues appear to saturate at some finite ${\rm Re} \enskip \lambda$ value,
the DA and D3A ones do not. We also notice that the D3A captures extremely well the central part of the CMF spectrum,
while the DA deviates quite soon from it. Despite these differences in this case the collective dynamics of the system is essentially
controlled by the two most unstable modes, typical of a Hopf bifurcation, that practically coincide within the three approaches.

In Fig. \ref{fig3s} (b) we report the CMF spectrum for a situation
where the oscillatory regime is definitely due to the finitness of the synaptic
stimulations and not captured at all by the DA and D3A. In this case, we observe that a large part of the eigenmodes are now practically undamped, compare the scales over which  ${\rm Re} \enskip \lambda_j$ varies in Fig. \ref{fig3s} (a) and (b).
Therefore, we expect that the collective dynamics is no longer dominated by only the 2 most unstable modes as usually observable in the DA, but that also the marginally stable or slightly unstable modes will contribute in the coherent dynamics, see the inset of panel (b).

In summary, the shot-noise promotes the emergence of weakly damped eigenmodes that have a relevant role in
the instability of the asynchronous regime at sufficiently small $i_0/g_0^2$ and in-degrees and that are neglected in the DA.

\subsection{Weakly Non-Linear Analysis}

As for the linear stability, also for the weakly non-linear analysis required to characterize the nature of the observed Hopf
bifurcations, one must handle $z_n^\mathrm{re}$ and $z_n^\mathrm{im}$ as independent real-valued variables. In particular, it is important to stress that the dynamical system~(\ref{eqGPD01},\ref{eqGPD02}) is quadratic with respect to the variables $z_n^\mathrm{re}$ and $z_n^\mathrm{im}$. The absence of cubic terms makes the calculation of the coefficients of the amplitude equations  quite simple
as we will see in the following.

Let us now analyse the Hopf instability of the stationary solution $\mathbf{z}^{(0)}=\{z_1^{(0)},z_2^{(0)},z_3^{(0)},\dots\}$ in proximity to the critical value $K_\mathrm{cr}$, where the in-degree $K$ acts as  the control parameter
and $i_0$ and $g_0$ are maintained constant. In order to perform this analysis we have considered
{\it finite} perturbations $\mathbf{x}=\{x_1,x_2,x_3,x_4,\dots\}$ of the stationary solution, where $x_m$ are real-valued variables defined as follows: $z_n^\mathrm{re} \equiv (z_n^{(0)})^\mathrm{re}+x_{2n-1}$, $z_n^\mathrm{im}\equiv (z_n^{(0)})^\mathrm{im}+x_{2n}$. To obtain an {\it exact} evolution equation for the perturbations,
we have substituted $z_n=z_n^{(0)}+x_{2n-1}+ix_{2n}$ into the system~(\ref{eqGPD01},\ref{eqGPD02}).
Furthermore, by remembering that $\mathbf{z}^{(0)}$ solves the right-hand side of the system~(\ref{eqGPD01},\ref{eqGPD02}), and rearranging the terms, one can arrive to the following
system of differential equations ruling the dynamics of $\mathbf{x}$:
\begin{equation}
\frac{\mathrm{d}x_n}{\mathrm{d}t}=F_{nm}x_m+\widetilde{F}_{nml}x_mx_l\,,
\label{eqGPD04a}
\end{equation}
where we imply the Einstein summation rule over repeated indices, odd-indexed(even-indexed) $x_n$ are the perturbations of the real (imaginary) part of $z_m$.
All the linear in $\mathbf{x}$ terms appearing in \eqref{eqGPD04a} correspond to those obtained within the linear stability analysis,
therefore, the coefficients~(\ref{eqGPD03}) calculated at $\mathbf{z}^{(0)}$ correspond to the coefficients $F_{nm}$ as follows:
\begin{align}
F_{2n-1\;2m-1}=\left.\frac{\partial F_n^\mathrm{re}}{\partial z_m^\mathrm{re}}\right|_{\mathbf{z}^{(0)}},
\quad&\quad
F_{2n\;2m-1}=\left.\frac{\partial F_n^\mathrm{im}}{\partial z_m^\mathrm{re}}\right|_{\mathbf{z}^{(0)}},
\nonumber
\\
F_{2n-1\;2m}=\left.\frac{\partial F_n^\mathrm{re}}{\partial z_m^\mathrm{im}}\right|_{\mathbf{z}^{(0)}},
\quad&\quad
F_{2n\;2m}=\left.\frac{\partial F_n^\mathrm{im}}{\partial z_m^\mathrm{im}}\right|_{\mathbf{z}^{(0)}}.
\nonumber
\end{align}

Further, we need to collect all the terms quadratic in $\mathbf{x}$ which can only originate
from the following term in Eq.~(\ref{eqGPD01}):
\[
K\left(\widetilde{\nu}-\pi^{-1}\right)\sum_{m=1}^{+\infty}(I_{nm}-\delta_{nm})z_m =
\sum_{m,m^\prime}\frac{2K(-1)^{m^\prime}}{\pi}(I_{nm}-\delta_{nm})z_m z_{m^\prime}^\mathrm{re}  \,;
\]
substituting $z_n=z_n^{(0)}+x_{2n-1}+ix_{2n}$ into the latter expression and taking the real and imaginary parts, one can collect all the quadratic $x_mx_l$-terms:
\begin{align}
\widetilde{F}_{2n-1\,ml}x_mx_l &=
 \sum_{m,m^\prime}\frac{2\sqrt{i_0}K^\frac54(-1)^{m^\prime}}{\pi}
%\nonumber\\
%&\qquad
%\times
\big[\left(I_{nm}^\mathrm{re}-\delta_{nm}\right)x_{2m-1} -I_{nm}^\mathrm{im}x_{2m}\big]x_{2m^\prime-1}
\,,
\nonumber\\
\widetilde{F}_{2n\,ml}x_mx_l &= \sum_{m,m^\prime}\frac{2\sqrt{i_0}K^\frac54(-1)^{m^\prime}}{\pi}
%\nonumber\\
%&\qquad
%\times
\big[\left(I_{nm}^\mathrm{re}-\delta_{nm}\right)x_{2m} +I_{nm}^\mathrm{im}x_{2m-1}\big]x_{2m^\prime-1}
\,.
\nonumber
\end{align}
Since the original equation system~(\ref{eqGPD01},\ref{eqGPD02}) contained only quadratic in $\mathbf{z}$ terms, we will also only have linear and quadratic in $\mathbf{x}$ terms.

For the weakly non-linear analysis of a bifurcation, we have to consider dynamical system~(\ref{eqGPD04a}) in a small vicinity of the bifurcation point $K_\mathrm{cr}(i_0,g_0)$, where
\begin{equation}
\frac{\mathrm{d}x_n}{\mathrm{d}t}=F_{nm}(K_\mathrm{cr})\,x_m+\widetilde{F}_{nml}(K_\mathrm{cr})\,x_mx_l
 +(K-K_\mathrm{cr})\left(\frac{\partial F_{nm}}{\partial K}\right)_{i_0,g_0}x_m\,,
\label{eqGPD04}
\end{equation}
and $\left(\frac{\partial\dots}{\partial K}\right)_{i_0,g_0}$ indicates the derivative with respect to $K$ for fixed $i_0$, $g_0$. Below we will omit the argument $K_\mathrm{cr}$ for coefficients $F_{nm}$ and $\widetilde{F}_{nml}$ for brevity.

At a Hopf instability, the most unstable modes are associated to two purely imaginary and complex
conjugate eigenvalues. Therefore the linear stability matrix $(F_{nm})$ has a pair of eigenvalues
$\pm i\Omega \equiv \pm i 2 \pi f^{H}$ with eigenvector $\mathbf{Y}$  and $\mathbf{Y}^\ast$, namely:
\begin{equation}
i\Omega Y_n=F_{nm}Y_m\,,
\qquad
-i\Omega Y_n^\ast=F_{nm}Y_m^\ast\,,
\label{eqEigSol}
\end{equation}
and the Hermitian conjugate problem possesses eigenvalues $\mp i\Omega$ with eigenvectors $\mathbf{Y}^+$ and $(\mathbf{Y}^+)^\ast$ :
\begin{equation}
-i\Omega Y_n^+=F_{mn}Y_m^+\,,
\qquad
i\Omega (Y_n^+)^\ast=F_{mn}(Y_m^+)^\ast\,.
\label{eqEigSolHC}
\end{equation}

\begin{figure}[!t]
\centerline{
\hspace{-19pt}
\includegraphics[width=0.47\textwidth]{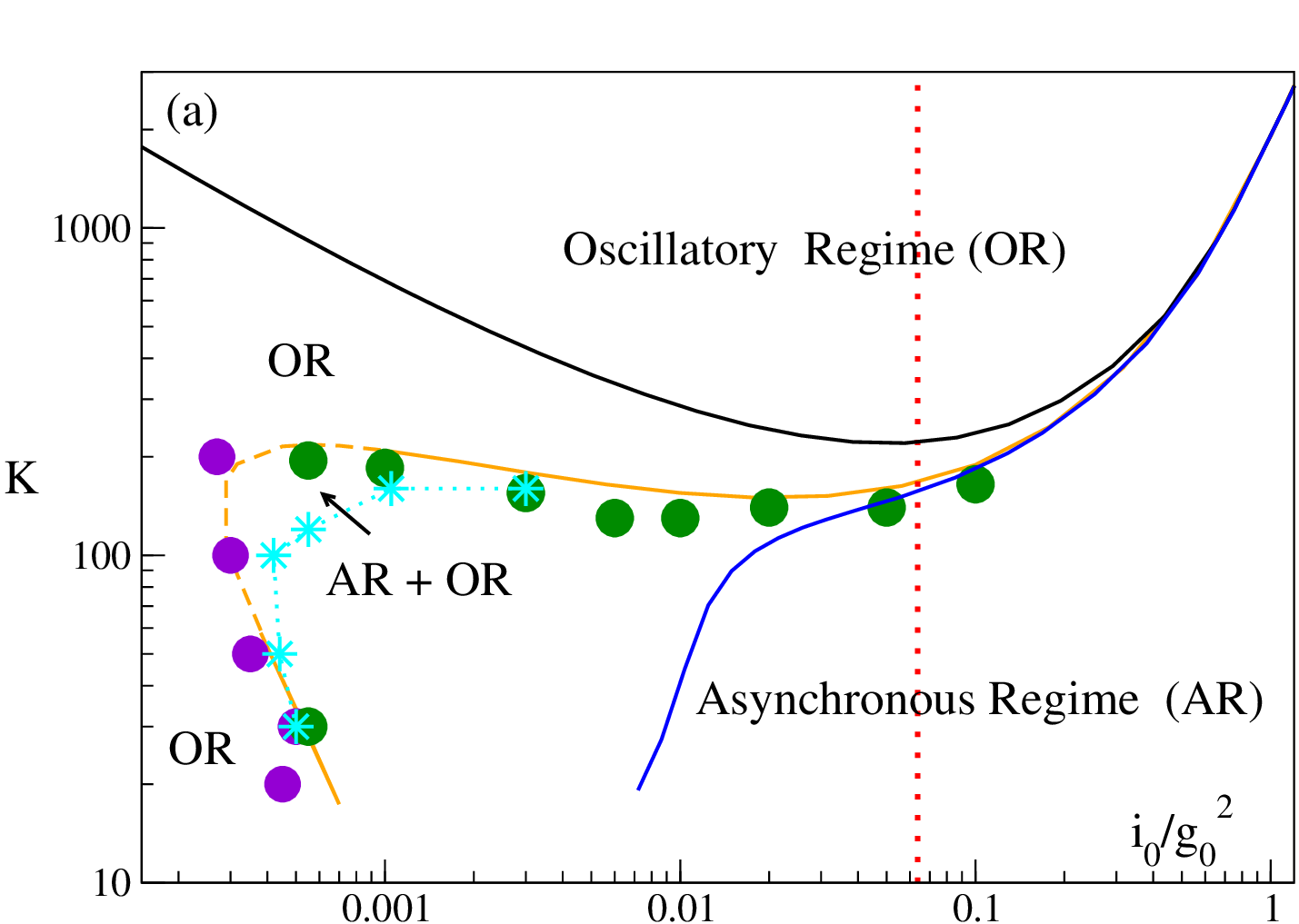}
\includegraphics[width=0.47\textwidth]{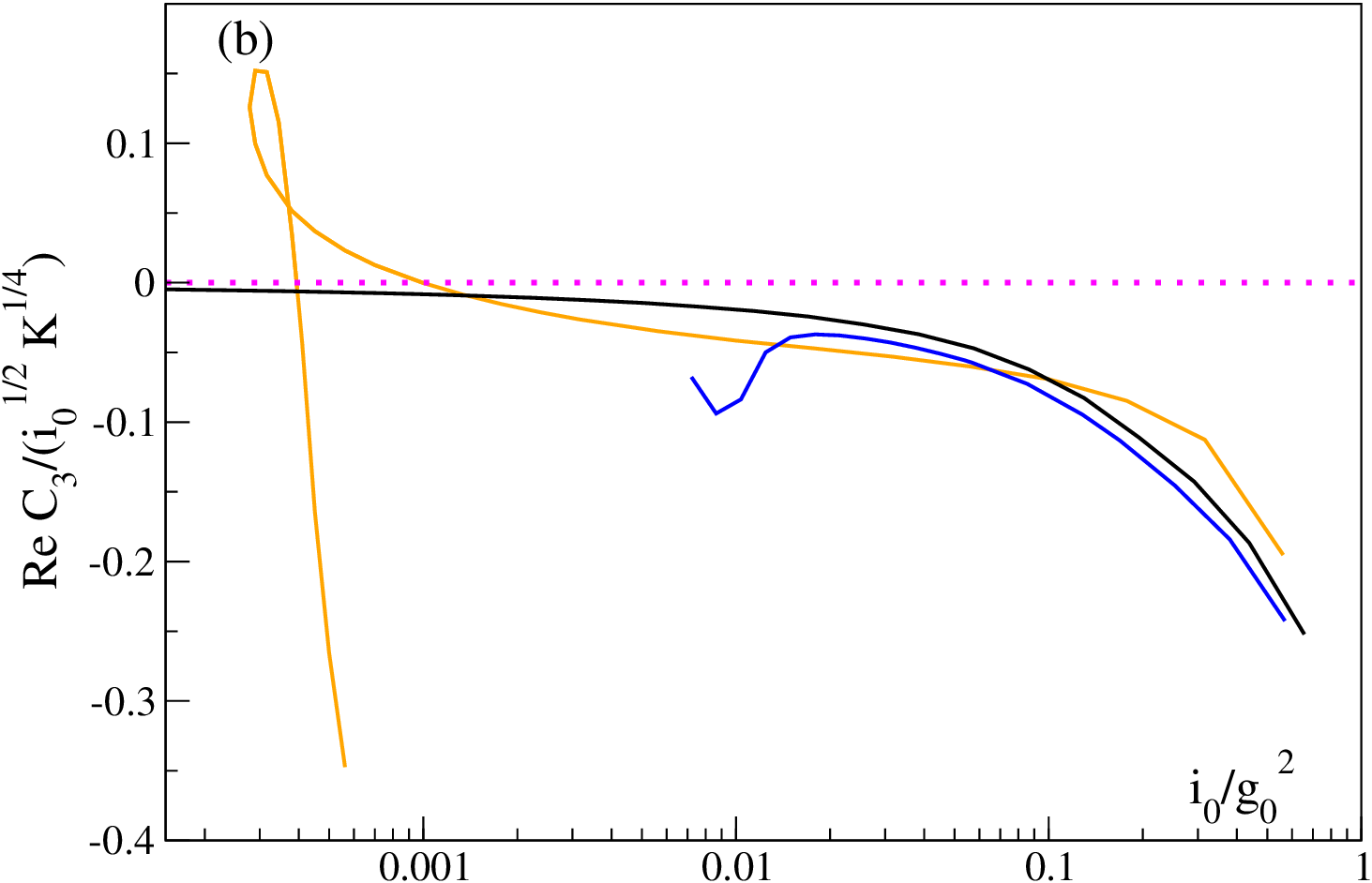}}
\caption{(a) Phase diagram for the QIF network in the plane
$(i_0/g_0^2, K)$: the black (blue) solid line is the super-critical HB line
obtained within the DA (D3A); the orange solid (dashed) line is the
super- (sub-) critical HB line given by the CMF; the symbols refer to numerical estimations of the HBs and SNBs.
The green (violet) circles denote HBs obtained by performing
quasi-adiabatic simulations by varying $K$ ($i_0$) for constant $i_0$
($K$) values; the cyan stars indicate SNBs. The dimension of symbols takes in account for
the error bar in the transition point evaluation. The vertical red dotted line is the critical
value defined in \cite{noi} within the DA for the emergence of fluctuation-driven balanced asynchronous
dynamics (for more details see the text).
(b) The coefficient $\mathrm{Re}\,C_3$ in rescaled units at the critical points $K=K_{cr}$ versus
$i_0/g_0^2$: CM orange solid line), DA (black line) and D3A (blue line).
In all the simulation $g_0=1$.
}
  \label{figphasediag}
\end{figure}

To find the amplitude equation we employ the standard multiple scale method \cite{kuramoto2012,nayfeh2024}  with formal small parameter $\varepsilon$\,;  $\frac{\mathrm{d}}{\mathrm{d}t}=\frac{\partial}{\partial t_0} +\varepsilon\frac{\partial}{\partial t_1} +\varepsilon^2\frac{\partial}{\partial t_2} +\dots$,
$x_n=\varepsilon x_n^{(1)}+\varepsilon^2 x_n^{(2)}+\varepsilon^3 x_n^{(3)}+\dots$, where $t_0$ is the time scale related to the period of the emergent oscillation and $t_1$, $t_2$ are the ``slow'' time scales--- associated to the modulation of the oscillation amplitude. In proximity of a Hopf bifurcation one expects the following scaling $|\mathbf{x}|\sim(K-K_\mathrm{cr})^{1/2}$
for the oscillation amplitude, thus suggesting to expand the control parameter as $K=K_\mathrm{cr}+\varepsilon^2 K_2$, where $\varepsilon^2 K_2$ is the deviation from the bifurcation point.

By substituting these expansions in Eq.~(\ref{eqGPD04}) we obtain at the first order in $\varepsilon$ the
following expression
\[
x_n^{(1)}=A(t_1,t_2,\dots)Y_ne^{i\Omega t_0}+c.c.\,.
\]

And at the order $\varepsilon^2$:
\begin{align}
&\frac{\partial x_n^{(2)}}{\partial t_0}+\frac{\partial A}{\partial t_1}Y_ne^{i\Omega t_0}+c.c.=F_{nm}x_m^{(2)}
%\nonumber\\
%&{}
+\widetilde{F}_{nml}\big[A^2Y_mY_le^{i2\Omega t_0}+c.c.+|A|^2(Y_mY_l^\ast+Y_m^\ast Y_l)\big]\,.
\nonumber
\end{align}
One finds $\partial A/\partial t_1=0$  and
\[
x_n^{(2)}=A^2\chi_ne^{i2\Omega t_0}+c.c.+|A|^2\phi_n\,,
\]
where $\phi_n$ is the solution of the linear equation system
\[
F_{nm}\phi_m=-\widetilde{F}_{nml}(Y_mY_l^\ast+Y_m^\ast Y_l)\,,
\]
and
$\chi_n$ is the solution of
\[
(F_{nm}-i2\Omega\delta_{nm})\chi_m=-\widetilde{F}_{nml}Y_mY_l\,.
\]
At the order $\varepsilon^3$, we find the solvability condition,
\begin{align}
&\frac{\partial A}{\partial t_2}(Y_{n^\prime}^+)^\ast Y_{n^\prime}
=K_2 A(Y_n^+)^\ast\left(\frac{\partial F_{nm}}{\partial K}\right)_{i_0,g_0}\!Y_m
\nonumber\\
&
+A|A|^2(Y_n^+)^\ast\widetilde{F}_{nml}\left(Y_m^\ast\chi_l+\chi_mY_l^\ast +Y_m\phi_l+\phi_mY_l\right)\,.
\nonumber
\end{align}
The latter is the so-called amplitude equation and it can be recast in the standard form 
\begin{equation}
\frac{\partial A}{\partial t}
=(K-K_\mathrm{cr})\frac{\mathrm{d}\lambda}{\mathrm{d}K}A +C_3A|A|^2\,,
\label{eqGPD05}
\end{equation}
where $A$ is the complex amplitude of oscillations, $x_n=AY_ne^{i\Omega t}+c.c.+\mathcal{O}(A^2)$; $\lambda$ is the exponential growth rate of the leading mode. The expression of the coefficient $C_3$ being the following:
\begin{equation}
C_3=\frac{(Y_n^+)^\ast\widetilde{F}_{nml}\left(Y_m^\ast\chi_l+\chi_mY_l^\ast +Y_m\phi_l+\phi_mY_l\right)}{(Y_{n^\prime}^+)^\ast Y_{n^\prime} }\; .
\label{eqGPD06}
\end{equation}
The sign of the real part of $C_3$ identifies the Hopf bifurcation as sub-critical (super-critical) for $\mathrm{Re} \enskip C_3 >0$ ($\mathrm{Re} \enskip C_3 < 0$).

The Hopf bifurcation lines obtained in the plane $(i_0/g_0^2,K)$
with this approach are reported for CMF (orange line), DA (black line) and D3A (blue line) in Fig. \ref{figphasediag} (a).
For the DA and D3A the Hopf lines are always super-critical, this is not the case
for the CMF, where they can be either super-critical (solid orange line) or sub-critical
(dashed orange line). The sub- or super-critical nature of the Hopf bifurcation is
decided on the basis of the sign of  $\mathrm{Re} \enskip C_3$ displayed in Fig.~\ref{figphasediag} (b)
as a function of $i_0/g_0^2$ at the critical points $K=K_{cr}$.

A peculiarity of the CMF results is that the HB line is re-entrant, thus in a certain range of
$i_0/g_0^2$ we have an asynchronous regime  only within a finite interval of in-degrees and
GOs at sufficiently small and large $K$. As explained in the following these
two oscillatory regimes are due to  different mechanisms.

Furthermore, there is a dramatic difference among the CMF and D3A approaches and the DA at small $i_0$ (large $g_0) $: within the DA  GOs are observable only above a critical $K$ diverging to infinity for $i_0/g_0^2 \to 0$, while
for the CMF (D3A) analysis GOs are present at any $K$ value for $i_0/g_0^2 < 0.00029$ ($i_0/g_0^2 < 0.007$).

\begin{figure}
\includegraphics[width=0.32\textwidth]{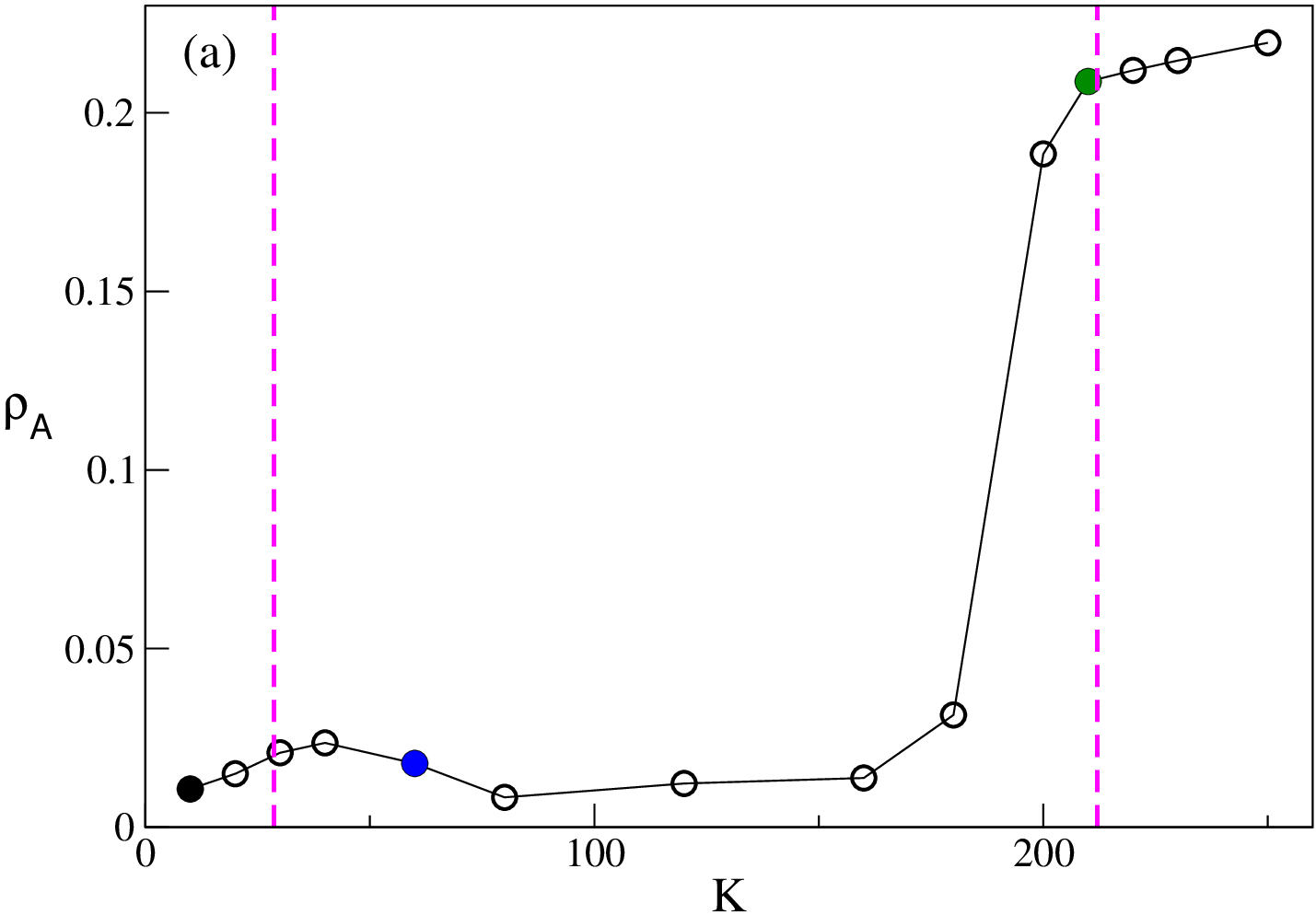}
\includegraphics[width=0.32\textwidth]{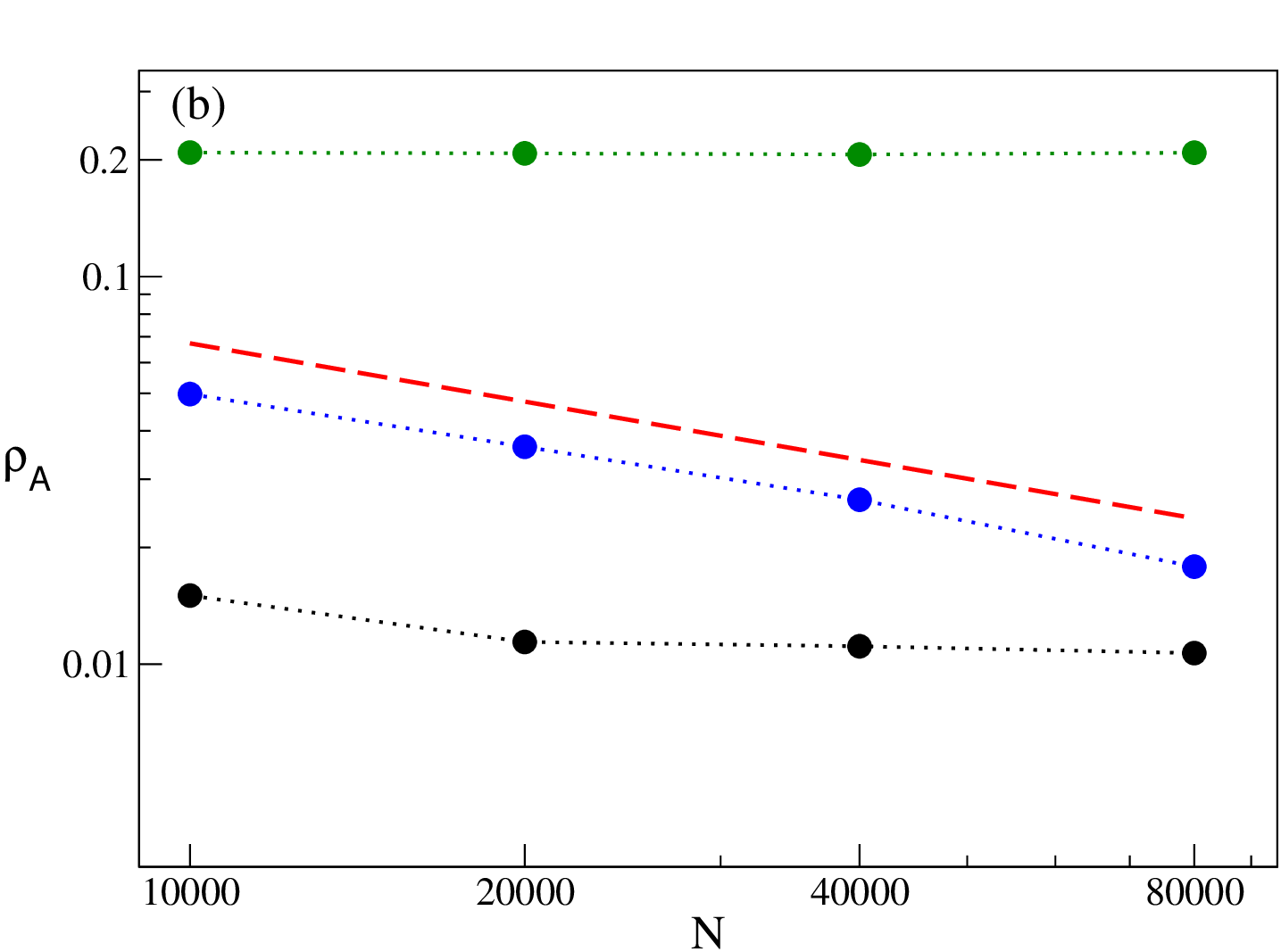}
\includegraphics[width=0.32\textwidth]{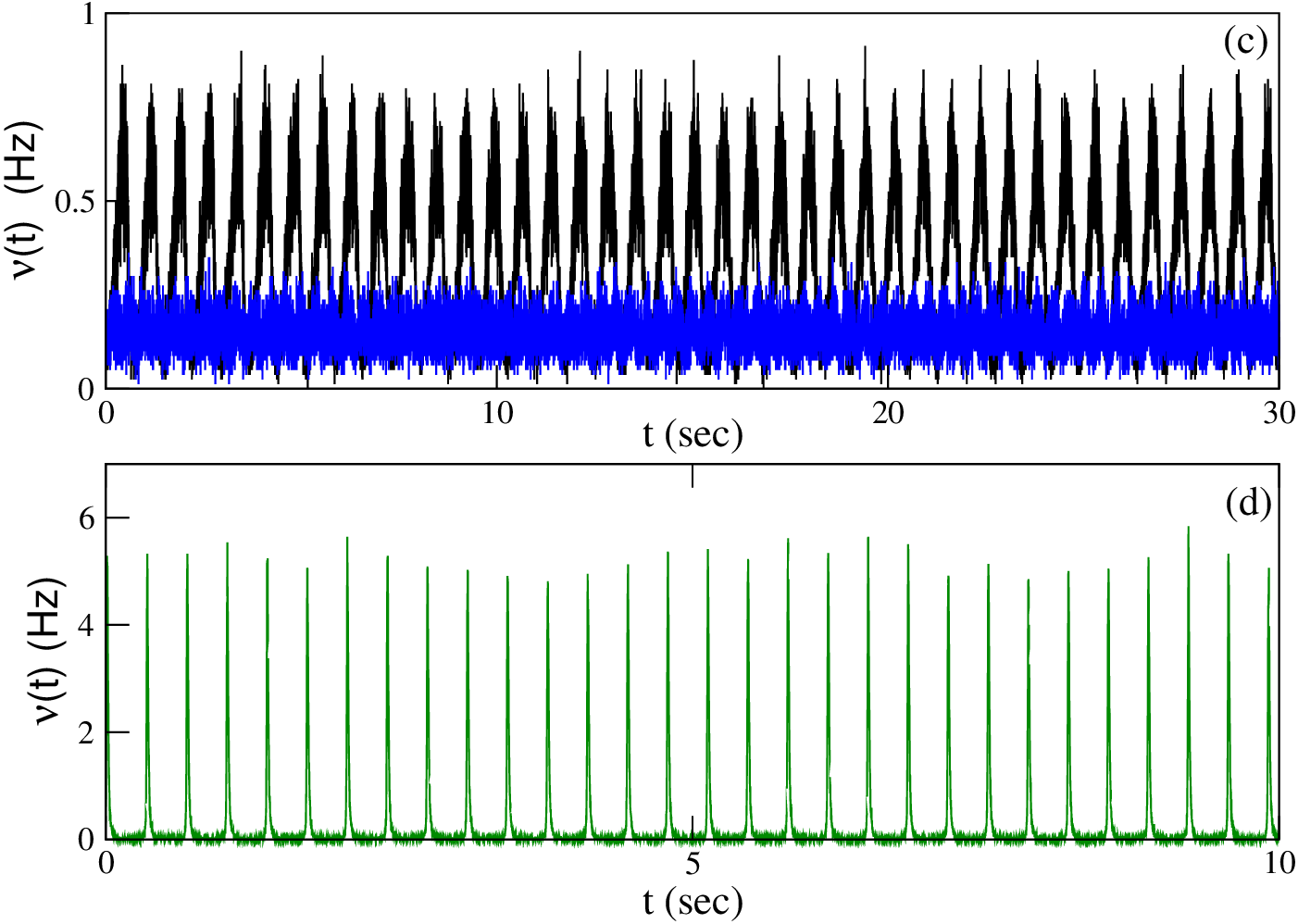}
\caption{Average order parameter $\rho_A$ versus $K$ (a) and versus the system size $N$ (b) for $K=10$ (black circles),
$K=60$ (blue circles) and $K=210$ (green circles). In (a) the vertical magenta lines denote
the HBs obtained within the CMF approach. In (b) the dotted lines are a guide for the eyes, while the red dashed line
denotes a power-law decay $\propto N^{-1/2}$. The values of $\rho_A$ have been
averaged over 5-20 network realizations for a time interval $t_E=20 - 30$ s
following a transient of $t_T= 20-30$ s. (c-d) Firing rates $\nu(t)$ versus time $t$ for $K=10$ (black line)
and $K=60$ (blue line) in (c) and $K=210$ (green line) in (d).
The data in (a) refer to quasi-adiabatic simulations for fixed $i_0$
obtained by increasing (decreasing) $K$ in steps $\Delta K = 10-20$.
All data refer to $i_0 = 0.00055$ and $g_0=1$,
the simulations in (a) and (c-d) concern networks of size $N=80000$.
}
\label{figS0}
\end{figure}

\section{Network Simulations}

In order to verify the CMF predictions we have performed extremely accurate numerical
simulations of QIF networks, according to \eqref{eq:1},  by employing an event-driven integration scheme (see Appendix A for more information), which allowed us to follow the network dynamics for long times, up to $60-100$ sec, for systems of size $N = 10000 - 80000$.
The CMF (\ref{eqGP03},\ref{eqGP12}) is able to reproduce reasonably well network simulations (black curves)
as shown by the data reported in Fig. \ref{fig1} (b-d) for various values of $K$ and $i_0$. Furthermore, also the agreement between the Langevin results with shot-noise (violet curve) and the CMF (blue curve) is definitely good for $K > 20$, as shown in Fig. \ref{fig1} (b-d). These comparisons confirm the validity of the considered CMF approach.

Let us now compare the bifurcation diagrams obtained within the CMF approach and via extensive numerical simulations.
In particular, to characterize the macroscopic evolution of the network we measured the
indicator $\rho_A$ \eqref{indicator} averaged over several different network realizations.
As we will show in the following a finite-size analysis of this order parameter has allowed us to identify numerically the
Hopf (HBs) and the Saddle-Node Bifurcations (SNBs). This analysis is essentially based on the fact that
a coherent macroscopic activity is characterized by a value of $\rho_A$ remaining finite in the thermodynamic limit
(irrespective of its value), while asynchronous dynamics is associated to a vanishing $\rho_A$, indeed
from the central limit theorem  one expects that $\rho_A \propto N^{-1/2}$ for $N \to \infty$
\cite{ullner2016,matteo}.

The results of this analysis are reported in Fig. \ref{figphasediag}(a) green (magenta) circles
refer to HBs identified via quasi-adiabatic simulations by varying $K$ ($i_0$) for constant $i_0$ ($K$) values;
while the cyan stars indicate SNBs. Numerical simulations are in good agreement with
the CMF results and allowed us also the identification of a coexistence region for
asynchronous and oscillatory collective dynamics.

In large part of the phase diagram (namely, for $i_0/g_0^2 < 0.1$), both in the AR and OR
we observe an irregular firing activity of the neurons
associated to population averaged coefficient of variations  ${CV} \simeq 0.8-1.0$, as expected in sparse balanced networks.
This is consistent with the analysis reported in \cite{noi},
where within the DA a critical value $i_0/g_0^2 = 0.0637 \dots$ separating fluctuation-driven from mean-driven balanced asynchronous dynamics \cite{lerchner2006} has been identified in the thermodynamic limit. This value, shown in Fig. \ref{figphasediag} (a) as a red dotted vertical line, can be considered as a lower limit, below which the spiking activity is definitely irregular.

\begin{figure}
\includegraphics[width=0.4\textwidth]{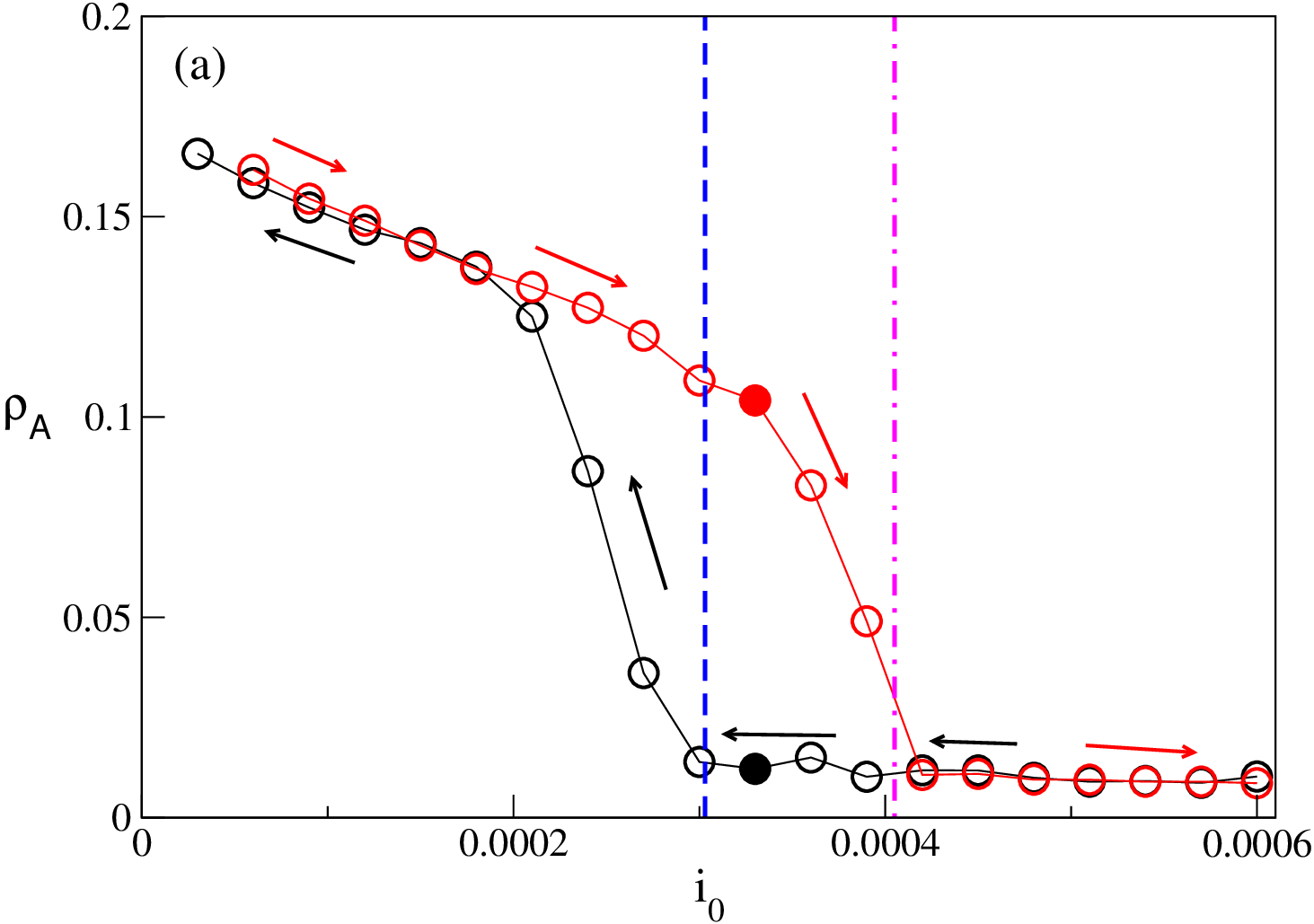}
\includegraphics[width=0.4\textwidth]{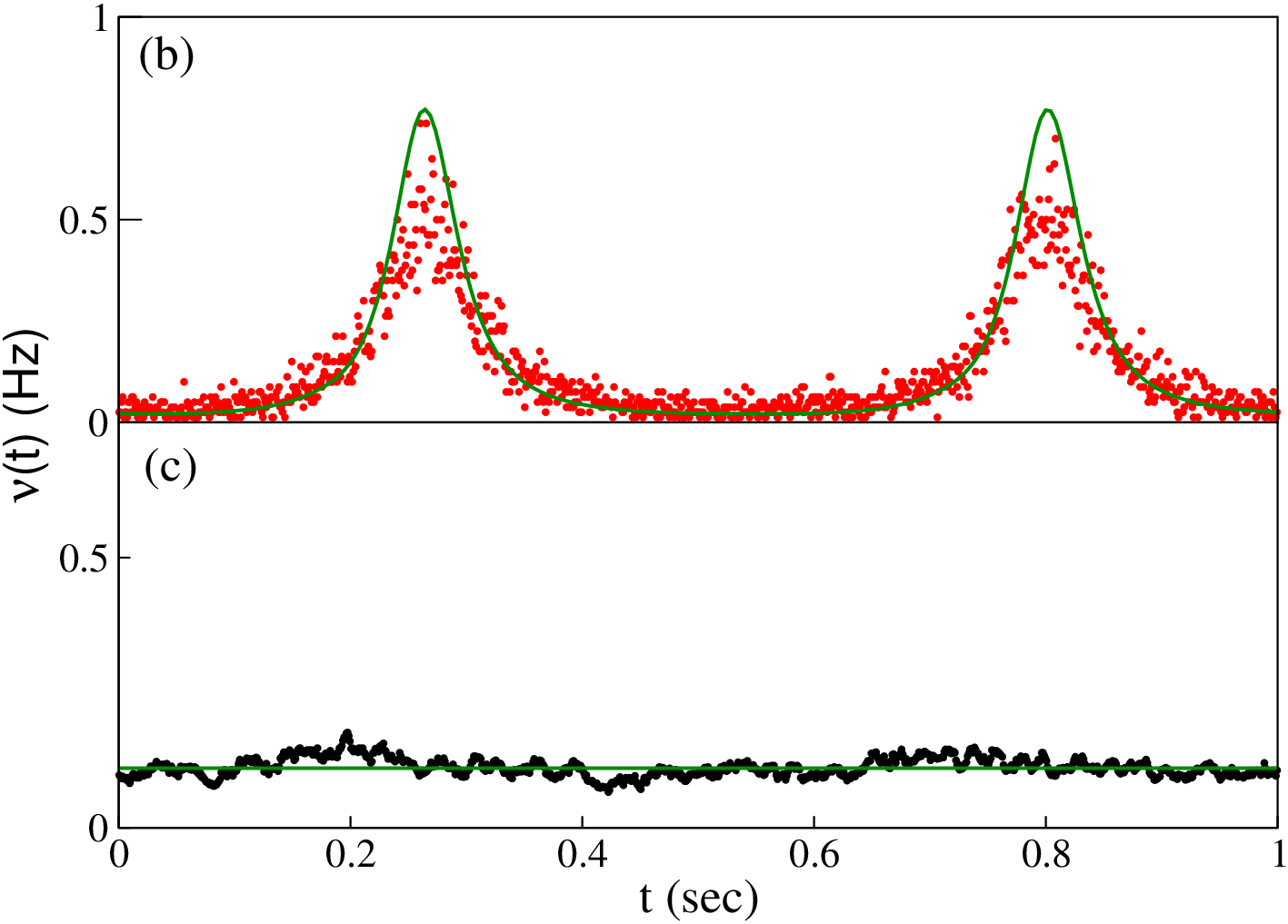}
\caption{(a) Average order parameter $\rho_A$ versus $i_0$
for $K=100$.  The blue dashed line in (a) denotes the sub-critical HB given by
the CMF and the magenta dot-dashed line: numerically estimated SNB. In
(b-c) the population firing rates $\nu(t)$ versus time are reported for the states indicated by
the corresponding colored filled circles, the results for the CMF are also shown as green solid lines.
The data in (a) refer to quasi-adiabatic simulations for fixed in-degree
obtained by increasing (decreasing) $i_0$ in steps $\Delta i_0 = 0.00003$.
The values of $\rho_A$ in (a) have been averaged over $5$ network realizations for
$t_E=30$ s following a transient of $t_T=20$ s. For all data $N=80000$ and $g_0=1$.
}
\label{fig7}
\end{figure}

\subsection{Finite size characterization of the observed transitions}

Let us now explain in details for two characteristic cases how we proceeded to
estimate numerically the transitions.
In order to compare simulations done for a finite network to the CMF results obtained
in the limit $N \to \infty$ it is necessary to average over different network realizations, to reduce
finite size effects. The HBs and SNBs reported in Fig. \ref{figphasediag} (a) have been obtained by
performing finite size analysis of the order parameter $\rho_A$
obtained during quasi-adiabatic simulations of the QIF network by varying $K$ ($i_0$) for constant
$i_0$ ($K$). In particular, the average value of the order parameter $\rho_A$
for each system size $N$ has been obtained by averaging over
a time $t_E = 20 - 30$ s after discarding a transient of $t_T = 20 - 30$ s
and over 5 to 20 different network realizations. Therefore, $\rho_A$ is obtained
as a double average over time and over different random realizations of the connections among the neurons
by maintaining $K$ constant, the latter amounts to
average over  realizations of the quenched disorder present in the network.
According to the central limit theorem and analogously to what done in \cite{matteo},  we have classified
the dynamical regimes by estimating the parameter $\rho_A$ for increasing system sizes $N$:
if the indicator $\rho_A$ decreases as $1/ \sqrt{N}$ (saturates to some constant value)
then the dynamics is identified as asynchronous (oscillatory). The actual value of $\rho_A$
is just related to the level of synchronization in the neuronal population not to the 
specific macroscopic regime (asynchronous or GOS) displayed by the network.

Furthermore, as shown in Fig. \ref{figS0} (a) for sufficiently small $i_0/g_0^2$ values,
GOs are observable at small ($K \le 30$) and large ($K \ge 200$) in-degrees, while
the AR is present only at intermediate in-degrees ($K \in [40,180]$).
The dynamics in these three intervals is visualized by reporting in Fig. \ref{figS0} (c-d) the firing rates $\nu(t)$ at $K=10$ (black line),
$K=60$ (blue line) (c) and $K=210$ (green line) (d).

In particular, we have estimated, for these three different values of the in-degree
$K=10,60$, and 210, the scaling of $\rho_A$ for system sizes $N=10000$, 20000, 40000, and 80000.
From Fig. \ref{figS0} (a) we observe that for $K=10$ and $K=210$ (black and green symbols, respectively)
$\rho_A$ saturates to some constant value and therefore we expect oscillatory dynamics
as confirmed by the evolution of the population firing rate $\nu(t)$ reported in Fig. \ref{figS0} (c-d).
The value to which the indicator converges is irrelevant for the
observation of GOs, it only indicates a different level
of synchronization among the neurons. Indeed the neurons are much more
synchronized for $K=210$ as  shown in Fig. \ref{figS0} (d) (green line), as compared
to the case $K=10$ reported in  Fig. \ref{figS0} (c) (black line).
For $K=60$ we observe that $\rho_A$ decreases as $1/ \sqrt{N}$
as shown in Fig. \ref{figS0} (b)(blue symbols) and as expected
the population firing rate displays irregular fluctuations around a constant value,
see Fig. \ref{figS0} (c)(blue line),
therefore the dynamics can be safely identified as asynchronous in this case.
In this case we do not observe any hysteretic effect, consistently with what predicted by the CMF theory.

A hysteretic transition from AR to OR is instead obtained by varying quasi-adiabatically $i_0$
as displayed in Fig. \ref{fig7} (a) for $K=100$ and $g_0=1$, the coexistence region can be clearly identified
between the sub-critical HB (blue dashed line) and the SNB of limit cycles (magenta dot-dashed line).
Two coexisting solutions are reported in Fig. \ref{fig7} (b-c)
confirming the good agreement between CMF (green lines) and the network simulations
(red and black circles).

\begin{figure}
\includegraphics[width=0.45\textwidth]{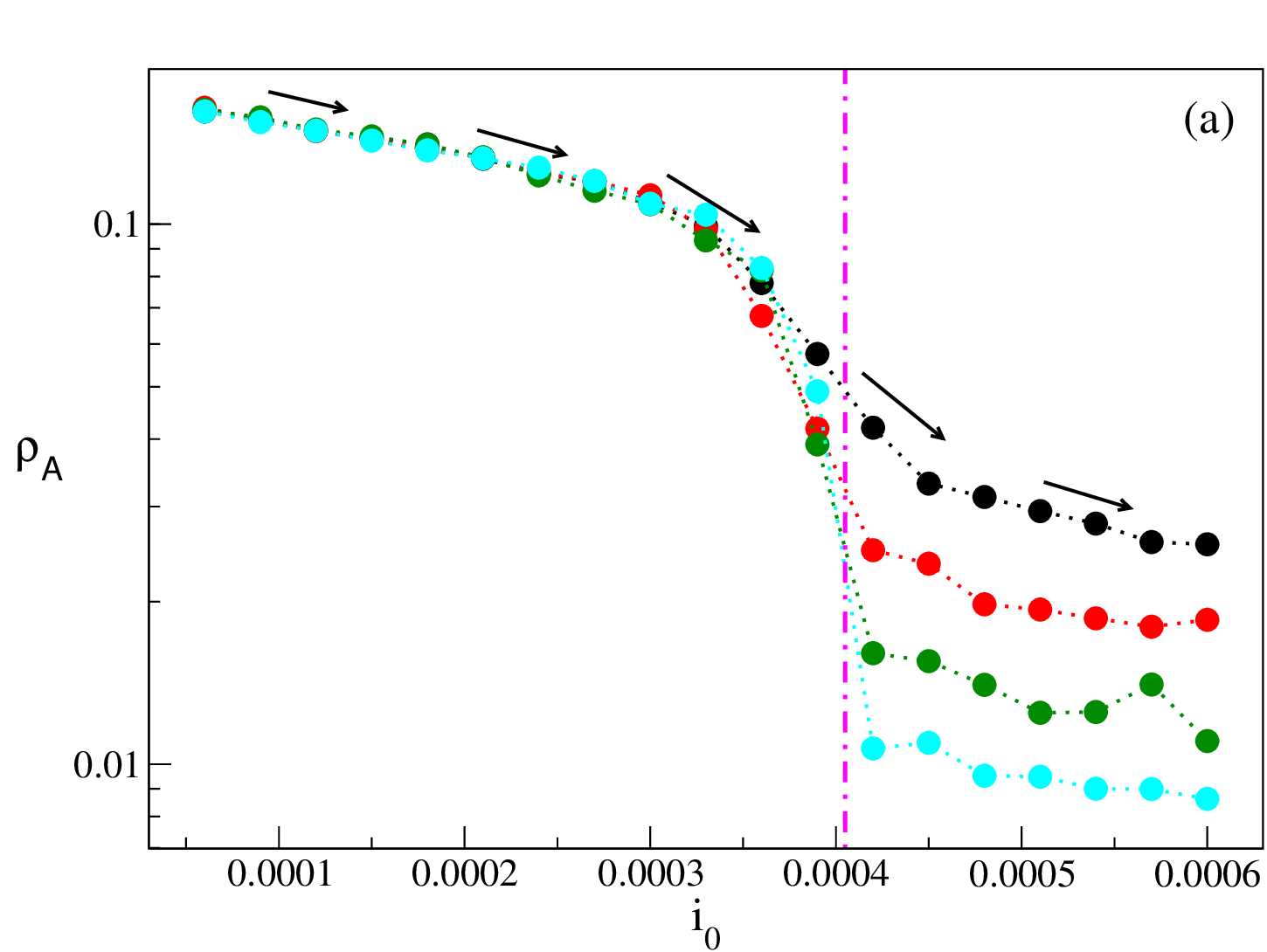}
\includegraphics[width=0.45\textwidth]{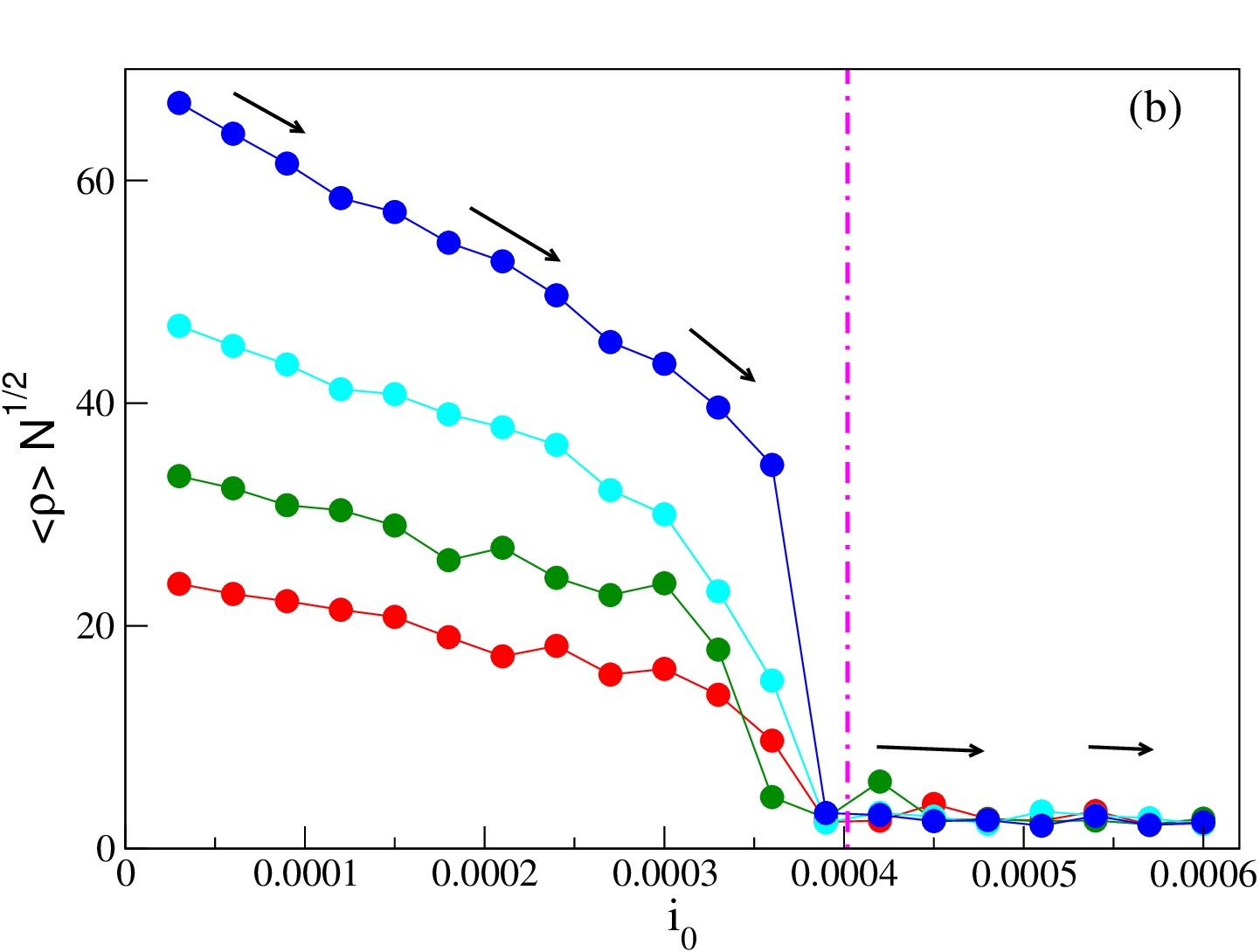}
\includegraphics[width=0.45\textwidth]{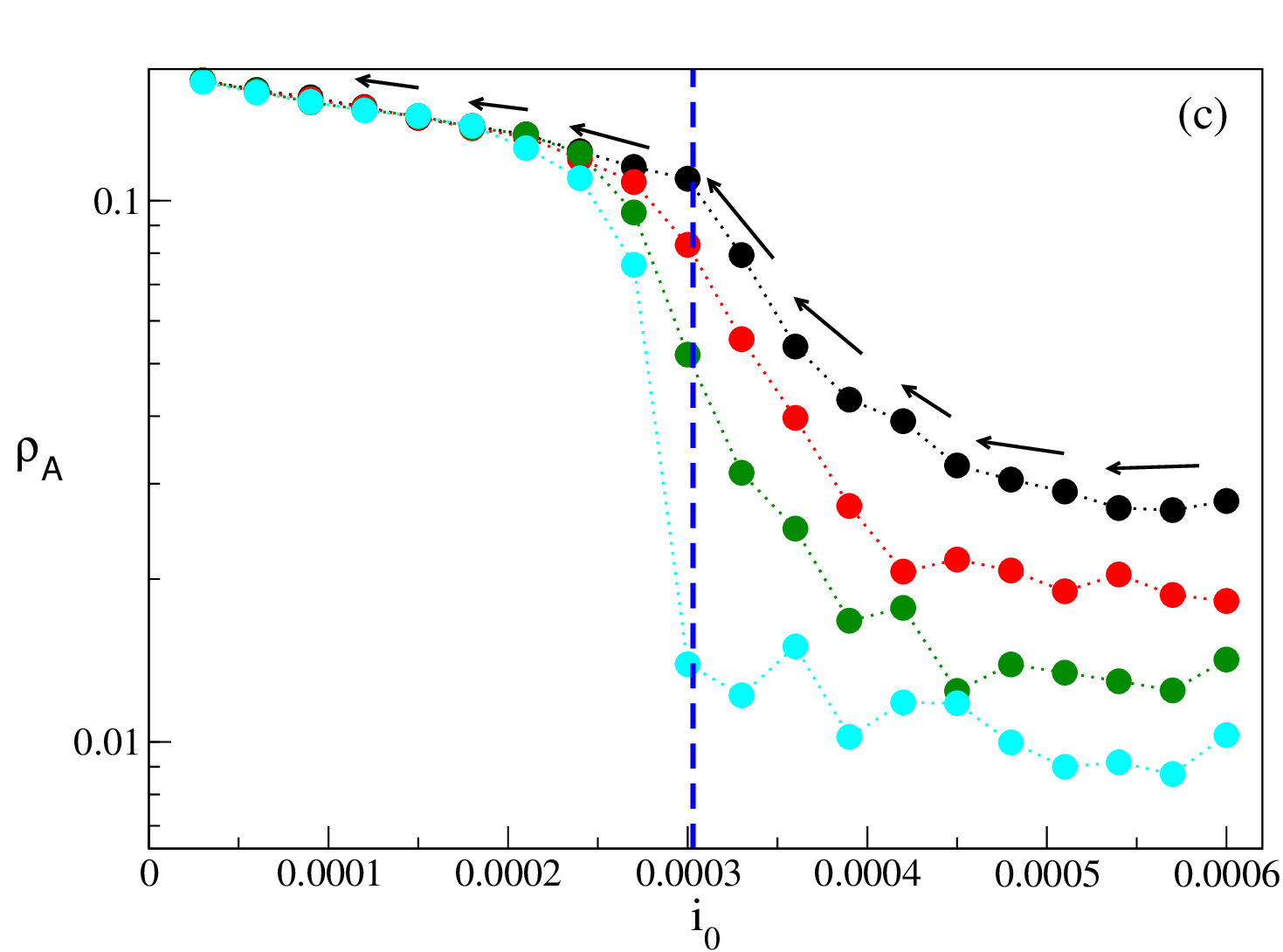}
\includegraphics[width=0.45\textwidth]{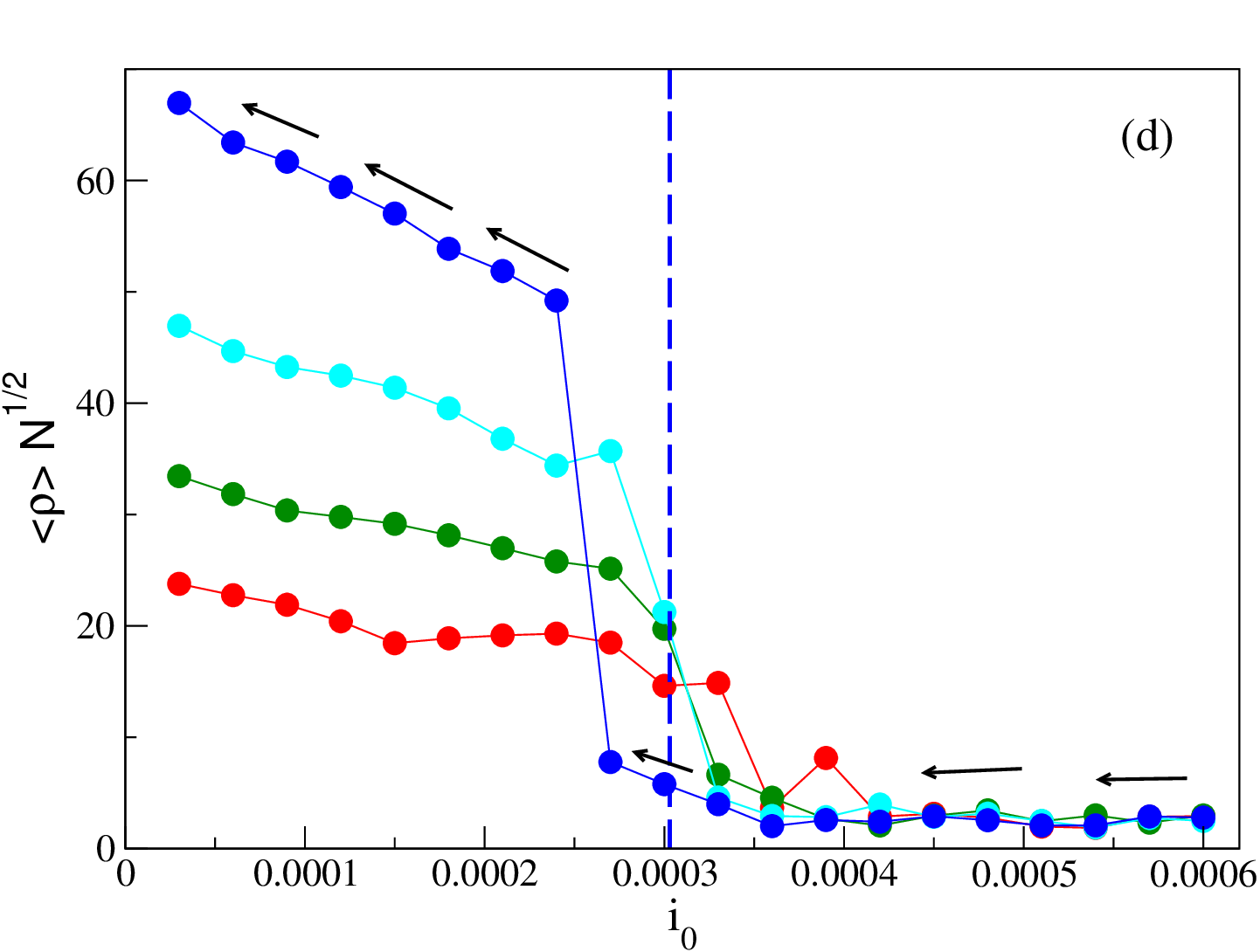}
\caption{(a,c) Average order parameter $\rho_A$ versus $i_0$ for different system sizes:
$N=10000$ (black circles),  20000 (red circles), 40000 (green circles) and 80000 (cyan circles).
The values of $\rho_A$ have been
averaged over 5-20 network realizations for a time interval $t_E=20 - 30$ s
following a transient of $t_T= 20-30$ s.
(b,d) Order parameter $\langle \rho \rangle$ for a single network realization multiplied by
$\sqrt{N}$ versus $i_0$
for different network sizes. The sizes are identified by the same symbols as in panel (a)  apart
$N=160000$ (blue circles).
The parameter $\rho$ has been averaged in time for a time interval $t_E=20$ s
following a transient of $t_T= 60$ s and refers to a single network realization.
The magenta dash-dotted line in (a) and (b) is the SNB  numerically estimated via finite size
scaling analysis of $\rho_A$, while the
blue dashed line in (c-d) is the sub-critical HB obtained from the CMF approach.
All data refer to quasi-adiabatic simulations of QIF networks with $g_0=1$ for fixed in-degree $K = 100$ by varying $i_0$:
the data in (a) and (b) (in (c) and (d)) have been obtained by increasing (decreasing) $i_0$ in steps $\Delta i_0 = 0.00003$.}
\label{figS1}
\end{figure}

The bifurcations reported in Fig. \ref{fig7} (a) have  been identified as a sub-critical HB and a SNB of limit cycles,
since they were associated to a hysteretic transition from asynchronous to oscillatory dynamics.
The sub-critical HB, as identified from the CMF approach, is located at $i_0^{(HB)} \simeq 0.00303$,
while the SNB at $i_0^{(SN)} \simeq 0.0040(1)$ has been identified numerically by finite size scaling of the order parameter
$\rho_A$ as explained in the following.

 It is known that sub-critical HBs (SNBs of limit cycles) are characterized by an abrupt transition from an
 asynchronous to an oscillatory state (from an oscillatory to a stationary state), however from the Fig. \ref{fig7} (a) both transition appear
as smoothed over a finite interval of the current $i_0$.
This is due to two facts: 1) in the present case
the asynchronous regime corresponds to a focus, therefore damped oscillations forced
by finite size fluctuations are present in the
asynchronous state; 2) the indicator $\rho_A$ is averaged over different network
realizations. This average allows us to give a better estimate of the transition points, however
it has also the drawback to smooth out the transition over a finite interval, since different networks will
have slightly different transition points. However, we expect that for increasing system sizes
the transition region will become narrower, since the network realization will be
less statistically different and finite size fluctuations will decrease.

Let us describe in details the finite size analysis we have performed.
As a first analysis we consider $\rho_A$ versus $i_0$ as obtained
by quasi-adiabatic simulations by increasing (decreasing) $i_0$ from the value $6 \times 10^{-5}$ ($6 \times 10^{-4}$) to
$6 \times 10^{-4}$  ($6 \times 10^{-5}$) in steps
of $\Delta i_0 = 3 \times 10^{-5} $ for system sizes $N=10000$, 20000, 40000 and 80000
(the data are shown in Fig. \ref{figS1} (a) and (c)).
It is evident from the figures that the transition region shrinks for increasing $N$
and at the same time  for $i_0 > i_0^{(SN)}$
($i_0 > i_0^{(HB)}$) the value of the parameter decreases drastically with $N$
indicating that the dynamics above such current is asynchronous.
On the other hand for $i_0 < i_0^{(SN)}$ ($i_0 < i_0^{(HB)}$)
$\rho_A$ either remains essentially
constant or does not present a clear decrease with $N$, in this range of currents
we can affirm that we have a coherent behaviour characterized by GOs.
Furthermore by increasing $N$ the jumps of the order parameter at the transition become higher and steeper,
but still at $N=80000$ they seem not to be abrupt. As previously stated, we believe that the origin of this smoothing
is  related not only to the finite size effects, but also to the fact that
$\rho_A$ is obtained as an average over different network realizations, typically presenting
slightly different values of the bifurcation points  $i_0^{(SN)}$ and  $i_0^{(HB)}$.

To verify this conjecture we have considered the time average of the indicator $\langle \rho \rangle$ for only one realization of
the network and for systems sizes increasing from $N=20000$ to $N=160000$, the results of this analysis are reported in
Fig. \ref{figS1} (b) and (d). To make more evident the transitions we have now reported
$\langle \rho \rangle$ multiplied by $\sqrt{N}$ in the figures, in this case
for increasing $N$ one will observe an almost constant (growing) value of $\langle \rho \rangle \sqrt{N}$
for asynchronous (oscillatory) dynamics. As a matter of fact, in Fig. \ref{figS1} (b)
we observe that  for $i_0 \le 0.00039 < i_0^{(SN)}$ the quantity
$\langle \rho \rangle \sqrt{N}$  grows as $N^{1/2}$, while above such value it is almost constant.
Furthermore, at $N=160000$ we see an abrupt transition from a value $\langle \rho \rangle \sqrt{N} \simeq 34$
at $i_0 = 0.00036$ to a value $\langle \rho \rangle \sqrt{N} \simeq 2$ at $i_0 = 0.00039 $.
Similar effects are observable in Fig. \ref{figS1} (d) where the sub-critical Hopf is characterized,
and also in this case at $N=160000$ the jump from asynchronous to oscillatory dynamics is abrupt.
Furthermore, the single realization gives values of the bifurcation points slightly different from
that obtained from the analysis of the average $\rho_A$ as expected.
These indications are consistent with the interpretation we have given of the
smoothing of the transitions observable in this case over a finite interval of $i_0$ and
of the lack of abrupt variations of the measured order parameter.

\subsection{Frequency range of the global oscillations}

\begin{figure}
\begin{center}
\includegraphics[width=0.9 \linewidth]{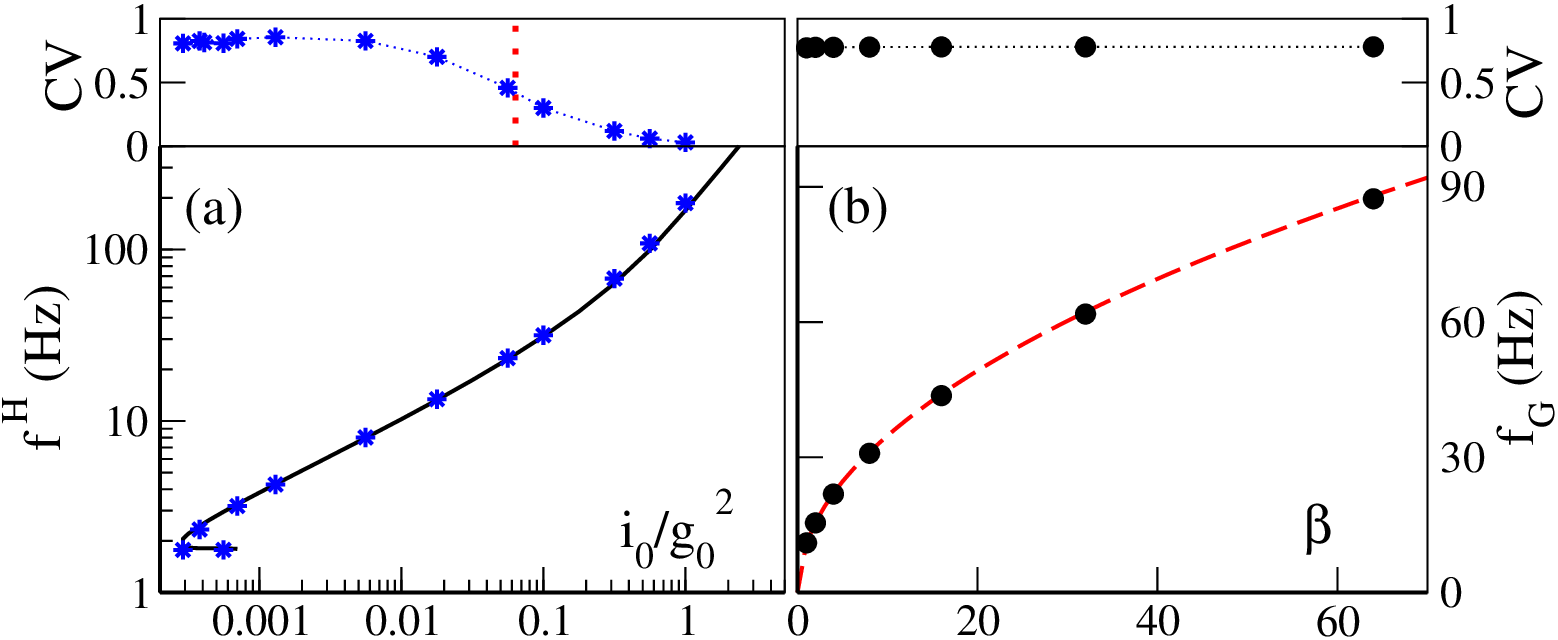}
\end{center}
\caption{(a) Frequency $f^H$ of the GOs at the HB versus $i_0/g_0^2$: symbols are simulations for $N=20000$ and the solid line is the CMF result. (b) Frequency $f_{C}$ of the GOs as a function of the parameter $\beta$,
where $i_0=\beta \times 0.01$, $g_0=\sqrt{\beta}$, $K=200$. Circles are network simulation data with $N=20000$ and the red dashed line
represents the curve $\nu_{GO}= 11 \sqrt{\beta}$ Hz. The corresponding population averaged coefficients of variation CVs are shown above the two panels with the same symbols employed to denote simulation data.
The vertical red dotted line shown in the upper part of panel (a) is the critical
value defined in \cite{noi} within the DA for the emergence of fluctuation-driven balanced asynchronous
dynamics (for more details see the text). All data have been obtained by considering a time interval $t_E=100$ s, 
after discarding a transient of $t_T=20$ s.  In (a) we set $g_0=1$.
}
\label{fig8s}
\end{figure}

At the HBs, GOs emerge with a frequency $f^H$ that is reported as a function of $i_0/g_0^2$ in Fig. \ref{fig8s} (a). The comparison between the CMF results (solid line) network simulations with $N=20000$ (blue stars) is very good along the whole bifurcation line predicted by the CMF. Furthermore, $f^H$ covers a wide range
of frequencies ranging from $1.77$ Hz ($\delta$ band) to $\simeq 100$ Hz ($\gamma$ band),  according to the usual classification of the brain rhythms\cite{buzsaki2006}.  As visible in the upper part of the
panel the population averaged coefficient of variation $CV$ is significantly greater than zero for $i_0/g_0^2 < 0.1$, where,
as previously discussed, the firing activity is driven by fluctuations, as expected in the balanced regime. The fluctuation-driven regime is associated to quite high values of the ratio $R_G \simeq 6-11$, indicating that only $9-16 \%$ of neurons
participates to each population burst and that their activity is definitely slower than the collective one.
Instead, in the mean-driven regime $R_G \to 2$ indicating that we observe the emergence of a peculiar
cluster synchronization characterized by two groups (clusters) of neurons alternating their participation to the population bursts 
with low irregularity. This dynamics is quite similar to the one recently reported
in \cite{feld2024} for globally coupled QIF networks subject to additive Gaussian noise.
Furthermore, across the examined range, $R_0 \simeq 0.9-1$, suggesting that the GOs are associated to those
neurons receiving few/no inhibitory PSPs during the global oscillation period.

As predicted by the CMF, the same dynamics should be observable at fixed $K$
by maintaining the ratio $i_0/g_0^2$ constant. We verified this by considering a state in the oscillatory regime corresponding to $(K,i_0/g_0^2) = (200,0.01)$ and by
varying, as a function of  a control parameter $\beta$, the synaptic coupling and the current
as $g_0 = \sqrt{\beta}$  and $i_0 = \beta \times 0.01$, while $K$ stays constant.
In particular, we observed irregular dynamics characterized by an average $  \overline {CV} \simeq 0.78$,
as shown in the upper part of Fig. \ref{fig8s} (b),
associated to GOs in the whole examined range $\beta \in [1,64]$.
As expected, only the time scale varies, decreasing as $1/\sqrt{I}$, as previously shown,
consequently the frequency $f_G$ of the GOs grows proportionally to $\sqrt{\beta}$
(as shown in Fig. \ref{fig8s} (b)) as well as the population average of the firing rates $F$.
Obviously, the ratio of the two frequencies remains
constant, in the present case the ratio $R_G \simeq 7.05$ indicating that one has always a slow firing
activity of the neurons with respect to the GOs. Moreover, also $R_0$ stays almost constant around $\simeq 0.91$
implying that the frequency of the GOs essentially coincides with the frequency of the free neuron $\nu_0$ \eqref{free},
Thus one can observe GOs with the same dynamical features in a wide frequency range by simply
varying the parameter $\beta$.

\begin{figure}
\begin{center}
\includegraphics[width=0.45 \linewidth]{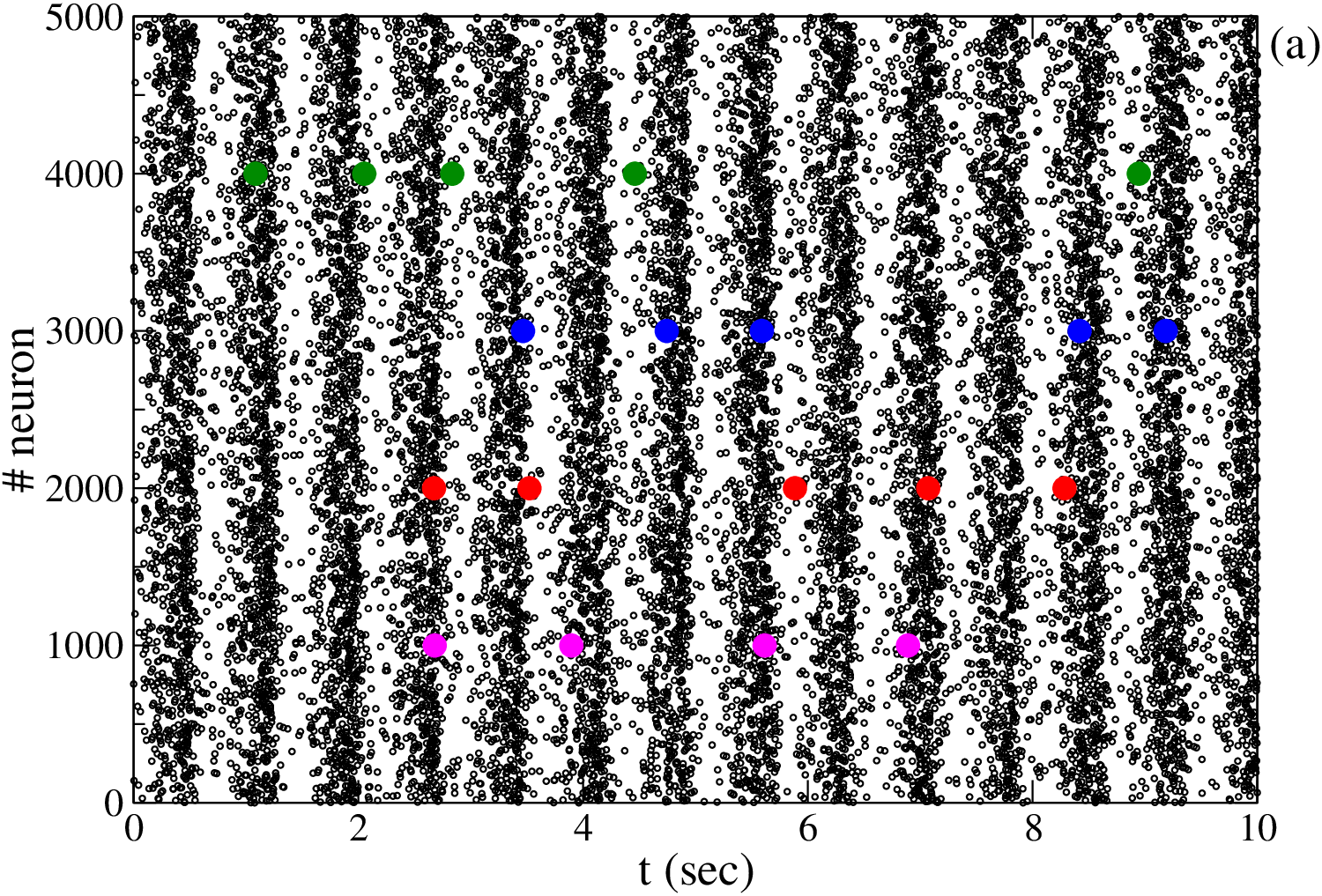}
\includegraphics[width=0.45 \linewidth]{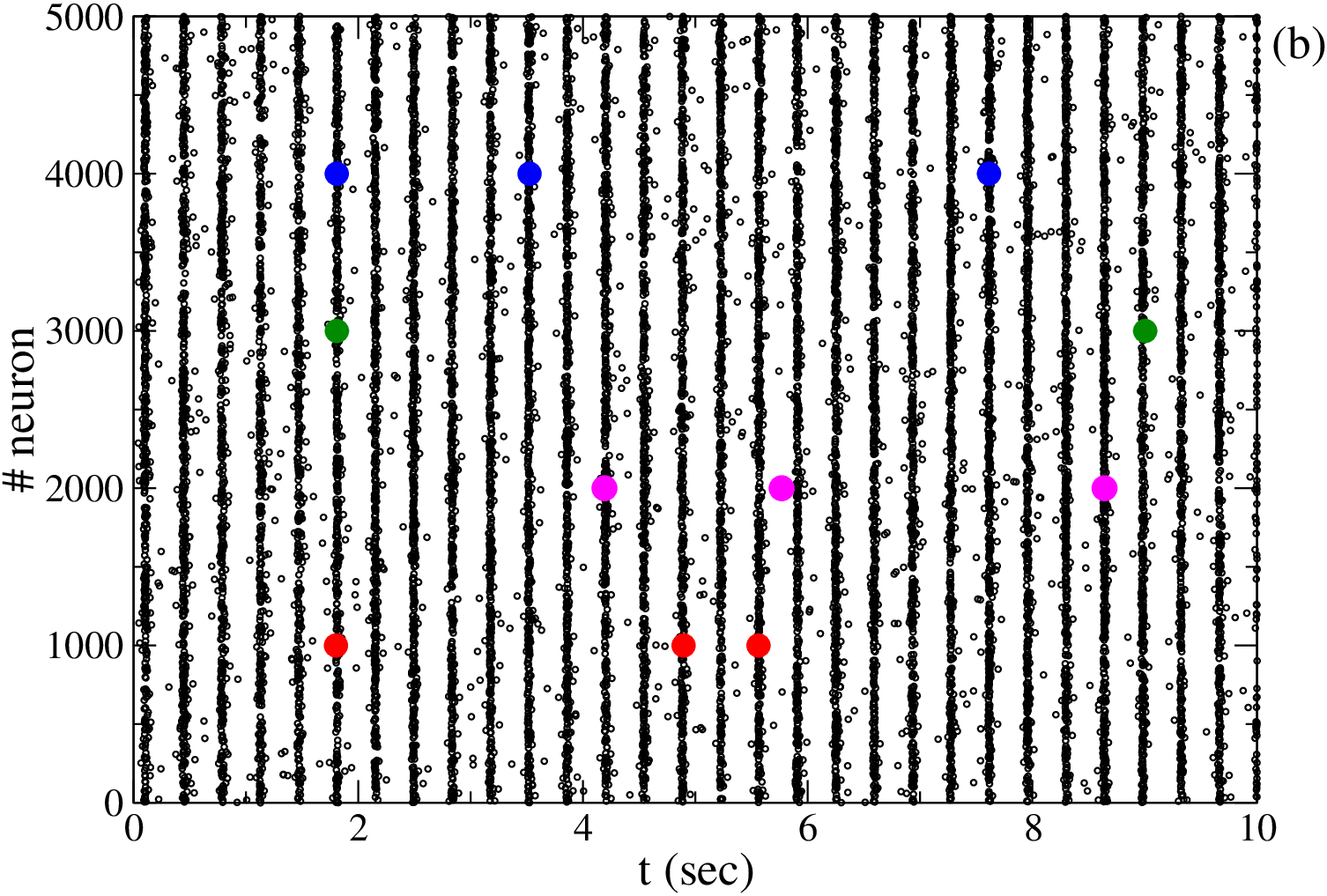}
\end{center}
\caption{Raster plots of 5000 neurons for $K=10$ (a) and $K=250$ (b).
In both cases the firing times of 4 neurons are evidenced by colored circles.
Parameters $g_0=1$, $i_0=0.00055$, and $N=20000$.
}
\label{fig10}
\end{figure}

\subsection{Two different types of GOs}

As previously mentioned, we can identify two
classes of GOs induced by discrete synaptic events
in the interval $i_0/g_0^2 \in [0.00036 : 0.00070]$.
These two classes correspond to GOs observed at sufficiently low $K$ and high $K$,
in particular for $i_0/g_0^2 = 0.00055$, we have the first type for $K \le 30$ and
the second one for $K \ge 200$, and asynchronous dynamics at intermediate in-degrees
(as shown in Fig. \ref{figS0}). Raster plots corresponding to $K=10$ ($K=250$)
are reported in Fig. \ref{fig10} (a) (Fig. \ref{fig10} (b)).  The GOs are characterized
by sparser (less synchronized) activity at low $K$, while at higher $K$ the neurons
contributing to the population bursts fire almost at the same moment, with few outliers.
However, the neurons participate in a quite random manner to the bursts in both
cases (see the red circles in Fig. \ref{fig10}). Since the percentage of neurons
contributing to a burst is $25 \%$ (since $R_G=4$) for $K=10$ and $10 \%$ for $K=250$ (here
$R_G= 9.6$), in the first case each neuron takes part in many more bursts than in the second case.

\begin{figure}
\begin{center}
\includegraphics[width=0.45 \linewidth]{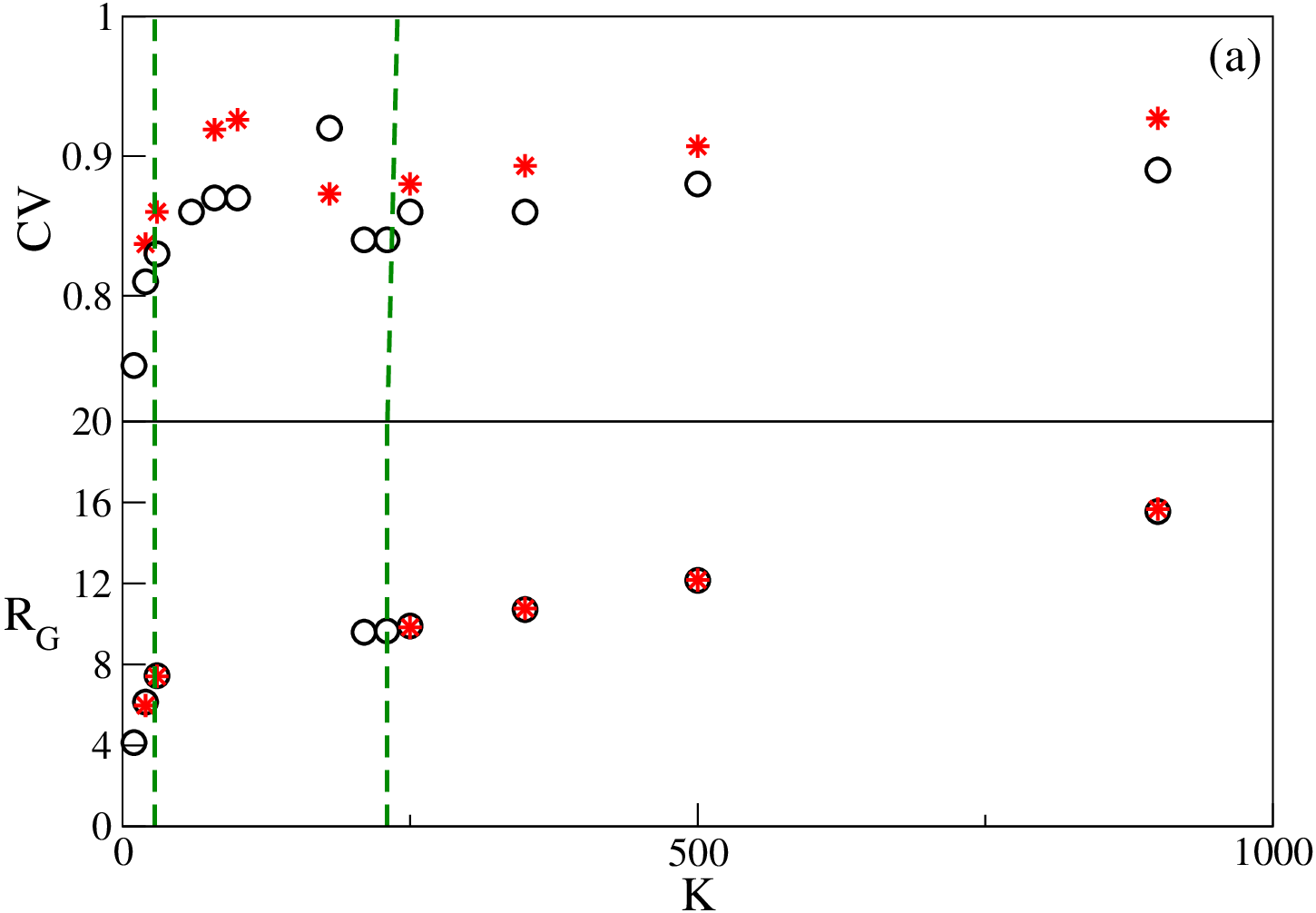}
\includegraphics[width=0.45 \linewidth]{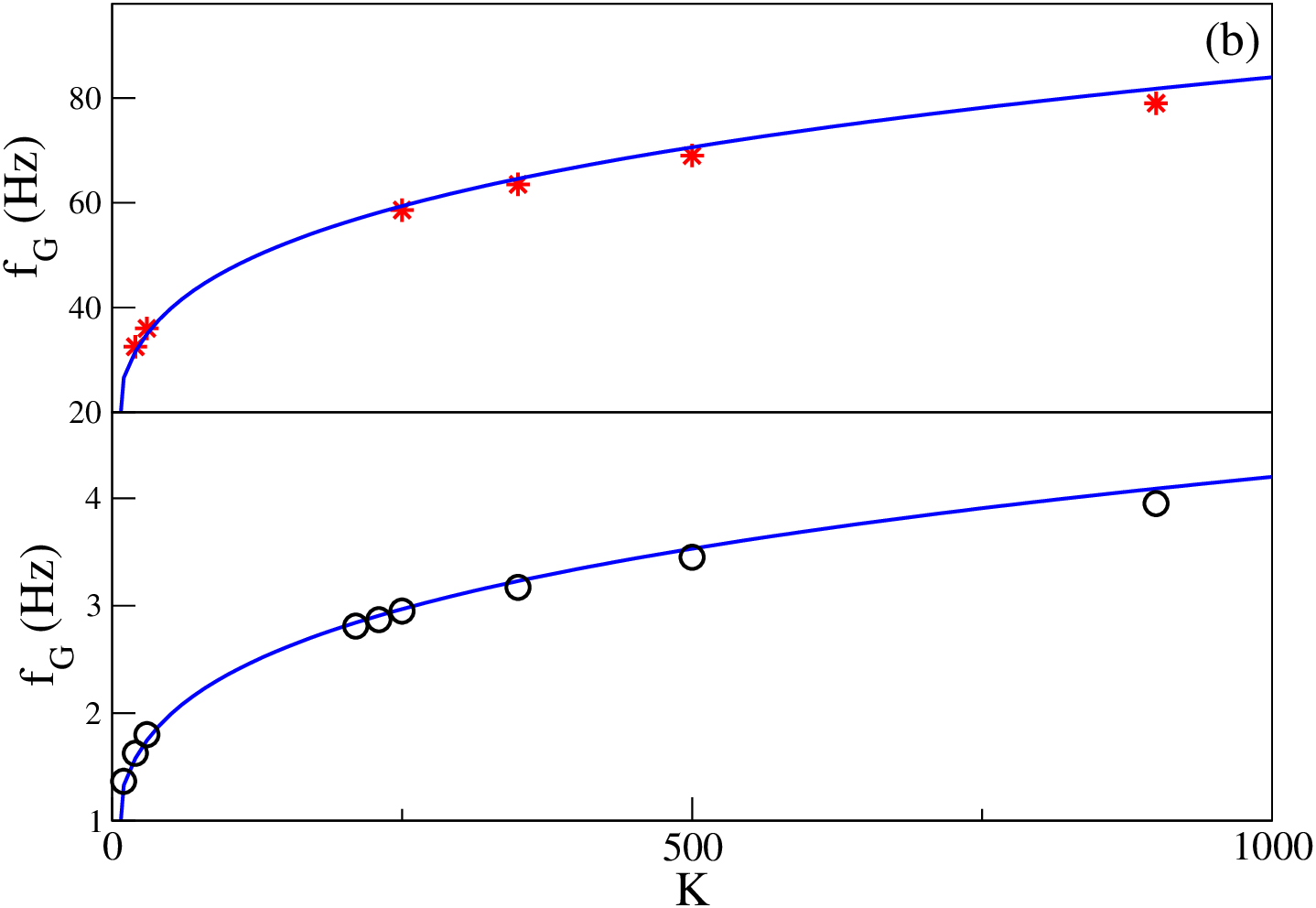}
\end{center}
\caption{(a) $R_G$ versus $K$ in the lower panel, while in the upper panel is reported the population averaged coefficients of variation  $CVs$ versus $K$.
The green dashed lines refer to the HBs found within the CMF: GOs are expected for $K \le 28$ and $K \ge 230$ in this case. (b) $f_G$ versus $K$, the blue solid lines denote $\nu_0$.
Parameters $g_0=1$, $i_0=0.00055$, and $N=40000$ for the empty black circles, $g_0=20$, $i_0=0.22$, and $N=20000$
for the red stars.
}
\label{fig11}
\end{figure}

Moreover, GOs at $K \le 30$ are associated to $R_G = 4-7$,
while those at $K \ge 200$ to much larger $R_G > 10$, since $R_G$ grows with $K$,
as evident from the lower panel of Fig. \ref{fig11} (a) where data for $i_0/g_0^2 = 0.00055$
are reported for different values of $g_0=1$ (black circles) and $g_0=20$ (red stars).
At the same time the spiking activity of the neurons is always quite irregular being $CV \simeq 0.75-0.92$
in all the examined range of $K$ both in the asynchronous and in the oscillating regime and for $g_0=1$ and $g_0=20$, as
shown in the upper panel of Fig. \ref{fig11} (a).

 As shown in Fig. \ref{fig11} (b) and as noticed in the previous sub-section, the frequency of the GOs
$f_G$ is well approximated by the frequency of the free neuron $\nu_0$ \eqref{free},
both for $g_0=1$ (lower panel) and $g_0=20$ (upper panel). The only difference
among these two cases being the range of explored frequencies: for $g_0=1$ the frequencies are in the $\delta$-range between 1-4 Hz, smaller than 2 Hz for $K \le 30$ and among 3-4 Hz for $K > 200$, while for $g_0=20$ the frequencies are slow $\gamma$ ($<40$ Hz) for $K \le 30$ and fast
$\gamma$ ($ 60 -80$ Hz) for large $K$ \cite{colgin2010}.

The difference among these two GOs becomes clearer
by considering the mean-field membrane potential evolution $V_0(t)$ given by the
following  {\it zeroth-order} Langevin equation:
\begin{equation}\label{eq:langevin_sn0}
\dot{V_0}(t) = V_0^2 + \sqrt{K} [i_0 - g_0 \nu(t)] = V_0^2 + A(t) \qquad ;
\end{equation}
where $A(t)$ is the effective input current, $\nu(t)$ is the population firing induced by the shot-noise
and where current fluctuations have been neglected.
Whenever $A < 0$ ($A>0$) the QIF neuron will display excitable dynamics (periodic firing) \cite{gutkin2022}.

\begin{figure}
\begin{center}
\includegraphics[width=0.6 \linewidth]{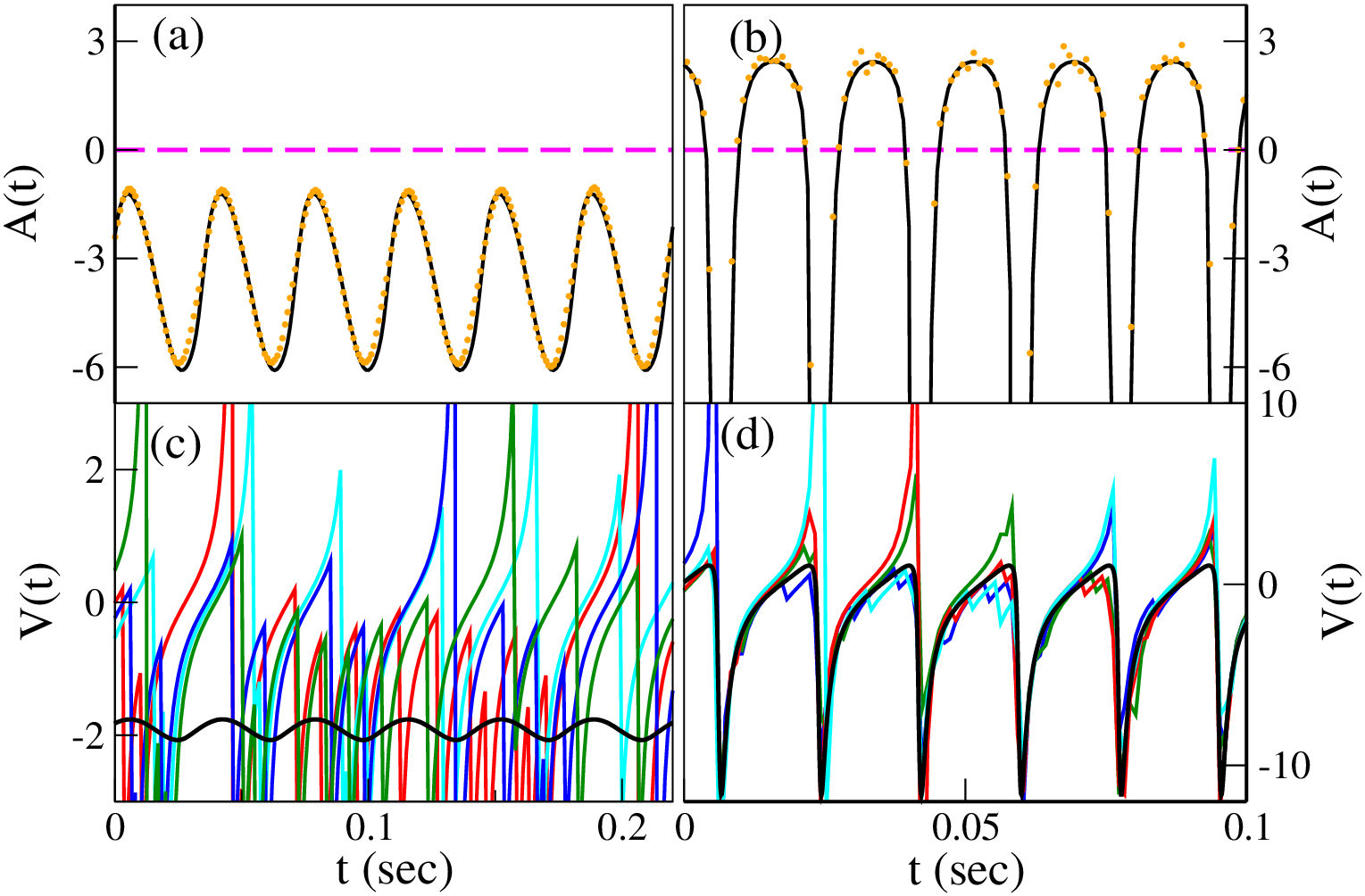}
\end{center}
\caption{(a-b) Effective input currents $A(t)$ versus time : black lines (orange dots) refer to CMF (network simulations) results. (c-d) Membrane potential evolution in time: black lines (other colors) refer to $V_0(t)$ (single neuron
dynamics). Data correspond to $i_0/g_0^2 = 0.00055$ (with $g_0 = 20$) and to $K = 10$ (a-c),
$K = 210$ (b-d), network simulations realised for $N = 80000$
}
\label{fig12}
\end{figure}

The GOs reported in Fig. \ref{figS0} (c) corresponding to $K=10$ are
characterized by $A(t)$ always negative, while the GOs shown in Fig. \ref{figS0} (d) for $K=210$
display an effective current that is positive for a large part of the oscillation period,
as shown in Fig. \ref{fig12} (a) and Fig. \ref{fig12} (b), respectively.
Therefore, for $K=10$ $V_0(t)$ displays sub-threshold oscillations,
while for $K=210$ it has large excursions from negative to positive values driven by $A(t) >0$,
as evident from the signal denoted by the black solid line in Fig. \ref{fig12} (c) and Fig. \ref{fig12} (d), respectively.
In the same figures we report the time evolution of the membrane potential
$v_i(t)$ of 4 generic neurons (characterized by colored lines), it is immediately evident
that low (high) in-degree GOs are characterized by a low (high) level
of coherence among the neurons. For $K=210$ the single neurons $v_i(t)$ essentially follow the 
men-field evolution $V_0(t)$ for a large part of the evolution. However, current fluctuations induce irregular firings of the neurons in proximity
of the threshold, indeed $CV \simeq 0.85$. In contrast the dynamics of the neurons is quite uncorrelated
over the whole evolution from reset to threshold for $K=10$ and also in this case
the coefficient of variation is reasonably large $CV \simeq 0.75$.

\begin{figure}
\begin{center}
\includegraphics[width=0.45 \linewidth]{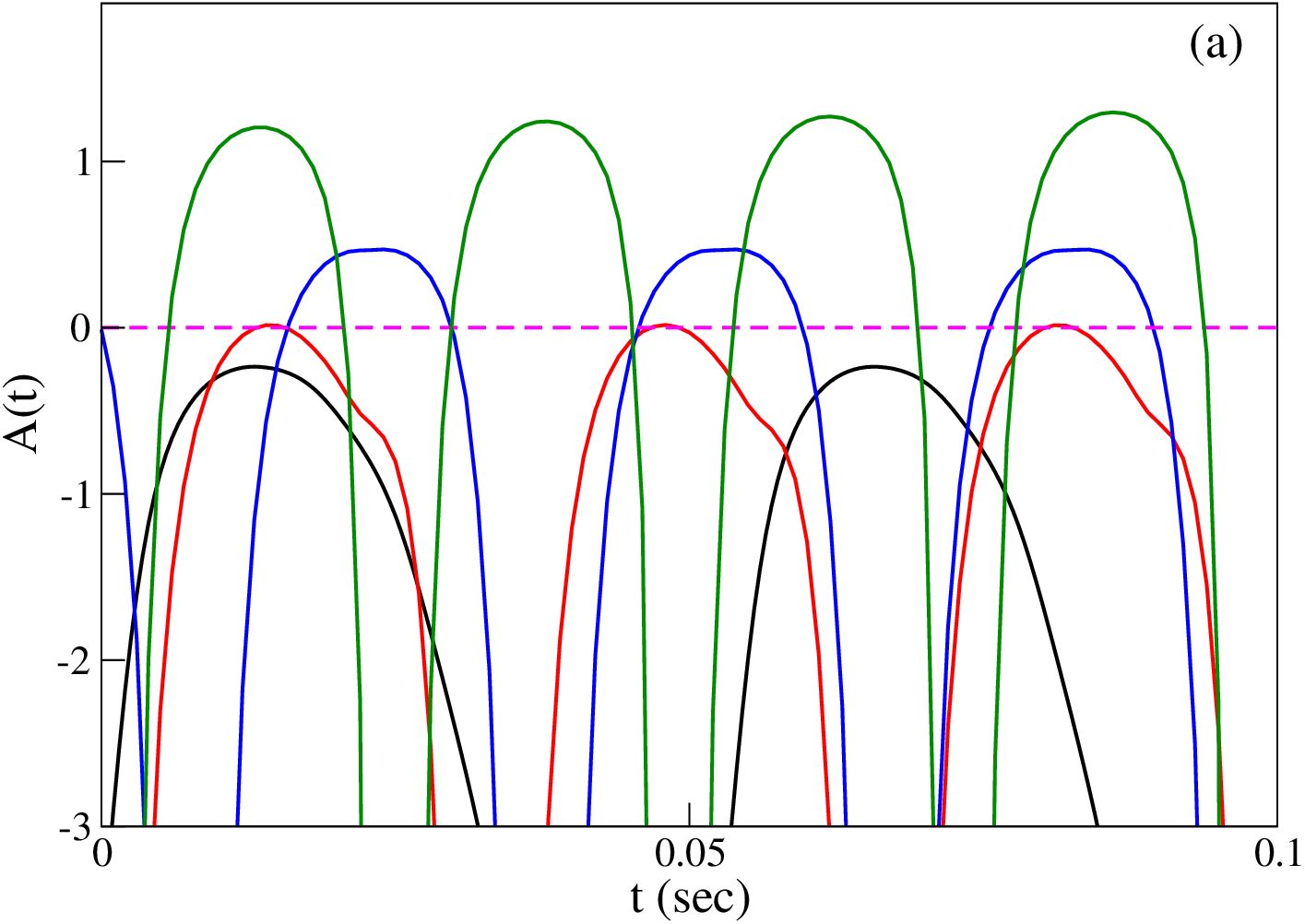}
\includegraphics[width=0.45 \linewidth]{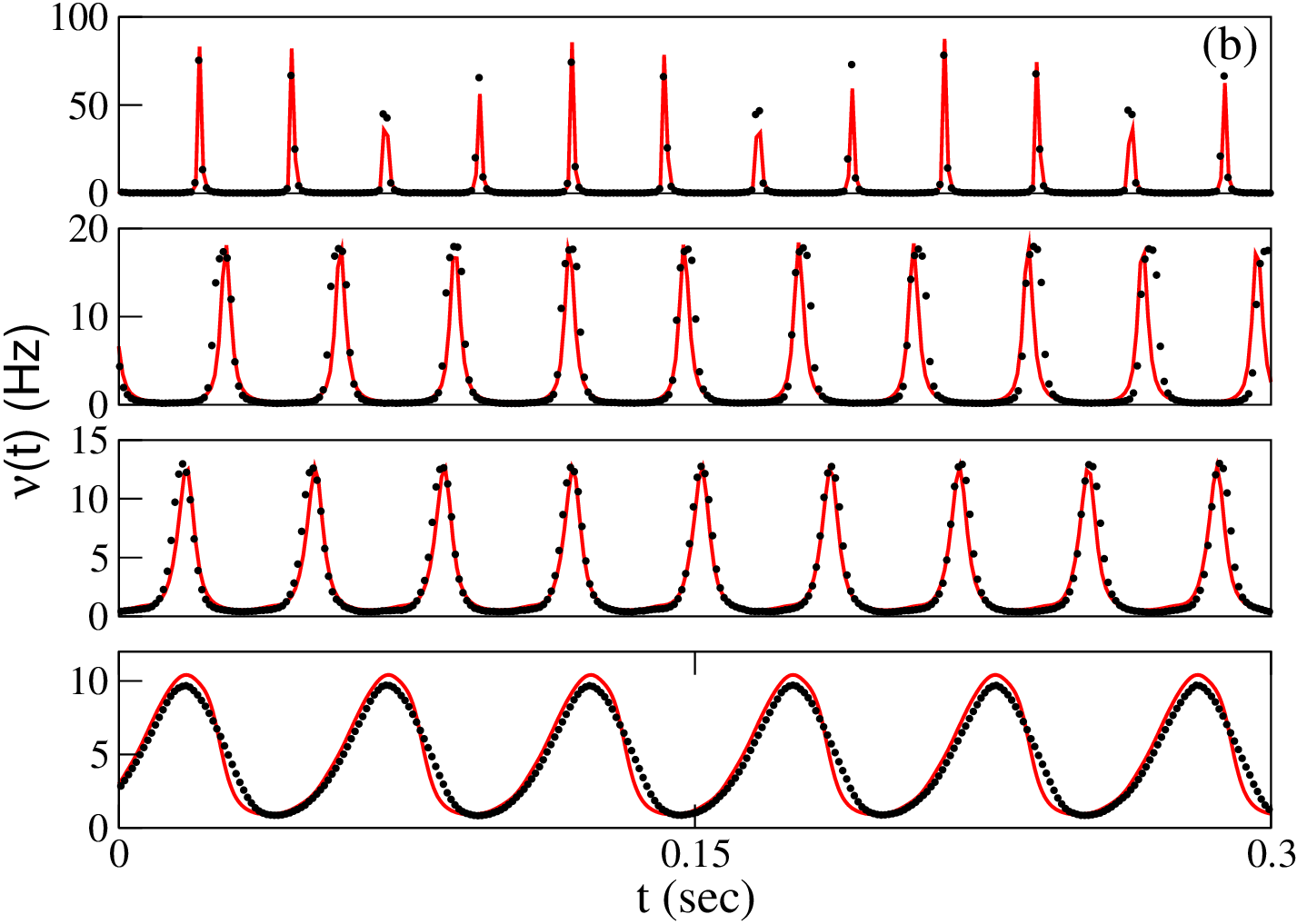}
\end{center}
\caption{(a) Effective input currents $A(t)$ versus time : black line ($K=10$),
red line ($K=60$), blue line ($K=100$) and green line $K=250$. (b)
Population firing rates $\nu(t)$ versus time the panels from bottom to top refer to $K=10$, $K=60$, $K=100$ and $K=250$.
Solid lines are CMF results obtained via the integration of Eq. \eqref{CE},
black symbols denote network simulation for $N=80000$.
Data correspond to $i_0/g_0^2 = 0.00027$ (with $g_0 = 20$).
}
\label{fig13}
\end{figure}

These behaviours can be somehow explained by 2 different mechanisms once remembering that
for both cases the GO frequency  $f_G$ is extremely close to the firing frequency
of an isolated neuron $\nu_0 = 1/T_0$. This suggests that at a first approximation
the GOs are due to the neurons not receiving any inhibitory PSP from reset to threshold.
For low $K$, whenever a neuron fires large amplitude inhibitory PSPs are delivered,
since their amplitude is $g_0/\sqrt{K}$. These lead the membrane potentials of
the $K$ post-synaptic neurons to have quite similar values, and therefore
the pre-synaptic neuron spiking induces
a transient synchronization of the $K$ post-synaptic neurons.
A sub-group of these, not receiving any further PSPs,
can eventually reach threshold together at a time $\simeq T_0$. This transient synchronizing effect of small clusters
of neurons (termed {\it cluster activation}) is at the origin of the GOs observable for $K=10$.
For increasing $K$, the amplitude of the PSPs decreases, therefore above
some critical in-degree ($K \simeq 30$ in this case) a single inhibitory PSP is no longer able
to induce a sufficiently strong synchronizing effect on the post-synaptic neurons and
the dynamics become asynchronous (as shown in Fig. \ref{figS0} (c) (blue line)).

For larger $K$, the post-synaptic neurons receive many small inhibitory PSPs at each population burst,
whenever $K$ is sufficiently large a non negligible part of the neurons
can get synchronized by the discharge of inhibitory PSPs.  As shown in Fig \ref{fig13} (d), the time courses of the
membrane potentials are now extremely coherent by approaching the threshold, where
fluctuations lead to irregular firing of the neurons. However, a sufficient percentage of these 
drift-driven neurons is always able to fire together with a period $\simeq T_0$
giving rise to the GOs.

The one described above is the scenario in the interval $i_0/g_0^2 \in [0.00036 : 0.00070]$, where
for fixed $i_0/g_0^2$ one has GOs at low and large $K$ separated by a region of intermediate $K$
where the dynamics is asynchronous. For larger $i_0/g_0^2 > 0.00070$ we have only {\it drift-driven} GOs,
since oscillatory dynamics is observable only for sufficiently large $K$.
At low $K$ either the value of $g_0$
is too small to promote transient synchronization among the post-synaptic neurons or
the period $T_0= \pi/\sqrt{I} = \pi/\sqrt{i_0 \sqrt{K}}$ becomes too long to allow the cluster of post-synaptic neurons to reach threshold
without receiving other PSPs.

However, it is unclear what happens for $i_0/g_0^2 < 0.00036$, in this region there are GOs for any value of the
in-degree, as shown in Fig. \ref{figphasediag} (a). The main question is if we pass from one mechanism to the other in a continuous or discontinuous way. Therefore, we have analyzed in detail the behaviour in this
region by fixing $i_0/g_0^2 = 0.00027$ and by exploring the collective dynamics at different in-degrees.
In Fig. \ref{fig13} (a) we report for various $K$ the effective current $A(t)$ obtained within the CMF approach, estimated via the integration
of Eq. \eqref{CE} with the pseudo-spectral methods (see Appendix C for more details). From the figure it is
evident that $A(t)$ begins to become partially positive for $K \ge 60$, therefore we could
speak of the {\it cluster activation} mechanism for lower $K$ and of the {\it drift-driven} mechanism
for larger $K$.  At the same time the neurons become more and more synchronized within each population burst,
as shown in Fig. \ref{fig13} (b),
thus indicating that the effect of a positive $A(t)$ leads them to be more and more coherent as expected.
Despite the number of neurons participating to a population burst declines with $K$, analogously to what seen already
for $i_0/g_0^2 = 0.00055$, indeed $R_G$ passes from value 4 at $K=10$ to value 10 at $K=250$. At the same time the firing activity remains always irregular with 
$CV \simeq 0.75-0.85$ in the whole examined range. However, all the indications suggest that the transition from 
one type of GO to the other occurs smoothly by increasing $K$ in this case.

\section{Summary and Discussion}

In this paper we have presented a mean-field theory for balanced QIF neural networks encompassing
synaptic Poisson noise and sparsity in the network connections. This approach, termed Complete Mean-Field (CMF),
has allowed us to perform linear and weakly non-linear stability analysis of the asynchronous state, revealing sub- and super-critical Hopf instabilities leading to global oscillations (GOs). 
Therefore, besides asynchronous and oscillatory regimes, there is also a region of co-existence of these two solutions near the hysteretic
transitions associated with the sub-critical Hopf bifurcations. As a further result of this analysis it emerges that the phase diagram depends only on two parameters: 
the in-degree $K$ (number of pre-synaptic neurons) and the ratio $i_0/g_0^2$ between the external excitatory drive and the square of the synaptic coupling. 
A peculiarity of the Hopf bifurcation line is that it is re-entrant at low $i_0/g_0^2$: in a limited range of $i_0/g_0^2$ one has GOs at low $K$ and high $K$-values, and asynchronous dynamics at intermediate in-degrees. 
Furthermore, the GOs observable at low and high $K$, despite being both associated with the irregular firing of single cells,
are promoted by two different mechanisms: "cluster activation" at low in-degree and a "drift-driven" mechanism at high in-degree. 

The network dynamics is well captured by the CMF, but not by the diffusion approximation (DA)
\cite{noi} valid in the limit of small PSPs (large $K$) and high neuronal activity (large $i_0$ and/or small $g_0$). Indeed, the DA works well for large $i_0/g_0^2 > 0.5$, but fails at low $i_0/g_0^2$ where it is completely unable to reproduce the scenario. In particular, the DA predicts the emergence of GOs only via super-critical Hopf bifurcations and that
in the limit $i_0/g_0^2 \to 0$  the critical in-degree $K_c$ required to observe GOs should
diverge to infinity. Instead, at sufficiently low $i_0/g_0^2 < 0.00036$ GOs are observable for any $K$ value, as predicted by the CMF and in agreement with simulations.

An intermediate approximation going in the direction of the CMF can be obtained by considering an expansion to the third order of the source term in the continuity equation. This gives rise to the
D3A approximation that we have studied in this paper. This approximation also predicts the emergence of GOs only via super-critical Hopf bifurcations and it is unable to reproduce the re-entrant Hopf bifurcation line found within the CMF. 
On the positive side, D3A captures well the network dynamics already for $i_0/g_0^2 > 0.02$ (i.e. for 
a parameter value an order of magnitude smaller than the DA) and it forecasts correctly that the GOs are present for sufficiently low  $i_0/g_0^2$ for any in-degree, even if it gives a higher bound for this value with respect to CMF.

To understand the mechanisms responsible for the emergence of GOs at sufficiently low and high in-degree
one should consider the effective input current $A(t)$ driving the mean-field dynamics in the two cases:
at low $K$ $A(t)$ always induces sub-threshold dynamics, while at large $K$
this is mostly supra-threshold. In the first case GOs are promoted by small groups of neurons transiently 
synchronized by a large inhibitory PSP traveling together towards threshold, without being affected 
by further inhibitory PSPs. This is the origin of the name of the mechanism: "cluster activation".
At large $K$ a large part of neurons, highly synchronized due to the previously received barrage of small inhibitory PSPs, are "drift driven" towards the threshold by the positive effective input current,
however only a small part of them, receiving few or no PSPs, will reach threshold almost at the same time
giving rise to the GO. The Hopf bifurcations leading to these two types of GOs are characterized by
distinct features. For large $K$ the collective dynamics is clearly dominated by the two most unstable
(complex conjugate) eigenvalues. At low $K$, the spectrum is not only characterized by a pair of complex-conjugate eigenvalues responsible for the instability, but also by additional undamped pairs of eigenvalues that lie very close to the unstable ones.
This suggests that, in this case, the instability is no longer controlled only by the two most unstable modes, but also by additional modes, which contribute to the coherent dynamics. Indeed, the
population bursts are not generated by highly synchronized neurons, as is the case at high in-degrees.

From analytic investigations performed within the DA in \cite{noi}, it emerges that the asynchronous balanced dynamics could be either fluctuation-driven (mean-driven) for sufficiently small (large) values of $i_0/g_0^2$ \cite{lerchner2006,renart2007}. 
Indeed, we have verified that in a large region of the phase space (namely, for $i_0/g_0^2 < 0.1$) the single neuron dynamics remains highly irregular, characterized by $CV > 0.75$, regardless of the in-degree and 
whether the global activity is asynchronous or oscillatory.

As mentioned in the Introduction, this paper can be considered as the natural completion of the DA analysis 
reported in two well-known papers devoted to the emergence of GOs in sparse inhibitory \cite{brunel1999}
and excitatory-inhibitory neural networks \cite{brunel2000}. However, the model we consider is the quadratic integrate-and-fire, while the one analyzed in \cite{brunel1999,brunel2000} is the leaky integrate-and-fire. As already 
reported in \cite{matteo}, GOs can be observed in QIF sparse networks even for instantaneous synapses in the absence
of any synaptic delay. This is not the case for LIF. Indeed, for inhibitory LIF networks the period of GOs
is two/three times the synaptic delay (assumed to be on the order of few ms), thus the observable frequencies $f_G$ are limited 
to a range of 100-200 Hz. Instead, in our case $f_G$ can range from the $\delta$-band (0.5-3.5 Hz) to the 
$\gamma$-range (30-100 Hz), being $f_G$ essentially
proportional to the square root of the external excitatory drive $I$. One would naively expect that slow GOs
observable for low $I$ are associated with irregular spiking dynamics, while fast GOs at high $I$ are linked to
almost regular neuronal spiking. However, the microscopic dynamical behaviour (fluctuation-driven or mean-driven) 
is controlled by the value of the ratio $i_0/g_0^2$. Irregular or regular spiking 
is observable at fixed in-degree $K$ by maintaining the ratio constant, the only effect of rescaling
$i_0$ and $g_0$ being the modification of the time scale of the model.
Therefore, GOs of any frequency (fast or slow) can be obtained by conveniently rescaling $i_0$ and $g_0$ with either irregular
or regular firing.

In \cite{brunel2000}, Brunel, by considering an excitatory-inhibitory network in the inhibition-dominated regime was
able to identify slow ($f_G \simeq 20-60$ Hz) and fast ($f_G \simeq 120-180$ Hz) GOs separated by an asynchronous
regime. This scenario closely resembles the one we have identified at low $i_0/g_0^2$. However, the analysis in \cite{brunel2000},
besides considering excitatory and inhibitory neurons, is performed within the DA, and for purely inhibitory networks
only fast oscillations have been reported in \cite{brunel1999}. 
Furthermore, the frequency of the slow (fast) oscillations is controlled by the membrane time constant (synaptic time scale)
while in our case it is always related to the external excitatory drive. As a further difference, the parameter $R_G$ (which is the
ratio of $f_G$ with respect to the population firing rate) takes values around 3-4 in both the slow and fast
oscillation regimes, while in our case  $R_G \simeq 4-6$ for low $K$ and $R_G > 10$ for high $K$.
Therefore, it is worth extending our approach to excitatory-inhibitory networks in order to understand in more detail the origin of
these differences and similitudes and to investigate how the scenario will be modified by the presence of excitatory neurons.

Our theoretical results are rigorously derived for the continuity equation \eqref{CE} obtained in the thermodynamic limit $N \to \infty$. 
This equation is valid even for a finite network as long as (i)~the firing events of the pre-synaptic neurons can be treated as independent and (ii)~all emitting and receiving neurons 
can be considered to be statistically equivalent. This introduces a lower and upper bound on the in-degree $K$. On one side, the network must be sparse, $K\ll N$, otherwise, the pre-synaptic spike trains will no longer be independent. 
On the other side, for the statistical equivalence any neuron should be connected to any other (no closed loop should emerge) and this is achieved in an Erd\"os-Renyi network, 
analogous to the set-up we considered, when the in-degree is larger than the percolation threshold, i.e. $K > 1$ \cite{albert2002}. The role of finite $N$ effects for the macroscopic behaviour of sparse LIF networks
has been considered in \cite{brunel1999,brunel2000} and more recently in \cite{klinshov2022,klinshov2023} for globally 
coupled QIF networks. Future studies should be devoted to the inclusion of finite size effects
in the framework of the CMF for balanced neural networks. 

As stated above, GOs with tunable frequencies can emerge even in extremely sparse inhibitory networks thanks to the mechanism of "cluster activation". This mechanism can be at the basis of the $\gamma$-oscillations 
observed in the hippocampus and generated by a population of interneurons with low in-degree $K \simeq 30-80$ \cite{sik1995,buzsaki2012}.

Clustering instabilities observed in globally coupled neural networks in the presence of additive noise for LIF \cite{brunel2006} 
and QIF neurons \cite{feld2024} should not be confused with the mechanism of "cluster activation" here reported. 
The dynamics induced by the clustering instability is characterized by GOs with a frequency that is a multiple of the single neuron
frequency and by low irregularity in the firing activity, whereas the GOs emerging due to the "cluster activation" mechanism 
exhibit stochastic synchrony (in contrast to regular synchrony where neurons resonantly lock with the oscillatory input) 
with associated high values of the coefficient of variation $CV \simeq 0.75-0.85$ denoting a quite irregular firing activity.
Clustered dynamics can be observed in this model only in the mean-driven regime,
for large $i_0/g_0^2 > 0.1$ and large $K > 200$. The clustered phase is characterized by population bursts where neurons are split in two equally populated clusters firing in alternation. Although the global activity appears regular, the single neurons display switching 
between the two clusters due to fluctuations in the input currents, but with a low level  of irregularity in their dynamics. This clustering phenomenon has been already reported for heterogeneous
and homogeneous globally coupled QIF networks subject to weak additive noise \cite{feld2024}. 

As shown in Appendix D, the introduction of finite synaptic times does not alter substantially the scenario here
depicted for instantaneous synapses. Indeed, also in this case for sufficiently low $i_0/g_0^2$ and $K$,
the DA is unable to capture the network dynamics and the inclusion of discrete synaptic events is fundamental in order to reproduce the network simulations. These results call for the inclusion in the CMF of
more biologically realistic features, such as the delay in the pulse transmission or the finite duration of the PSPs.
These generalizations are technically feasible, however they require a careful derivation that goes beyond the scopes of the present analysis and that will be reported in future studies.

\begin{acknowledgments}
We acknowledge stimulating interactions with Alberto Bacci, Nicolas Brunel, Pau Clusella, Rainer Engelken, Thierry Huillet, Nina La Miciotta,
Lyudmila Klimenko, Ernest Montbri\'o, Gianluigi Mongillo, Simona Olmi, Antonio Politi, Magnus JE Richardson. A.T. received financial support by the Labex MME-DII (Grant No.\ ANR-11-LBX-0023-01) and by CY Generations with the project SLLOWBRAIN (Grant No ANR-21-EXES-0008). M. d. V. also received
support by the Labex CORTEX (Grant No. ANR-11-LABX-0042) of Universit\'e Claude Bernard Lyon 1 and by the ANR via the Junior Professor Chair in Computational Neurosciences Lyon 1.
\end{acknowledgments}

\appendix

\section{Event driven simulations of the QIF network}

The QIF network model \eqref{eq:1} can be integrated by employing an event driven
method, since we always considered supra-threshold neurons $I >0$ \cite{tonnelier2007}.
In particular, the method consists in integrating exactly the network
evolution between one spike emitted by neuron $j$ at time  $T_1 = t^{(n)}_j$
and the next one emitted by the $q$-th neuron at time $T_2 = t^{(n+1)}_q$.

The first step consists in identifying the next firing neuron,
therefore one should evaluate the time ${\cal T}_i$
needed to each neuron $i$ to reach
the firing threshold $v_{th}=+\infty$. These times can be evaluated
on the basis of the value of the neuron membrane potential at time $T_1^+$ (immediately after the spike emission), as follows:
\begin{equation}
{\cal T}_i(n+1) = \frac{1}{\sqrt{I}} \left[\frac{\pi}{2} - \arctan\left(B_i(T_1^+)\right) \right] \qquad {\rm where}
\qquad B_i(t) \equiv \frac{v_i(t)}{\sqrt{I}} \enskip .
\end{equation}
The next firing neuron $q$ will be the one associated with the smallest time ${\cal T}_q(n+1)$
and the next firing time will be $T_2 = T_1 + {\cal T}_q(n+1)$.

Once the next firing neuron is identified, the membrane potential evolution between $[T_1^+,T_2^-]$ is given by
\begin{equation}
v_i(T_2^-) = \sqrt{I} \tan\left(\frac{T_2 - T_1}{\sqrt{I}} + B_i(T_1^+)\right)
\end{equation}
and the membrane potentials of the post-synaptic neurons connected to neuron $q$ should be modified as follows,
due to the spike arrival,
\begin{equation}
v_k(T_2^+) = v_k(T_2^-) - g  \qquad {\rm if} \qquad k \in {\rm post}(q) \enskip ;
\end{equation}
while the others are left unchanged. At this point the adjoint variables $B_i(T^+_2)$ are evaluated by following
their definition, apart the one associated to the spiking neuron $q$ that it is simply set
to $B_q(T^+_2) = - \pi/2$ due to the reset condition for the membrane potential (namely, $v_q(T^+_2) = -\infty$).
The iterative procedure can be now repeated to find the next firing neuron and so on.

See also \cite{engelken2023} for a recently developed efficient event-driven simulations scheme for sparse random
networks. 

\section{Integration of the Langevin equations}

Regarding the integration of the mean-field Langevin equation \eqref{eq:langevin}, we considered $N=20000 - 80000$ replica of such equation, corresponding to $N$ {\it uncoupled} QIF neurons. In particular, we integrated such $N$ uncoupled evolution
equations \eqref{eq:langevin} by employing an Euler integration scheme with
$\Delta t$. The population firing rate $\nu(t)$ entering in Eq. (2) at
time $t$ is estimated self-consistently by counting the number of spikes $N_{sp}(\Delta t)$ emitted by the $N$ neurons
in the preceding time window $T_w$, as follows:
\begin{equation}
\nu(t) =  \frac{N_{sp}(\Delta t)}{T_w N} \quad .
\label{nnu1}
\end{equation}

For the shot-noise case, at the end of every time interval $T_w$, every neuron receives an inhibitory post-synaptic input of finite
amplitude $g$ with a probability $K\times\nu(t)\times \Delta t$ mimicking a Poissonian process with rate $K \nu(t)$.

To simulate the Langevin evolution \eqref{eq:langevin} within the DA, we considered $N$ uncoupled neurons
whose membrane potential obeys the following stochastic differential equation
\begin{equation}
\label{sde}
\dot{V}(t) = F(V) + I - g_0 \sqrt{K} \nu(t) + g_0 \sqrt{\nu(t)} \xi(t)
\end{equation}
where $\xi$ is the normalized Gaussian white noise.
The firing rate is estimated also in this case by the activity of the $N$ neurons
via \eqref{nnu1} by performing an average on a sliding time window of duration $T_w$
and the integration  is performed again via an Euler scheme.
We fixed $\Delta t = 5 \times 10^{-4} \tau_m$ and $T_w = 0.005 \tau_m$,
and an initial transient time $t_t = 10$ s is discarded in all the simulations.

\section{Integration of the continuity and Fokker-Planck equations}

In order to integrate the CMF and the FPE we transformed the membrane potential $V$ in the
angular variable $\theta = 2 \arctan V$ with $\theta \in [-\pi,+\pi]$, in the present case
we used the standard transformation to pass from the QIF to the $\theta$-model.

Accordingly, the PDF $P(V,t)$
for the membrane potentials is related to the PDF $R(\theta,t)$ of the corresponding
angular variables as follows
\begin{equation}
  P(V,T) = (1+\cos \theta) R(\theta,t) \quad.
\end{equation}

The CMF equation then takes the following form
\begin{equation}
\frac{\partial R(\theta,t)}{\partial t} =
- \frac{\partial}{\partial \theta} \left[ ((I+1)+(I-1)\cos\theta) R(\theta,t)\right]
+ K \nu(t) \left[ R(\theta^+,t) \frac{1+\cos\theta^+}{1+\cos\theta} - R(\theta,t)\right]
\label{CMFt}
\end{equation}
where
\begin{equation}
\theta^+ = 2 \arctan\left[\tan\left(\frac{\theta}{2}\right) + g \right] \quad.
\end{equation}

By expanding the term within square brackets  on the right-hand side in \eqref{CMFt} up to the
second order in $g$ we obtain the corresponding FPE in the angular space
\begin{equation}
\frac{\partial R}{\partial t} =
- \frac{\partial}{\partial \theta} \left\{ \left[(1-\cos\theta) +(A(t) + D(t) \sin\theta) (1+\cos\theta)\right]R
- D(t) (1+ \cos\theta)^2 \frac{\partial R}{\partial \theta} \right\}
\label{FPEt}
\end{equation}
where
$A(t)=\sqrt{K}[i_0-g_0 \nu(t)]$, $D(t)=[g_0^2 \nu(t)]/2$ and $R=R(\theta,t)$.

The Fokker--Planck formulation can be extended by including the third order term in the expansion,
this amounts to add on the right-hand side of \eqref{FPEt} the following term
\begin{equation}
H(t) \frac{\partial}{\partial \theta} \left[ (1+\cos\theta)^3 \frac{\partial^2 R}{\partial \theta^2}
-3 \sin \theta (1+\cos \theta)^2 \frac{\partial R}{\partial \theta}
+ (1+\cos \theta)(1-\cos \theta -2 \cos^2 \theta) R
\right]
\label{3rdt}
\end{equation}
where $H(t)=\frac{g_0^3}{3!\sqrt{K}}  \nu(t)$.

In all the cases listed above we assumed that the flux at the threshold determines
the instantaneous population firing rate, therefore this sets the following
boundary condition for the PDFs
$$\lim_{V \to \infty} V^2 P(V,t)= 2  R(\pi,t) = \nu(t) \enskip . $$

\subsection{Time splitting pseudo-spectral method}

The above partial differential equations can be integrated by employing time-splitting pseudo-spectral methods developed for spatially extended systems, such as the complex Ginburg-Landau equation \cite{torcini1997},
surface growth models  \cite{torcini2002} and asymmetrically coupled Swift-Hohenberg and
Cahn-Hilliard equations  \cite{schuler2014}. In particular, in order to perform
the numerical integration we considered a discrete grid for the angular values
of resolution $\Delta \theta$ and a discrete time evolution with a constant time step $\Delta t$.
The discretized PDF can be written as $R(i,n)$, where the integer indices $i$ and $n$
represent the angular and temporal variables, respectively. Periodic boundary conditions
are assumed for the PDF: $R(i,n)=R(i+I,n)$, where $I$ is the number of sites of the grid
namely $2 \pi = I \Delta \theta$.

In general the evolution equations for the PDF $R(\theta,t)$ can be rewritten as follows
\begin{equation}
\partial_t R(\theta,t) = ({\cal L} + {\cal N}) R(\theta,t)
\label{operators}
\end{equation}
where ${\cal L}$ and ${\cal N}$ are a linear and non-linear operator, respectively.
For a sufficiently small time step $\Delta t$ the formal solution of \eqref{operators}
is the following
\begin{equation}
R(\theta,t+\Delta t)= {\rm e}^{({\cal L +N})\Delta t}R(\theta,t) \quad .
\label{ev1}
\end{equation}
The evolution operator appearing in \eqref{ev1}
can be approximated via the Trotter formulas as follows
\begin{equation}
{\rm e}^{({\cal L +N}) \Delta t} \approx {\rm e}^{{\cal N} \Delta t/2} {\rm e}^{{\cal L}\Delta t} {\rm e}^{{\cal N} \Delta t/2} + {\cal O} (\Delta t^3) \quad .
\end{equation}

For a finite time interval $T = M \Delta t$ we can lump together $M-1$ integration steps
and approximate the evolution operator over the interval $T$ as
\begin{equation}
{\rm e}^{({\cal L +N}) T} \approx {\rm e}^{{\cal N} \Delta t/2}
\left( {\rm e}^{{\cal L}\Delta t} {\rm e}^{{\cal N} \Delta t}
\right)^{M-1}
{\rm e}^{{\cal L}\Delta t} {\rm e}^{{\cal N} \Delta t/2}
 \quad ;
\end{equation}
this amounts to perform successive integration of the linear and non-linear
part. The linear part
\begin{equation}
\partial_t R(\theta,t) = {\cal L} R(\theta,t)
\label{linear}
\end{equation}
is usually solved in Fourier space where it takes the form
\begin{equation}
\partial_t {\tilde R}(p,t) = f(p) {\tilde R}(p,t)
\label{linearFFT}
\end{equation}
where ${\tilde R}(p,t)$ is the Fourier transform of the PDF $R(\theta,t)$ over the angular variable, $p$ is an integer defined in the interval $p \in [-I/2,I/2-1]$ representing the
wave number and $f(p)$ is a polynomial in $p$.
The time evolution for ${\tilde z}$ over a time step $\Delta t$ is given by
\begin{equation}
{\tilde R}(p,t+\Delta t) = {\rm e}^{f(p) \Delta t} {\tilde R}(p,t)
\label{linearFFT2} \quad ;
\end{equation}
therefore the integration of \eqref{linear} in the original space can be written as
\begin{equation}
{\tilde R}(\theta,t+\Delta t) =
{\cal F}^{-1} {\rm e}^{f(p) \Delta t} {\cal F} { R}(\theta,t)
\label{linear_int} \quad ,
\end{equation}
where ${\cal F}$ (${\cal F}^{-1}$) is the direct (inverse) Fourier transform.

The integration of the non-linear part amounts to solve
\begin{equation}
\partial_t R(\theta,t) = {\cal N} R(\theta,t)
\label{nonlinear}
\end{equation}
this can be integrated by employing some standard integration method,
in the specific we have employed the Euler scheme, therefore the time
evolution of the PDF is given by
\begin{equation}
R(\theta,t + \Delta t) = R(\theta,t)+ [{\cal N} R(\theta,t)] \Delta t
\quad .
\label{linear2}
\end{equation}
In order to increase the precision of the integration, the angular derivatives of $R$ eventually
entering in ${\cal N} R(\theta,t)$ have been evaluated in the Fourier space.

In the following we will explicitly report the linear and non-linear operators
considered in the article.

\subsubsection{The CMF}

The linear operator for the CMF is the following
\begin{equation}
{\cal L} R(\theta,t) = - (I+1) \frac{\partial R(\theta,t)}{\partial \theta} - K \nu(t)  R(\theta,t)
\label{CMFt_lin}
\end{equation}
and the non-linear one is
\begin{equation}
{\cal N} R(\theta,t) =
- \frac{\partial}{\partial \theta} \left[ (I-1)\cos\theta R(\theta,t)\right]
+ K \nu(t) \left[ R(\theta^+,t) \frac{1+\cos\theta^+}{1+\cos\theta}\right]
\label{CMFt_nl} \quad .
\end{equation}

The linear integration can be performed as in \eqref{linear_int} with
$$
f(p) = -K \nu(t) -(I+1) { i} p
$$
where ${i}$ is the imaginary unit.

\subsubsection{The FPE}

The linear operator for the FPE can be written as
\begin{equation}
{\cal L}_{FP} R(\theta,t) = D(t) \left[ R(\theta,t) +
\frac{\partial^2 R(\theta,t)}{\partial \theta^2} \right]
 -(1+A(t))
\frac{\partial R(\theta,t)}{\partial \theta}
\label{FPE_lin}
\end{equation}
and the non-linear one as
\begin{eqnarray}
{\cal N}_{FP} R(\theta,t) &=& f(\theta,t) R(\theta,t) + g(\theta,t)
\frac{\partial R(\theta,t)}{\partial \theta} + h(\theta,t)
\frac{\partial^2 R(\theta,t)}{\partial \theta^2}
\label{FPE_nlin}
\\
f(\theta,t) &=&  \sin \theta (A(t)-1)-D(t)(2 \cos^2 \theta + \cos\theta) \\
g(\theta,t) &=& \cos \theta (1-A(t)) -3 D(t) \sin\theta (1+ \cos \theta) \\
h(\theta,t) &=& D(t) \cos \theta (2 + \cos \theta)
\end{eqnarray}

The term needed for the linear integration appearing  in \eqref{linear_int} has the following
expression
$$
f(p) = D(t)(1-p^2) -(1+A(t)) { i} p
$$

The extention to the third order is algebraically complicated,
but straightforward, we do no report here the expressions of the linear
and non-linear operators but they can be easily derived form Eqs. \eqref{FPEt} and \eqref{3rdt}.

\section{Synaptic Filtering}

In the paper we have always considered instantaneous synapses, however
a fundamental question is if the introduction of finite synaptic times
will alter the observed scenario. In particular, to generalize our findings we have considered a
network of QIF neurons with exponentially
decaying PSPs, this amounts to rewrite the model \eqref{eq:1} as follows
\begin{eqnarray}
\dot{v}_{i}(t) &=& v_{i}^2 + I - g S_i(t)  \quad i=1, \dots, N \nonumber \\
\tau_s \dot{S}_i(t) &=& -S_i(t) + \sum_{j=1}^{N} \sum_{n} \varepsilon_{ji} \delta(t - t_{j}^{(n)})
\label{eq:qif_exp}
\end{eqnarray}
where $S_i(t)$ represents the synaptic input current due to the linear super-position of the
exponential PSPs emitted
by the pre-synaptic neurons connected to neuron $i$ and $\tau_s$ is the synaptic decay time
of the PSP.

\begin{figure}
\includegraphics[width=0.45\textwidth]{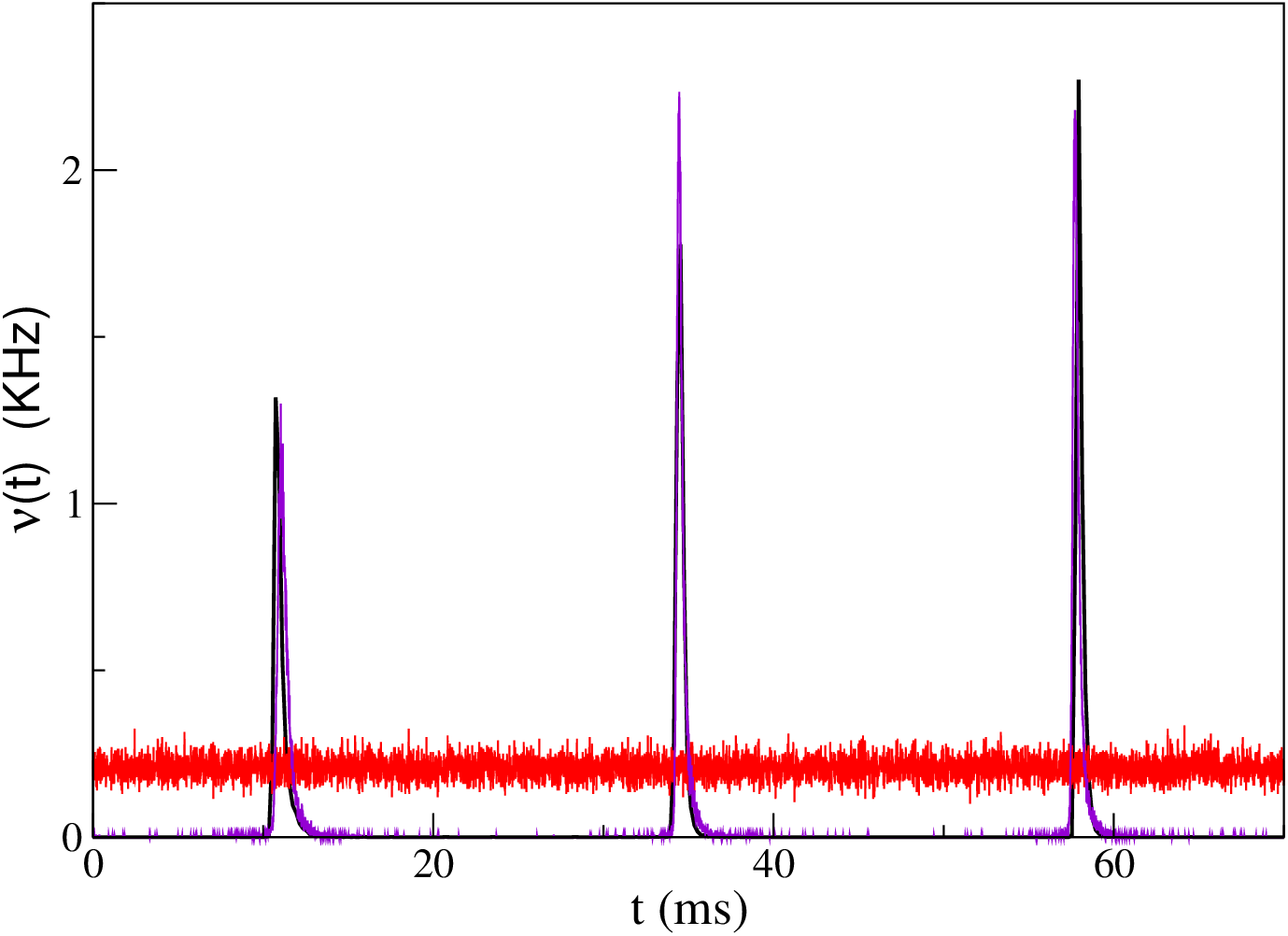}
\caption{Population firing rates $\nu(t)$ versus time for exponentially decaying PSPs.
The black line refers to network simulations with $N=20000$, while
the red (violet) line to Langevin simulations for diffusive (shot-) noise with $N=20000$ uncoupled neurons.
The data refer to QIF networks with exponentially decaying PSPs \eqref{eq:qif_exp}
with $K=200$, $i_0 = 0.16$, $g_0=4$, and $\tau_s = 0.1 \tau_m$.}
\label{figS3}
\end{figure}

We have considered the same case reported in Fig. 1 (b) in \cite{prl2024}, but this time
the PSPs have a finite duration, namely $\tau_s = 0.1 \tau_m$.
As shown in Fig. \ref{figS3}, analogously
to what found for instantaneous synapses the DA (red line)
fails in reproducing the network simulations (black line),
while the inclusion of shot-noise in the Langevin dynamics
allow to recover the correct evolution (violet line).

Therefore we can conclude that for finite $\tau_s$ we find the same scenario reported for instantaneous synapses, 
the detailed investigation of the role of the synaptic filtering is devoted to future analysis.

%%%%%%%%%%%%%%%%%%%%%%%%%%%%%%%%%%%%%%%%%%%%%%%%%%%%%%%%%%%%%%%%%%%%%%%%%%%%%%
%       References
%%%%%%%%%%%%%%%%%%%%%%%%%%%%%%%%%%%%%%%%%%%%%%%%%%%%%%%%%%%%%%%%%%%%%%%%%%%%%%

%%\bibliographystyle{apsrev-nourl}

%%\bibliographystyle{apsrev4-1}

%%\bibliography{lif_ost_cd}

\begin{thebibliography}{75}
\bibitem [{map()}]{maple}%
  \BibitemOpen
  \href@noop {} {}\bibinfo {note} {For numerical calculations in the Maple
  package, whenever needed the length of the digital mantissa of the variables
  was increased significantly beyond $16$ digits (environmental variable
  `Digits') in order to maintain this accuracy. We adopted the following
  technical criterion for accuracy: the elevations of the decimal mantissa
  length by $20$ and $M$ by a factor of $1.5$ induce the relative changes of
  the calculated quantity of interest below $10^{-12}$.}\BibitemShut {Stop}%
\end{thebibliography}

%merlin.mbs apsrev4-1.bst 2010-07-25 4.21a (PWD, AO, DPC) hacked
%Control: key (0)
%Control: author (72) initials jnrlst
%Control: editor formatted (1) identically to author
%Control: production of article title (-1) disabled
%Control: page (0) single
%Control: year (1) truncated
%Control: production of eprint (0) enabled
%

\end{document}